\documentclass[useAMS,usenatbib]{mn2e}

\usepackage{color}
\usepackage{graphicx}
\usepackage{ifpdf}
\usepackage{epstopdf}
\usepackage{url}
\usepackage{multirow}

\usepackage{subfig}

\usepackage{amssymb}
\usepackage{amsmath}
\let\oldAA\AA
\renewcommand{\AA}{\text{\normalfont\oldAA}}

\usepackage{natbib}

\title[Emission-line diagnostics of H\,II regions]{Emission-line diagnostics of nearby H\,II regions including interacting binary populations}
\author[Lin Xiao, E. R. Stanway \& J.J. Eldridge]{Lin Xiao$^{1}$\thanks{E-mail: lin.xiao@auckland.ac.nz}, Elizabeth R. Stanway$^{2}$\thanks{E-mail: e.r.stanway@warwick.ac.uk} and J.J. Eldridge$^{1}$\thanks{E-mail: j.eldridge@auckland.ac.nz}\\
$^{1}$Department of Physics, Private Bag 92019, University of Auckland, NZ,\\
$^{2}$Department of Physics, University of Warwick, UK}
\begin{document}

\pagerange{\pageref{firstpage}--\pageref{lastpage}} \pubyear{2017}
\maketitle

\label{firstpage}

\begin{abstract}
We present numerical models of the nebular emission from H\,II regions around young stellar populations over a range of compositions and ages. The synthetic stellar populations include both single stars and interacting binary stars. We compare these models to the observed emission lines of 254 H\,II regions of 13 nearby spiral galaxies and 21 dwarf galaxies drawn from archival data. The models are created using the combination of the Binary Population and Spectral Synthesis (\textsc{bpass}) code with the photoionization code \textsc{cloudy} to study the differences caused by the inclusion of interacting binary stars in the stellar population. We obtain agreement with the observed emission line ratios from the nearby star-forming regions and discuss the effect of binary star evolution pathways on the nebular ionization of H\,II regions. We find that at population ages above 10\,Myr, single-star models rapidly decrease in flux and ionization strength, while binary-star models still produce strong flux and high ${\rm [O\,III]}$/$ {\rm H\beta} $ ratios. Our models can reproduce the metallicity of H\,II regions from spiral galaxies but we find higher metallicities than previously estimated for the H\,II regions from dwarf galaxies. Comparing the equivalent width of $ {\rm H\beta} $ emission between models and observations, we find that accounting for ionizing photon leakage can affect age estimates for H\,II regions. When it is included, the typical age derived for H\,II regions is 5\,Myr from single-star models, and up to 10\,Myr with binary-star models. This is due to the existence of binary-star evolution pathways which produce more hot WR and helium stars at older ages. For future reference, we calculate new BPASS binary maximal starburst lines as a function of metallicity, and for the total model population, and present these in Appendix \ref{sec:maximal}.

\end{abstract}

\begin{keywords}
binaries: general  $ - $  H\,II regions $ - $ galaxies:general
\end{keywords}

\section{Introduction}
Classical H\,II regions are large, low-density clouds of fully ionized gas in which star formation has recently taken place ($ < $ 15\,Myr) \citep{2009A&A...507.1327H}. The short-lived OB stars forged in these regions emit ultraviolet light that ionizes the surrounding gas. Extragalactic H\,II regions have been studied observationally using narrow band imaging centred on the H$ \alpha $ emission line, in order to obtain their luminosity functions and radii \citep[e.g.][]{{1984ApJ...287..116K},{1990ARA&A..28..525S}}. Optical spectroscopy has yielded electron density and temperature, metallicity, and extinction \citep[e.g. ][]{{1985ApJS...57....1M},{1995AJ....109..990V},{1996ibms.conf..563V}, {2000ApJS..128..511O}, {2000ApJ...539..687O}}, while broad-band images allow study of their stellar population \citep[e.g.][]{2010A&A...521A..41G}.
 
Observing H\,II regions is a mature method to indirectly obtain information on the stellar ionizing continuum of hot stars by comparing observed emission properties (e.g. flux and equivalent width) to predictions from photoionization codes such as \textsc{cloudy} \citep{1998PASP..110..761F} or \textsc{mapping iii} \citep{2008ApJS..178...20A} when combined with irradiating spectra from stellar population synthesis codes. For example \cite{2010AJ....139..712L} used \textsc{starburst99} \citep{{1999ApJS..123....3L}, {2005ApJ...621..695V}} population synthesis models with the Mappings III photoionization code \citep{{1985ApJ...297..476B}, {1993ApJS...88..253S}} to produce model galaxy spectra that were compared to nearby (z $\leq$ 0.1) star-forming galaxies on a variety of optical emission-line ratio diagnostic diagrams. Adopting a continuous SFH they could reproduce metallicity-sensitive line ratios observed, including in a low-metallicity sample, but they also found the ionizing spectra produced by the \textsc{starburst99} models were not able to reproduce the observed [S\,{\sc II}] fluxes. An alternate approach was taken by \cite{2017ApJ...840...44B} who used the Flexible Stellar Population Synthesis (FSPS) code together with \textsc{cloudy} photoionization models and found they were able to match the properties of observed H\,II regions when stellar rotation was incorporated in the models.

Metallicity is a key parameter in the study of star-forming environments. The so-called `direct method' for determining metallicities involves measurement of temperature-sensitive line ratios such as [O\,III]$\lambda$4363/5007, [N\,II]$\lambda$5755/6584, [S\,II]$\lambda$6312/9532 \citep{2003ApJ...591..801K,2005A&A...441..981B,2007ApJ...656..186B,2006A&A...448..955I,2006A&A...454L.127S}). These provide an estimate of the electron temperature $ T_e $ of the nebular gas. The electron temperature can then be converted into a metallicity estimate using models of ionization structure in the region, since metal-poor stars are typically hotter, with a harder spectrum to irradiate the gas. However, when these weak lines are not available, a method involving the ratios of strong emission lines must instead be used as an indirect metallicity indicator. Various calibrations, using a variety of optical emission-line ratio diagnostics, have been presented and employed to determine the metallicity of H\,II regions and their host galaxies. These are often empirically calibrated, based on robust direct measurements of a subset of H\,II regions/galaxies, but can also be model-based, combining stellar population synthesis and photoionization models to predict line ratios \citep[e.g.][]{2002ApJS..142...35K}. These techniques are applicable at all metallicities, with the possible exception of temperature-sensitive methods since these involve intrinsically faint emission lines which are further weakened at low metallicities. However, as discussed in \cite{2008ApJ...681.1183K}, strong line ratio methods can show large systematic differences in their metallicity estimates, which translate into a considerable uncertainty in the absolute metallicity scale. 

An important determining factor in the properties of a star-forming galaxy or H\,II region is the nature of the young stellar population creating it. Multiple studies both from theoretical \citep[e.g.,][]{1999ApJS..123....3L} and observational work \citep[e.g.,][]{2015A&A...574A..47S,2012A&A...546A...2S} indicate that the equivalent widths of the strongest hydrogen recombination lines - $ {\rm H\alpha} $ or $ {\rm H\beta} $ - are excellent indicators of a young stellar population's age. \cite{2003MNRAS.340...29C} verified from observations that EW($ H\beta $) does indeed decrease with starburst age. \cite{2013AJ....146...30K} further extended these age diagnostics to more complex star formation histories (SFHs). They conclude that even when modelling multiple starbursts, EW($ H\beta $) serves as a good age indicator for the youngest burst, up to $ \sim $ 10\,Myr after the onset of star formation. 

Since the ionization of a H\,II region depends on the stellar population irradiating it, it is critical to examine the effects of stellar evolution pathways on the ionizing spectrum. \cite{2009MNRAS.400.1019E} began to investigate the affects of binary evolution on the emitted spectra of stellar populations in their \textsc{BPASS} models, showing that interacting binaries cause a population of stars to appear bluer at an older age than predicted by single-star models. At low metallicities interacting binary stars produce a modified ionizing photon output and may well help to reproduce the observed high [O\,{\sc III}]/H$ {\rm \beta} $ emission line ratios in the sub-solar, low-mass star-forming galaxies of the distant Universe \citep{2014MNRAS.444.3466S}, as well as results seen in more local H\,II regions \citep{2015MNRAS.447L..21Z}. As \cite{2016MNRAS.456..485S} demonstrated, the predicted ionizing flux from distant star-forming galaxies depends sensitively on their star formation history and metallicity. However, this work was not fully calibrated and tested against local star formation populations. 

In this paper, we investigate the nebular emission of nearby H\,II regions in light of these theoretical stellar populations that include interacting binaries. Our main aim is to characterize their emission lines and reveal the connection between the ionization conditions and the properties of the underlying stellar population. In this work we present new models of the optical emission from nebular gas in theoretical H\,II regions at a range of metallicities, and include irradiation by interacting binaries, investigating how these affect the predicted ionizing flux. We validate these against well-established local samples of H\,II regions, exploring how their interpretation differs given the new models. The paper is organized as follows: in Section 2, we describe the characteristics of the observations and compilation of the sample; in Section 3 we describe the methods used to create our synthetic spectra and summarize the effect of binary interactions; in Section 4, comparison between our models and the observational sample are made in BPT diagrams and best-fitting models are presented; in Section 5, we discuss the effect of ionizing photon leakage due to low gas density and dust attenuation and derive how many ionizing photons are leaked; finally in Section 6, we summarize and give our conclusions.\\

\begin{table}
\caption{H\,II region host galaxies from the van Zee sample of \citet{1998AJ....116.2805V} and \citet{2006ApJ...636..214V}. We quote the properties of the 13 spiral galaxies and 21 dwarf galaxies. Absolute B-band magnitude is given in the Vega system. The aperture size is derived from the 2\arcsec\ slit width in the observations and at the distance listed in the Table.}

\begin{center}
                      \begin{tabular}{l @{\hskip 0.05in} | @{\hskip 0.05in} c @{\hskip 0.05in} | @{\hskip 0.05in} c @{\hskip 0.05in} | @{\hskip 0.05in} c @{\hskip 0.05in} | @{\hskip 0.05in} c}
                        \hline
                        \hline
                        \rule{0pt}{3.ex}
                        Galaxy & Distance & Slit Width  & $ M_{B} $ & Number of \\
                        - spiral & (Mpc) & (pc) & & H\,II Regions\\
                        
                        \hline 
                         
                       NGC 0628 & 9.7 & 94.1 & -20.1 & 19 \rule{0pt}{2ex}\\ 
                       NGC 0925 & 9.29 & 90.1 & -19.8 & 44 \rule{0pt}{2ex}\\ 
                       NGC 1068 & 14.4 & 139.6 &-21.3 & 1 \rule{0pt}{2ex}\\ 
                       NGC 1232 & 21.5 & 208.5 & -21.2 & 16 \rule{0pt}{2ex}\\ 
                       NGC 1637 & 8.6 & 83.4 & -18.5 & 16 \rule{0pt}{2ex}\\ 
                       NGC 2403 & 3.25 & 31.5 &-19.1 & 17 \rule{0pt}{2ex}\\ 
                       NGC 2805 & 23.5 & 227.9 & -20.7 & 17 \rule{0pt}{2ex}\\ 
                       IC 2458 & 23.5& 227.9 &-16.9 & 3 \rule{0pt}{2ex}\\ 
                       NGC 2820 & 23.5 & 227.9 &-20.1 & 4 \rule{0pt}{2ex}\\     
                       NGC 2903 & 6.3 & 61.1 & -19.9 & 9 \rule{0pt}{2ex}\\      
                       NGC 3184 & 8.7 & 84.4 &-19.3 & 17 \rule{0pt}{2ex}\\ 
                       NGC 4395 & 4.5 & 43.6 & -17.7 & 12 \rule{0pt}{2ex}\\         
                       NGC 5457 & 7.4 & 71.8 &-21.2 & 13 \rule{0pt}{2ex}\\
                        \hline
                        \rule{0pt}{3.4ex}
                        Galaxy & Distance & Slit Width  & $ M_{B} $ & Number of \\
                        - dwarf & (Mpc) & (pc) & & H\,II Regions\\
                        \hline 
                        UGC 12894 & 7.9 & 76.6 & -13.38 & 3 \rule{0pt}{2ex}\\ 
                        UGC 290 & 12.7 & 123.1 & -14.48 & 2 \rule{0pt}{2ex}\\ 
                        UGC 685 & 4.79 & 46.5 &-14.44 & 4 \rule{0pt}{2ex}\\ 
                        UGC 1104 & 11.1 & 107.6 &-16.08 & 3 \rule{0pt}{2ex}\\ 
                        UGC 1175 & 11.3 & 109.6 & -14.13 & 3 \rule{0pt}{2ex}\\ 
                        UGC 1281 & 4.6 & 44.6 &-14.91 & 1 \rule{0pt}{2ex}\\ 
                        HKK L14 & 4.7 & 45.6 &-11.26 & 1 \rule{0pt}{2ex}\\ 
                        UGC 2023 & 10.2 & 98.9 &-16.54 & 9 \rule{0pt}{2ex}\\
                        UGC 3647 & 19.7 & 191.0 &-17.06 & 3 \rule{0pt}{2ex}\\
                        UGC 3672 & 12.7 & 123.1 &-15.43 & 1 \rule{0pt}{2ex}\\
                        UGC 4117 & 10 & 97.0 &-14.86 & 3 \rule{0pt}{2ex}\\
                        UGC 4483 & 3.21 & 31.1 & -12.55 & 2 \rule{0pt}{2ex}\\
                        CGCG 007-025 & 172.6 & 17.8 & -15.75 & 3 \rule{0pt}{2ex}\\
                        UGC 5288 & 5.3 & 51.4 &-14.44 & 6 \rule{0pt}{2ex}\\
                        UGCA 292 & 3.1 & 30.1 &-11.43 & 4 \rule{0pt}{2ex}\\
                        UGC 8651 & 3.01 & 29.2 &-12.96 & 1 \rule{0pt}{2ex}\\
                        UGC 9240 & 2.79 & 27.1 &-13.96 & 1 \rule{0pt}{2ex}\\
                        UGC 9992 & 8.6 & 83.4 & -14.97 & 3 \rule{0pt}{2ex}\\
                        UGC 10445 & 15.1 & 164.4 & -17.53 & 7 \rule{0pt}{2ex}\\
                        UGC 11755 & 18 & 174.5 &-17.14 & 2 \rule{0pt}{2ex}\\
                        UGC 12713 & 7.5 & 72.7 &-14.76 & 4 \\
                        \hline
 
\end{tabular}
\end{center}
\label{tab:vanZee_HIIs}
\end{table}

\begin{figure*}
\centering
\hspace*{-0.2cm}\includegraphics[width=18cm]{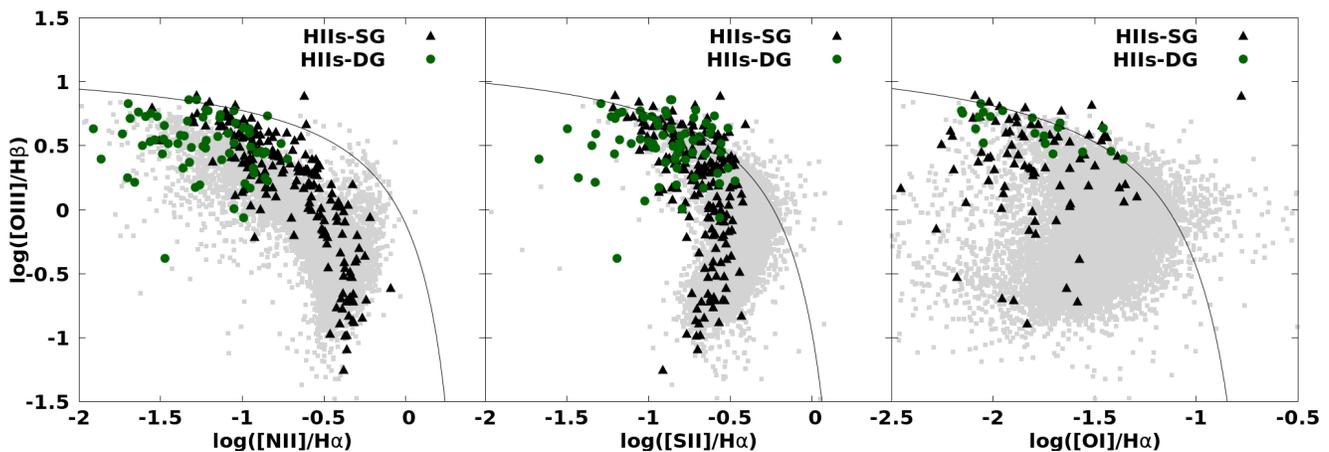}
\centering
\caption{Distribution of the van Zee sample H\,II regions in diagnostic BPT diagrams (coloured points) compared to the star-forming galaxies from the SDSS survey (shown in grey). The H\,II regions from spiral galaxies are represented by black triangles and those from dwarf galaxies by green circles. For consistency with the H\,II region sample, we cut off the SDSS galaxies at a H$\beta$ equivalent width of ${\rm 2\AA}$.} \label{fig:BPT}
\end{figure*}
\begin{figure*}
\centering
\hspace*{-0.2cm}\includegraphics[width=18cm]{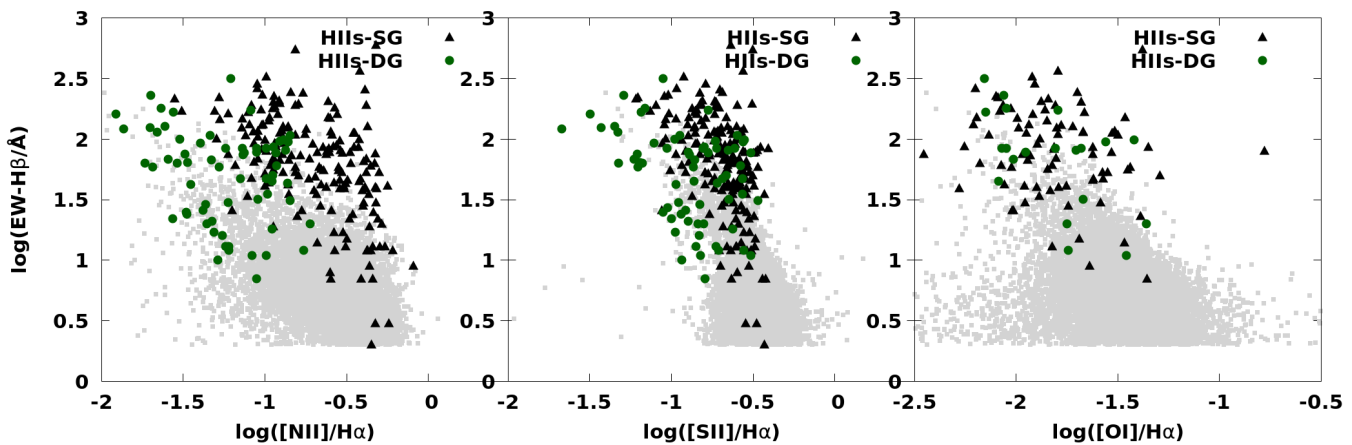}
\centering
\caption{As in figure \ref{fig:BPT} but now showing diagnostic line ratios against H$\beta$ equivalent width.} \label{fig:BPT_EW}
\end{figure*}


\section{Sample of Galaxies and Dataset Analysis}\label{sec:vanzee}
Observations of H\,II regions were selected from the optical imaging and spectroscopic samples of \cite{1998AJ....116.2805V} and \cite{2006ApJ...636..214V}, known collectively as `the Van Zee sample' hereafter. There are, in total, 254 H\,II regions, with 188 from 13 spiral galaxies and 66 from 21 dwarf galaxies, as listed in Table \ref{tab:vanZee_HIIs}. Details of the observations can be found in the source articles. Briefly, the spiral galaxies were presented in \cite{1998AJ....116.2805V}. H\,II regions were selected from optical imaging showing excess flux in a H$\alpha$ narrow band filter. Low resolution spectroscopy was obtained at the 5\,m Palomar telescope using a 2\arcmin $\times$ 2\arcsec\ slit, with individual slits centred on each HII region. The seeing was typically below 2\arcsec. Data were flux calibrated against standard star fields, and the continuum level in the spectroscopy checked against photometry. The data were corrected by the original authors for internal extinction using the observed Balmer line ratio. The data for dwarf galaxy H\,II regions was presented in  \cite{2006ApJ...636..214V}. The regions were selected and spectra obtained in an identical manner to the spiral galaxy study. The spectra for individual H\,II regions were extracted using an aperture extending to 10 per cent of the peak flux in the case of faint regions, or 3\arcsec\ in the case of bright regions which are typically more confused. Again, the catalogue data were extinction-corrected using the Balmer line ratios. The extinction was typically low, with $E(B-V)<0.2$ in most cases.

In any analysis of H\,II regions, it is important to consider whether the entire H\,II region has been extracted, whether more than one region may be represented and the effects of any aperture losses.  In the absence of integral field spectroscopy, which allows estimates of aperture losses on the parameters of entire galaxies  \citep[e.g.][]{2017A&A...599A..71D,2013A&A...553L...7I,2016ApJ...826...71I,2016A&A...586A..22G}, a limiting factor is the combination of a spectral aperture with the seeing at the time of observation. 
In Table \ref{tab:vanZee_HIIs} we provide an estimate of the projected physical aperture size of the 2\arcsec\ wide slit at the distance of each galaxy. These span from 20\,pc in the nearest example (contributing just three H\,II regions) to 230\,pc in the most distant sources. However these extreme cases are relatively rare in the sample. The mean projected slit width was 106\,pc for the spiral galaxy H\,II regions and 90\,pc for those in dwarf galaxies. The distribution of H\,II regions is biased to compact sources, and is flat in number density with size below 130\,pc before falling away rapidly with a power law slope of -4 \citep{2003AJ....126.2317O}. In the bulk of the sources targeted therefore, the slit width is well matched to the size of a typical H\,II region, while the spectral extraction aperture encompasses the whole region by construction.  In a few cases, the wide slit will allow contamination of the H\,II region spectrum by emission from diffuse interstellar gas which can act to bias the line measurements towards a low electron density regime.  In a few others, the slit width will result in the outskirts of the HII region being under-represented relative to the core (i.e. with the low density wings excluded in one spatial direction but not the other). The effects of this will be to push the line ratios towards a higher temperature, denser gas regime.  However, in all cases, the slit width is a reasonable match to the observational seeing. \cite{1998AJ....116.2805V} and \cite{2006ApJ...636..214V} demonstrated that their observational data provided a good measurement of individual H\,II region spectra, which can be well fit by photoionization models and we use their spectroscopic flux ratio catalogues without further correction.

In Fig \ref{fig:BPT} we show these data overplotted on three diagnostic diagrams defined by strong optical lines ratios, $ {\rm [O\,{\sc III}]\lambda5007/H\beta} $, $ {\rm [N\,{\sc II}]\lambda6584/H\alpha} $, $ {\rm [S\,{\sc II}]\lambda6724/H\alpha} $ and $ {\rm [O\,{\sc I}]\lambda6300/H\alpha} $, which are widely used in the literature to interpret observations of nebular emission \citep[commonly referred to as  the BPT diagrams,][]{1981PASP...93..817B}. In these diagrams, both individual H\,II regions and entire starburst galaxies (which include multiple populations and diffuse interstellar emission) occupy the lower left, while hard irradiation sources are required to move a region towards the top left. The curved lines indicate the position of the maximal starburst line predicted by \cite{2001ApJ...556..121K} for photoionization by young stellar populations. Sources lying above these lines either have a harder ionization source (e.g. accretion-powered X-ray binaries or Active Galactic Nuclei, AGN) or stellar populations not present in their model. In order to select the star-forming galaxies from the SDSS survey, a simple cut off on the H$\beta$ equivalent width at 2\,\AA\ is applied, which identifies sources with significant young stellar populations and matches the lowest values in the observed HII region catalogues used for comparison.

The H\,II regions from spiral galaxies (HIIs-SG) in the van Zee sample form a sequence close to these maximal lines and overlap the parameter space occupied by star-forming galaxies in the Sloan Digital Sky Survey (SDSS). However, H\,II regions from dwarf galaxies (HIIs-DG) mostly appear on the upper-left edges close to and beyond the \cite{2001ApJ...556..121K} lines. These regions are typically more metal poor than those in spiral galaxies (as determined through a number of diagnostics). At low metallicities, the typical radiation field is harder \citep{2014ApJ...795..165S,2014MNRAS.444.3466S}, raising the electron temperatures and thus boosting high ionization lines (such as [O\,III]) relative to those with lower potentials (such as [O\,I] or [S\,II]). 
           
We also show $ { \rm H\beta } $ equivalent width diagnostic diagrams in Figure \ref{fig:BPT_EW}. In these panels the $ {\rm H\beta} $ equivalent width is substituted for the emission line ratio of $ {\rm [O\,{\sc III}]\lambda5007/H\beta}$. We only include SDSS galaxies that have an equivalent width greater than ${\rm 2\AA} $. The H$\beta$ equivalent width is known to be a good proxy for the age of the youngest stellar components. Compared to galaxies with a similar average line ratio, the individual H\,II regions have a higher $ {\rm H\beta} $ equivalent width. At a given $ {\rm EW[H\beta]} $, the emission line ratio of ${\rm [N\,{\sc II}]\lambda6584/H\alpha}$ from HII-SGs is also higher than the ratios observed in HIIs-DG. This is because the [N\,{\sc II}] emission line is a low-ionization emission line that is more sensitive to metallicity than temperature, so there are stronger [N\,{\sc II}] lines in a higher-metallicity region. H\,II regions from spiral galaxies and dwarf galaxies have similar distributions in diagnostic diagrams using the ${\rm [S\,{\sc II}]/H\alpha}$ and ${\rm [O\,{\sc I}]/H\alpha}$ ratios.  \\


\section{Method}

\subsection{Binary population and spectral synthesis}

The synthetic spectra used in this paper are created using the Binary Population and Spectral Synthesis (BPASS) code\footnote{http://www.bpass.auckland.ac.nz}. Here we use the recent \textsc{bpass v2.1} models as described in \cite{2016MNRAS.456..485S} and \cite{arXiv:1710.02154}. In brief, the BPASS models are a set of publicly available stellar population synthesis models which are constructed by combining both single and binary stellar evolution models with the latest synthetic stellar atmosphere models. These BPASS models incorporate the evolutionary effects of mass transfer between members of a binary, and thus as well as the stellar initial mass function also require distributions in initial binary mass ratio and period, as detailed in \citep{arXiv:1710.02154}. These are tuned to reproduce the observed binary fractions in young, massive stellar populations in the local Universe. 

We consider stellar population models at 13 different metallicity mass fractions from Z = 0.00001 to 0.040 \citep[where we consider the metallicity of Z = 0.020 to be `Solar', after][for consistency with our empirical mass-loss rates which were originally scaled from this value -- see discussion in Eldridge, Stanway et al 2017]{1993oee..conf...15G}. For each metallicity we calculate the hydrogen mass fraction, X = 0.75 - 2.5Z, helium mass fraction, Y = 0.25 + 1.5Z. We use a scaled solar composition as detailed below. We treat the models evolving as a single instantaneous starburst with age varying from 1\,Myr to 10\,Gyr, in logarithmic time steps of 0.1 dex. 

We aim to investigate the effect of binary interactions on the properties of their corresponding nebular spectra. Binary interactions can prevent certain pathways and introduce others, such as preventing certain stars from becoming red giants by removing their hydrogen envelopes. These stars instead become helium-rich dwarfs and evolve with higher surface temperatures. At the same time their companion stars accrete material or the two stars may merge, in both cases creating a star that is more massive than it was initially. This upsets the simple relationship between the most massive or luminous star and the age of the population it is a member of. Adding to this complexity are the new evolutionary pathways introduced which disturb the relationship between the stellar population properties and nebular radiation.

\subsection{Model Nebular Emission of H\,II Regions}

Our nebular spectra are modelled using the predicted SED from BPASS models as a source flux for the photoionization code \textsc{cloudy 13.03} \citep{{1998PASP..110..761F},{2013RMxAA..49..137F}}. This makes our gas prescription flexible and accurate, and allows us to vary the physical conditions of the gas and evaluate their effects on spectral evolution. The result is a self-consistent prediction for the nebular continuum and emission line flux which reflects the temporal changes in the ionizing radiation field.

Throughout this paper, we will assume the gas nebula is without dust, spherical, ionization-bound and has a constant, non-evolving constant hydrogen density $ {\rm n(H)} $. We also assume that the nebular metallicity matches that of the irradiating stellar population. This is a highly ideal model but is a first approximation and thus useful to investigate the nebular emission. These assumptions affect the properties of the nebular gas and their impact on our conclusions, and are further discussed in Section 4. The main parameters of the photoionized gas and the range we create models over are as followed:

\textbf{Nebular Metallicity, Z}: we consider 13 values of $ Z $ between 0.05 per cent of solar ($ Z $ = 0.00001) and twice solar ($ Z $ = 0.040), corresponding to metallicities at which stellar population models are available. We scale the 9 basic elements' fraction according to the population metallicity $ Z $ using the opacity tables of \cite{2004MNRAS.348..201E} as shown in Table \ref{tab:Metallicity}. In our models we consider $ Z $ = 0.020 as Solar metallicity, while the unscaled Solar metallicity in a default CLOUDY model is closer to $ Z $ = 0.014 in our models. In addition, we assume a fixed N/O proportion in different metallicity models as shown in Table \ref{tab:Metallicity}. We note that there is strong evidence for evolution in the N/O ratio with either Fe/H or O/H. While this has a negligible effect on the stellar evolution models, it may be important in the nebular gas and this can lead to an offset in any diagnostic diagram involving N species.  See \citet{arXiv:1710.02154} for further discussion of this point. Given that a number of prescriptions exist for how N abundance varies, and that this may well depend on physical conditions, we do not attempt to implement a modified N/O abundance ratio prescription in this work, but rather make a comparison between single and binary stellar abundance ratios at fixed N/O \citep[for further discussion of this point, including using BPASS models, see e.g.][]{2017ApJ...836..164S}. 

\textbf{Ionization parameter at the Str$\mathbf{ \ddot{o}}$mgren radius, $ U $}: we compute models for 21 values of U logarithmically spaced in the range from $ -3.5 $ to $ -1.5 $, consistent with the observed range of ionization parameters in local starburst galaxies \citep{2004ApJ...606..237R}, in steps of 0.1 dex. 

\textbf{Hydrogen gas density, $ \mathbf{n_{H}}$}: we compute models for 7 hydrogen densities of the ionized gas, spanning logarithmically from 0 to 3 with 0.5 dex intervals. These span most of the range of the observed electron densities in extragalactic H\,II regions \citep{2009A&A...507.1327H}.

\textbf{ Stellar population age:} we use the BPASS population models with $ \log({\rm Age/yr}) = $ 6 to 8 in 0.1 dex step, leading to 21 independent ages.

These adjustable model parameters lead to a total of 80262 photoionization models, including separate models for single and binary star populations. 

\begin{table}
\caption{The metallicity steps and the mass fraction of the main elements used in the stellar evolution code.}
\begin{center}
                      \begin{tabular}{l @{\hskip 0.05in} | @{\hskip 0.05in} c @{\hskip 0.05in} | @{\hskip 0.05in} c @{\hskip 0.05in}| c}
                      \hline
                        \hline 
                          Identifier & \multicolumn{2}{c|}{Metallicity mass fraction} & Oxygen Abundance  \rule{0pt}{2.4ex}\\
                              & \multicolumn{2}{c|}{Z} & $ 12 + \log({\rm O/H}) $ \\
                       \hline                     
                       zem5 & \multicolumn{2}{c|}{$ 10^{-5}$} & 5.60  \rule{0pt}{2.4ex}\\
                       zem4 & \multicolumn{2}{c|}{$ 10^{-4}$} & 6.60  \rule{0pt}{2.4ex}\\
                       z001 & \multicolumn{2}{c|}{$ 0.001$} & 7.61    \rule{0pt}{2.4ex}\\
                       z002 & \multicolumn{2}{c|}{$ 0.002$} & 7.91    \rule{0pt}{2.4ex}\\
                       z003 & \multicolumn{2}{c|}{$ 0.003$} & 8.09    \rule{0pt}{2.4ex}\\
                       z004 & \multicolumn{2}{c|}{$ 0.004$} & 8.21    \rule{0pt}{2.4ex}\\
                       z006 & \multicolumn{2}{c|}{$ 0.006$} & 8.39    \rule{0pt}{2.4ex}\\
                       z008 & \multicolumn{2}{c|}{$ 0.008$} & 8.52    \rule{0pt}{2.4ex}\\
                       z010 & \multicolumn{2}{c|}{$ 0.010$} & 8.62    \rule{0pt}{2.4ex}\\
                       z014 & \multicolumn{2}{c|}{$ 0.014$} & 8.77    \rule{0pt}{2.4ex}\\
                       z020 & \multicolumn{2}{c|}{$ 0.020$} & 8.93    \rule{0pt}{2.4ex}\\
                       z030 & \multicolumn{2}{c|}{$ 0.030$} & 9.13    \rule{0pt}{2.4ex}\\
                       z040 & \multicolumn{2}{c|}{$ 0.040$} & 9.27     \rule{0pt}{2.4ex}\\
                        \hline  
                        \multicolumn{4}{c}{} \rule{0pt}{3.4ex}\\

                        \hline
                        \hline
                        \multicolumn{4}{c}{Hydrogen \& Helium} \rule{0pt}{2.4ex}\\
                        \hline
                       H & $ 0.75-2.5 \times Z $ & He & $ 0.25+1.5 \times Z $ \rule{0pt}{2.4ex}\\
                       \hline
                       \multicolumn{4}{c}{Metals} \rule{0pt}{2.4ex}\\
                       \hline                       
                       C & $ 0.173 \times Z $ & O & $ 0.482 \times Z $ \rule{0pt}{2.4ex}\\
                       N & $ 0.053 \times Z $ & Ne & $ 0.099 \times Z $ \rule{0pt}{2.4ex}\\
                       Mg & $ 0.038 \times Z $ & Si & $ 0.083 \times Z $ \rule{0pt}{2.4ex}\\
                       Fe & $ 0.072 \times Z $ & & \rule{0pt}{3.4ex}\\
                        \hline
 
\end{tabular}
\end{center}
\label{tab:Metallicity}
\end{table}

We note that rather than specifying $ U $ we could have specified a constant inner radius of the cloud. In our model the radius of photoionization models is set by specifying their ionization parameter $ U $, which is a function of $ Q(H) $, the number of ionizing photons per unit time, and $ R $, the distance to the ionization source:

\begin{eqnarray}
   U = \dfrac{Q({\rm H})}{4\pi R^{2}n({\rm H})c} 
\end{eqnarray}
Therefore the inner radius $ R $ of our models is determined by the radiation strength of the inner ionizing source and hydrogen density of the gas cloud.

An important caveat is that we omit dust from our CLOUDY model, as assuming a uniform dust geometry the \textsc{cloudy} output line emission and continuum will be suppressed equally leaving the line equivalent widths unaffected. However, without the attenuation by dust our models are relatively brighter than real stellar populations especially for younger populations ($ < 10 $\,Myr) that are strongly attenuated by their birth cloud as discussed by \cite{2000ApJ...539..718C} and \cite{2008ApJ...679.1192C}. By not including dust we make our models simpler allowing us to study the effects on the nebula emission alone without having to consider the increased complexity that varying dust properties and their parameters would require. The effect of dust may also lead to absorption of ionizing photons, alter the relation between electron density and hydrogen density and possibly deplete metals in the gaseous phase \citep[e.g.][]{2017PASP..129h2001P,1948ApJ...107....6S}.

For simplicity and to avoid proliferation of free parameters relating to dust grains and their composition, we assume these effects apply equally to all photoionization models, and so will not affect our conclusions regarding relative metallicities, stellar population age, and the differences between irradiating spectra in individual regions. Nonetheless we note that this is likely over-simiplistic. Dust is a very important component of H\,II regions and predictions of its effects will be included in further work.

\subsection{Ionizing Properties of the Model Spectra}
The observed emission from nebular gas is closely related to the evolution of a stellar population, since it is powered by the total ionizing spectrum that results from that population. A primary factor is the number of ionizing photons emitted by the stars shortwards of the 912\AA \,Lyman limit. It is these photons that interact with the nebular gas. The number of such photons varies with the parameters of the stellar population such as age, metallicity, number of interacting binaries and stellar rotation. Here we examine the impact of the first three. We do not consider rotation, which has been investigated by other authors such as \cite{2017ApJ...840...44B}, who calculate the ionizing spectra from stellar populations in which stars are rotating at a constant fraction of 40 per cent of their break-up speed. 

\subsubsection{Time Evolution of the Ionizing Spectra}
\begin{figure*}
\centering
\subfloat[][]{\includegraphics[width=17cm]{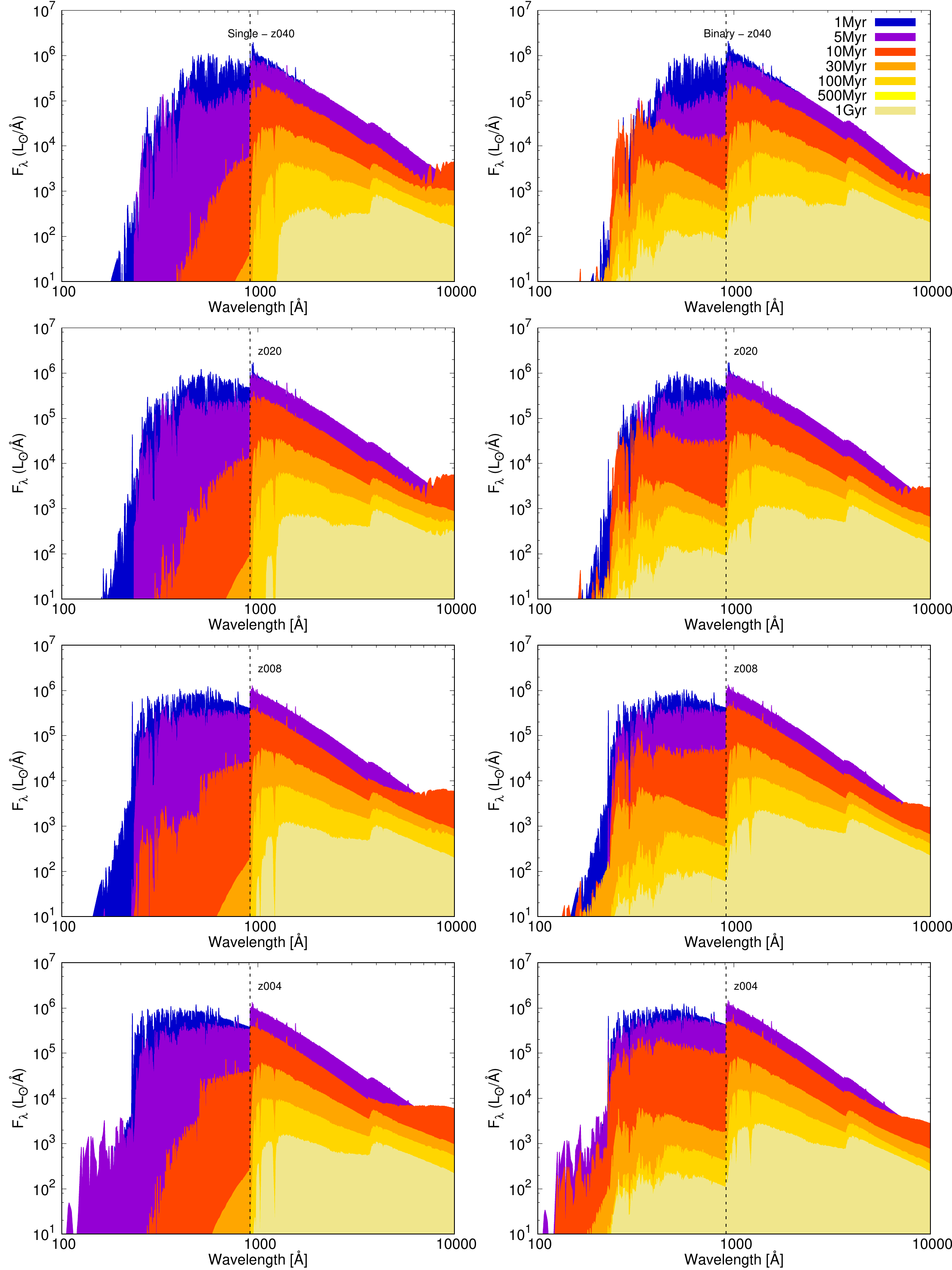}}
\centering
\caption{The flux contribution from different aged populations for both single-star (left) and binary-star (right) models at different metallicities. Only flux blueward of 912\AA\ shown by the dashed line is able to ionize hydrogen. \emph{(cont.)}} 
\end{figure*}

\begin{figure*}
\ContinuedFloat
\centering
\subfloat[][]{\includegraphics[width=17cm]{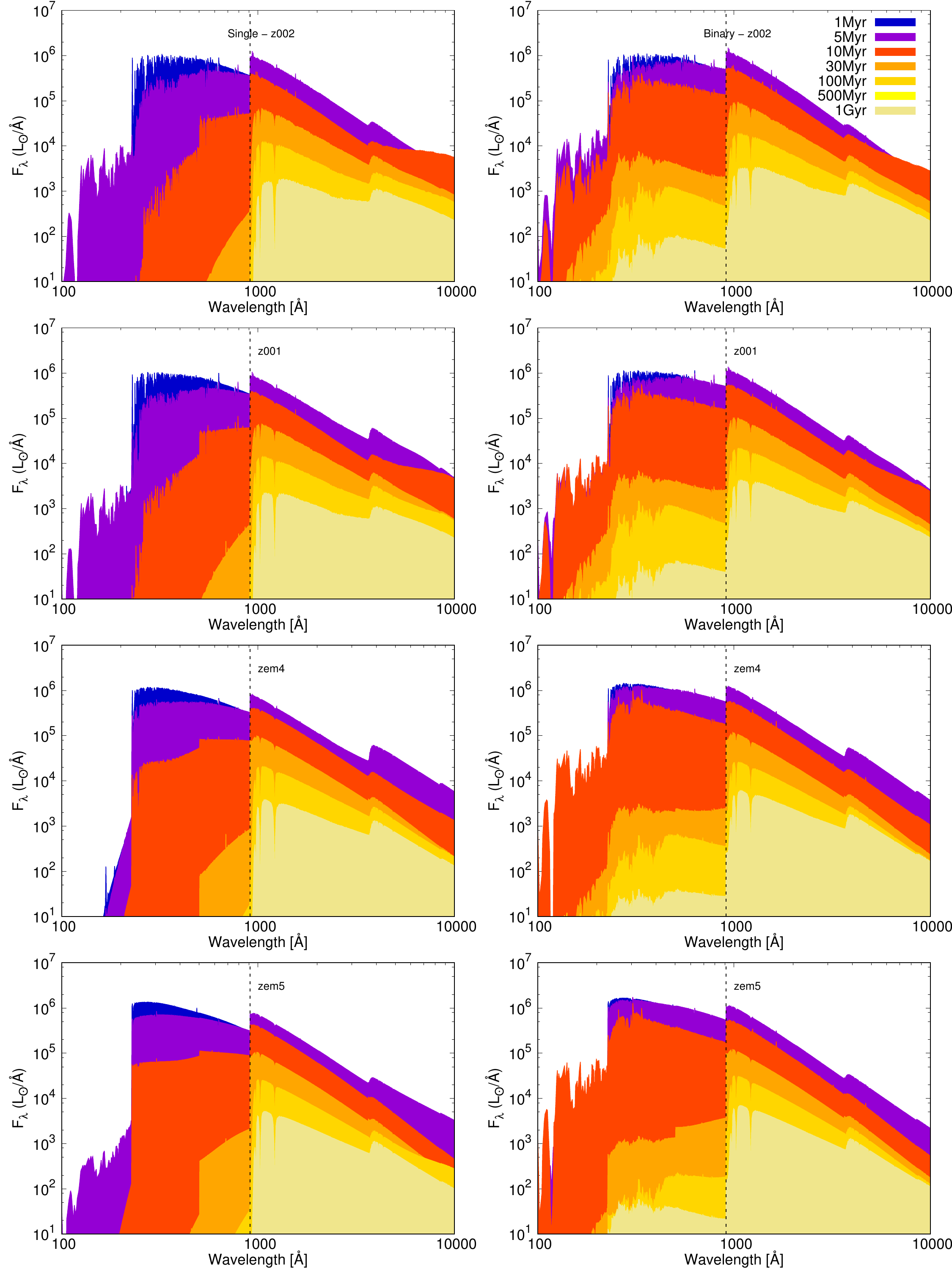}}
\centering
\caption{The flux contribution from different aged populations for both single-star (left) and binary-star (right) models at different metallicities. Only flux blueward of 912\AA\, shown by the dashed line is able to ionize hydrogen. } \label{fig:spec_sb}
\end{figure*}

\begin{figure}
\centering
\hspace*{-0.5cm}\includegraphics[width=9cm]{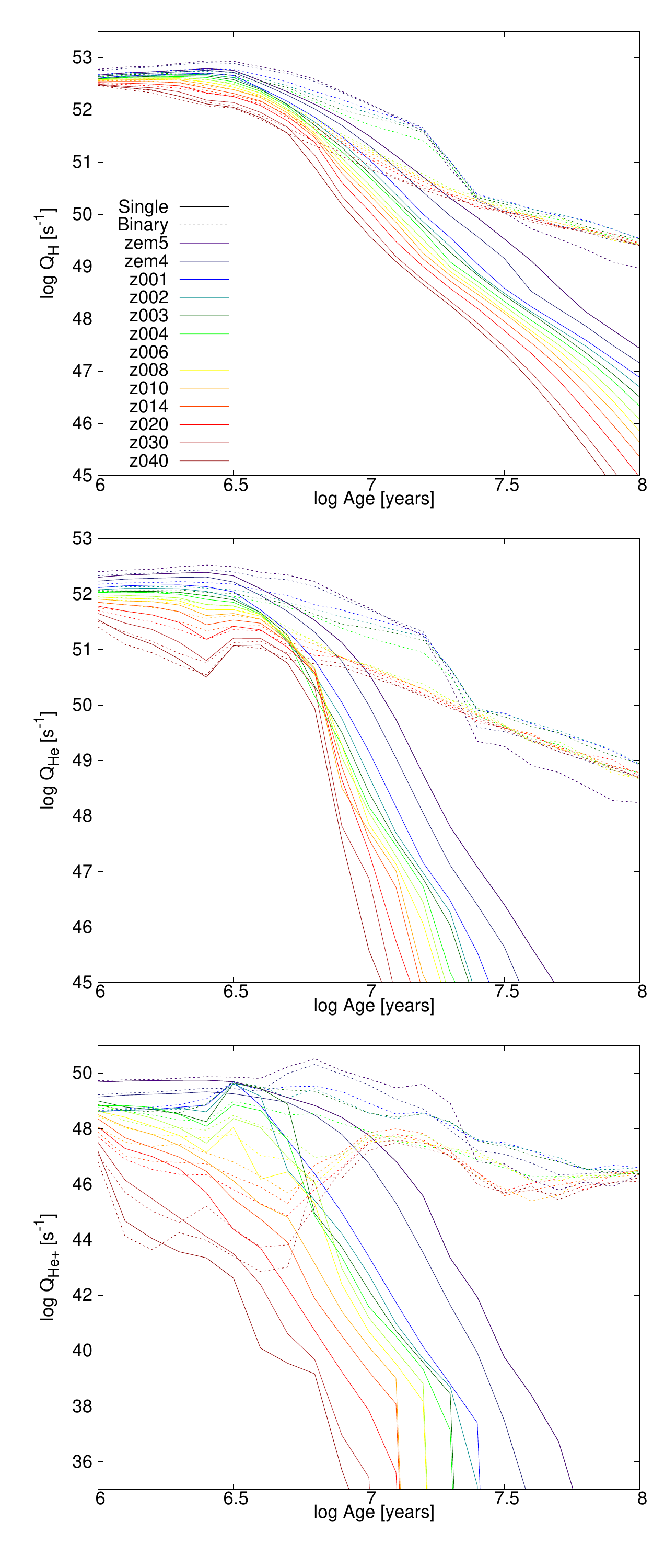}
\centering
\caption{The time evolution of the ionizing photon production rate at for H, He and He+, at different metallicities. The dashed lines indivate the binary-star populations and the solid lines are for the single-star populations. The metallicity increases from $ Z =10^{-5} $ (zem5) to $ Z =0.040 $ (z040) as the colour coding changes from blue to red.} \label{fig:QH_number}
\end{figure}

In HII regions the evolution of the ionizing spectra from the stellar populations is dominated by the most luminous, hottest and thus most massive stars in the population. Their efffective temperatures determine how much of their flux is emitted as ionizing photons. Figure \ref{fig:spec_sb} shows the spectral energy distribution (SED), including the ionizing spectrum, for our model starbursts over different ages. When considering the flux shortwards of 912\AA\, there are clear differences with metallicity. At lower metallicities the stars are hotter due to reduced opacities. There is a noticeable difference between the single-star and binary-star populations. For single-star populations there is more evolution of effective temperature of the hottest stars as less of the total flux is emitted at later ages. While for binary-star populations the primary change is in the luminosity of the stellar population, with binary interactions leading to on average hotter stars than in the single-star population. The most important factors determining the evolution therefore of a stellar population are the mass-loss history and metallicity of the stars that make up the population.

Firstly, mass loss by strong stellar winds and/or binary interaction can prevent massive stars cooling to become red supergiants during post-main sequence evolution. Instead they become hot Wolf-Rayet (WR) or helium stars and power a H\,II region at late times. The difference between single-star and binary-star population is the smallest in the youngest populations. In both cases stellar winds are strong enough to ensure that these hot stars exist even without binary interactions. It is the older single-star populations beyond 10\,Myr that lack such hot stars and so their ionizing flux almost disappears. A similar pattern occurs even if stellar rotation is included for the single-star population, as shown by \cite{2017ApJ...840...44B}. 

The shape of the ionizing spectrum gives us the distribution of ionizing photon energies. We quantify the total number of H, He and He+ ionizing photons as a function of age and metallicity from both single and binary stellar populations respectively in Figure \ref{fig:QH_number}. The youngest stellar populations produce the most hydrogen ionizing photons. As the population ages, less-massive and less luminous stars dominate the SED, decreasing the overall ionizing photon rate. 

In the first 3\,Myr, both single and binary populations have most of their massive stars on the main sequence and produce similar number of ionizing photons giving us a plateau. After that the ionizing photon number decays with time. There are noticeable departures between single and binary models after 3\,Myr. Single models undergo a steeper decay as its massive stars evolve off the main sequence, but binary models have a shallower drop-off and can produce about an order of magnitude more ionizing photons than single models at ages $>$10\,Myr. Compared among H, He and He+ ionizing photons, on average the single-star models have their He+ ionizing photons decay fastest, He in the middle and hydrogen for the slowest, since their corresponded ionizing spectra are from increasing higher frequency photons. These harder radiations decrease quickly with age. 

When a population ages, the stars dominating the spectrum change. The general trend is for a gradual softening of the far-UV spectrum. We find that our integrated light spectrum for single-star models begin to change at ages between 3 to 5\,Myr. This is similar to those found by \cite{2017ApJ...840...44B}. In both cases this is of the order of the lifetime of the most massive stars and marks the interval between when the first stellar deaths occur in the stellar population and when 40\,${\rm M_{\odot}}$ stars die. Before this epoch the number of stars and their luminous output has been mostly constant. Afterwards fewer and fewer O stars exist to provide most of the ionizing photons. While in binary-star models the ionizing flux remains until 100\,Myr and later. This is due to the binary interactions which can remove the hydrogen envelope of cool red supergiants and expose their hot helium cores to become a WR  star or low-mass helium star. These hot stars have harder ionizing flux than O stars due to their high temperature. We note in these binary populations every star is assumed to be in a binary which may not be correct for low-mass/old stellar populations \citep{2013ARA&A..51..269D} but is unlikely to affect the young populations that dominate ionizing flux.

\subsubsection{The Effect of Metallicity}
Increasing metals increases the opacity of stellar plasmas and makes stars less compact with lower surface temperature. This leads to the diminishing of the UV flux and decreasing ionizing photon production at higher metallicities. In stars with lower metallicities the stellar winds weaken and so the stars undergo less mass loss leading to slightly longer main-sequence lifetimes. Low metallicity stellar populations also have longer main sequence lifetimes in our models due to quasi-homogenous evolution (QHE). This is a consequence of the spin up of stars, which can result in rotational mixing of their layers and allow more efficient burning of hydrogen \citep{{2007A&A...465L..29C},{2011MNRAS.414.3501E},{2012MNRAS.419..479E},{2013ApJ...764..166D}}. Therefore lower metallicity models produce photons capable of ionizing hydrogen for longer, and are generally hotter, which hardens the UV spectrum for both single-star and binary-star models. 

We see a clear difference with metallicity in single-star population models, which exhibit a gradual decrease to their ionizing flux. Lower metallicity allows them to maintain a harder ionizing spectra for approximately 5\,Myr longer than the high metallicity populations. Again this is similar to the results found by \cite{2017ApJ...840...44B}. But the metallicity shows a weaker effect on binary populations, as binary interaction is less dependent on metallicity and dominates over the metal line-opacity-driven wind effects. 

As a result, we note the metallicity-dependent changes in ionizing photon numbers in Figure \ref{fig:QH_number}. The low metallicity models experience a slower decay than those of higher metallicities. Besides these two differences, the binary models with metallicity $ {\rm Z \leq 0.004} $ have the slowest decay and are separated from other binary models by forming a bump up to 20\,Myr  ($ {\rm \log(Age/yr) = 7.3} $) due to quasi-homogenous evolution (QHE). In a QHE scenario, stellar rotation causes mixing of the stellar interior and prevents shell burning, instead allowing additional hydrogen to be mixed into what would otherwise be a helium core. This extends the main sequence lifetimes of stars and has a significant effect at low metallicities \citep[see][for further discussion]{arXiv:1710.02154}. In addition, we notice for He ionizing photons higher metallicity models from both single- and binary-star models arise a bump just after the first 3\,Myr, due to the producing of hot WR stars. For the He+ ionizing photons, there are also a noticeable bump arising in binary-star models at later ages beyond 10\,Myr for higher metallicities and a bit earlier to nearly 10\,Myr for lower metallicities. This is attributed to the production of less-massive helium stars at later ages.

\subsection{Model Nebula Emission Populations}
With an understanding of the properties of the model ionizing spectra, we can better understand how their evolution causes changes in the nebular emission lines produced in the gas around a stellar population. Here we outline the properties of the nebular emission arising from our stellar populations, given different nebular gas assumptions.

\begin{figure*}
\centering
\subfloat[][]{\includegraphics[width=18cm]{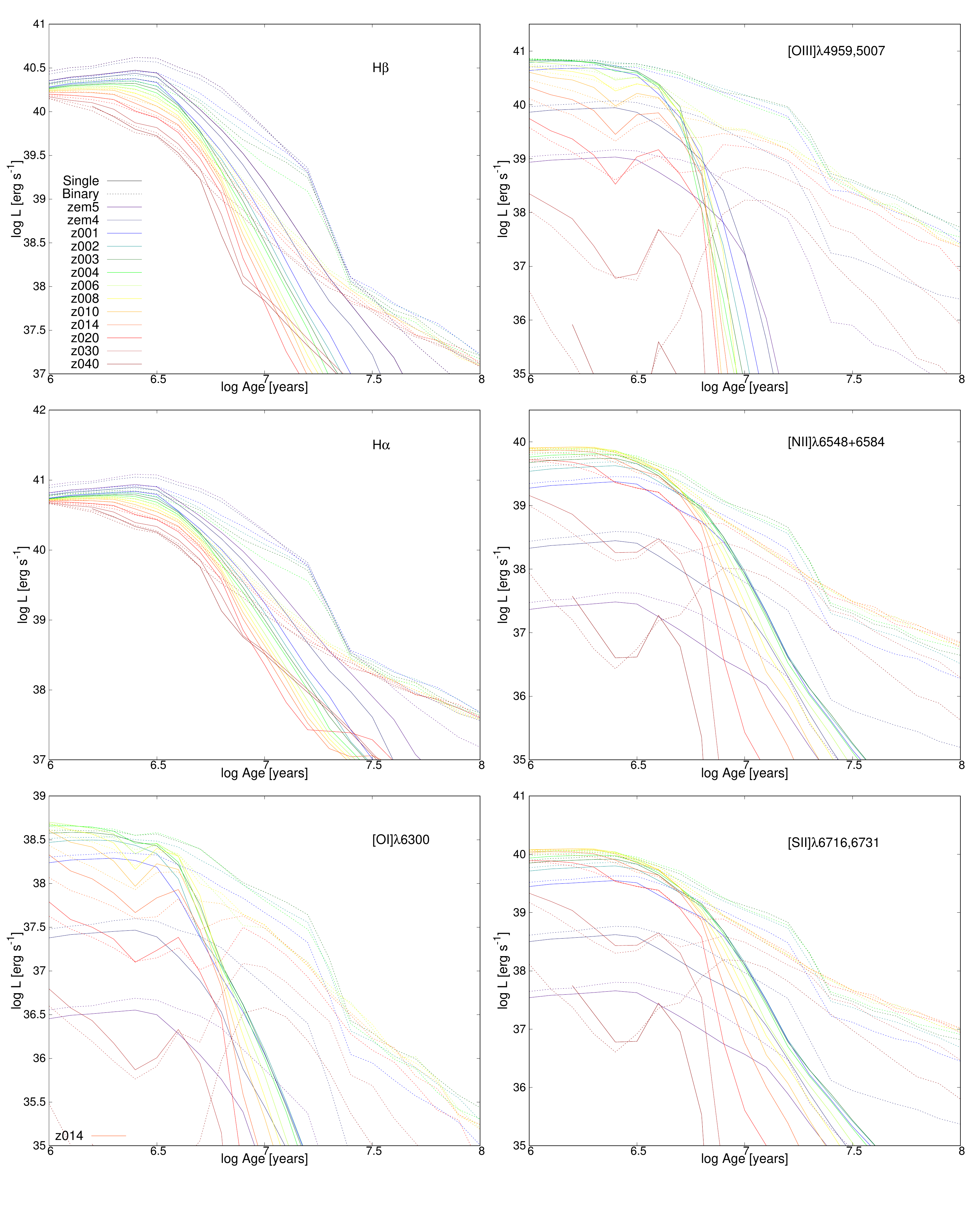}}
\centering
\caption{Emission line strength of H$ \beta  \, \lambda$4861, [O\,{\sc III}]$\, \lambda $5007, H$ \alpha \, \lambda$6563, [N\,{\sc II}]$ \, \lambda $6584 [S\,{\sc II}]$ \, \lambda $6713, [O\,{\sc I}]$ \, \lambda $6300, He\,{\sc II}$ \, \lambda $1640, He\,{\sc II}$ \, \lambda $4686, He\,{\sc II}$ \, \lambda $3965, He\,{\sc I}$ \, \lambda $4471, He\,{\sc I}$ \, \lambda $5876, and He\,{\sc I}$ \, \lambda $6678 
as a function of age, color-coded by model metallicity. The dashed lines indicate the binary-star populations and the solid lines are for the single-star populations. \emph{(cont.)}} 
\end{figure*}

\begin{figure*}
\ContinuedFloat
\centering
\subfloat[][]{\includegraphics[width=18cm]{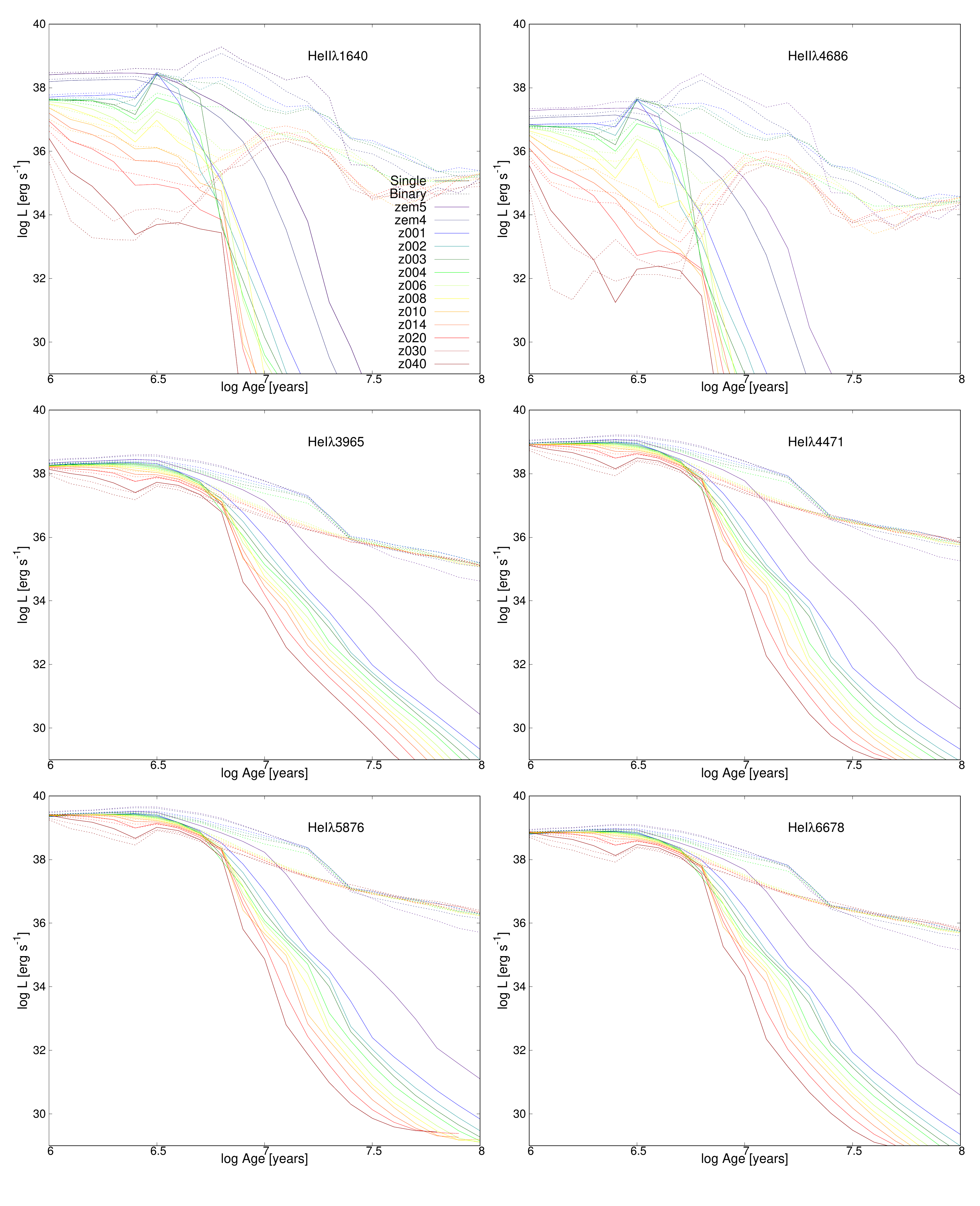}}
\centering
\caption{Emission line strength of H$ \beta  \, \lambda$4861, [O\,{\sc III}]$\, \lambda $5007, H$ \alpha \, \lambda$6563, [N\,{\sc II}]$ \, \lambda $6584 [S\,{\sc II}]$ \, \lambda $6713, [O\,{\sc I}]$ \, \lambda $6300, He\,{\sc II}$ \, \lambda $1640, He\,{\sc II}$ \, \lambda $4686, He\,{\sc II}$ \, \lambda $3965, He\,{\sc I}$ \, \lambda $4471, He\,{\sc I}$ \, \lambda $5876, and He\,{\sc I}$ \, \lambda $6678 
as a function of age, color-coded by model metallicity. The dashed lines indivate the binary-star populations and the solid lines are for the single-star populations.} \label{fig:lines} 
\end{figure*}

\begin{figure*}
\centering
\hspace*{-1.0cm}\includegraphics[width=18.5cm]{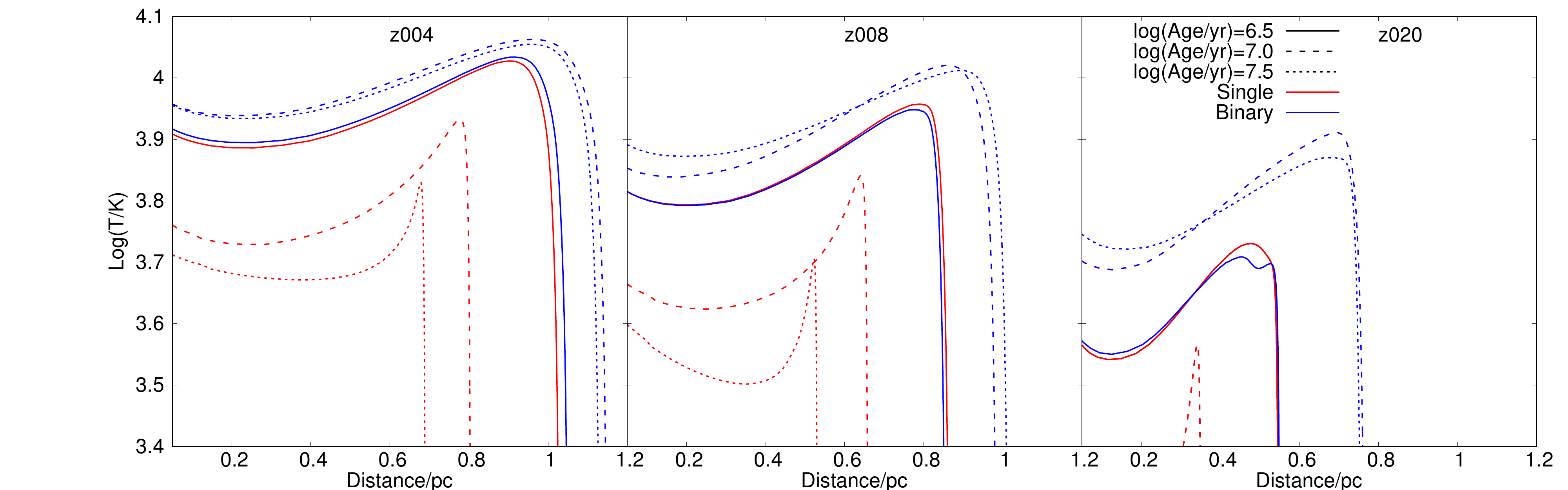}
\centering
\caption[\textsc{cloudy} model internal temperature structure of population models of metallicity $ Z = $ 0.04, 0.08, 0.020.]{\textsc{cloudy} model internal temperature structure of population models of metallicity $ Z = $ 0.04, 0.08, 0.020, with $ \log(U) = -2.5 $ at three different ages, 3\,Myr (solid lines), 10\,Myr (dashed lines) and 30\,Myr (doted lines). Red lines are from single-star models and blue lines are from binary-star models.} \label{fig:cloudy_temp}
\end{figure*}
\begin{figure*}
\centering
\hspace*{-1.0cm}\includegraphics[width=18cm]{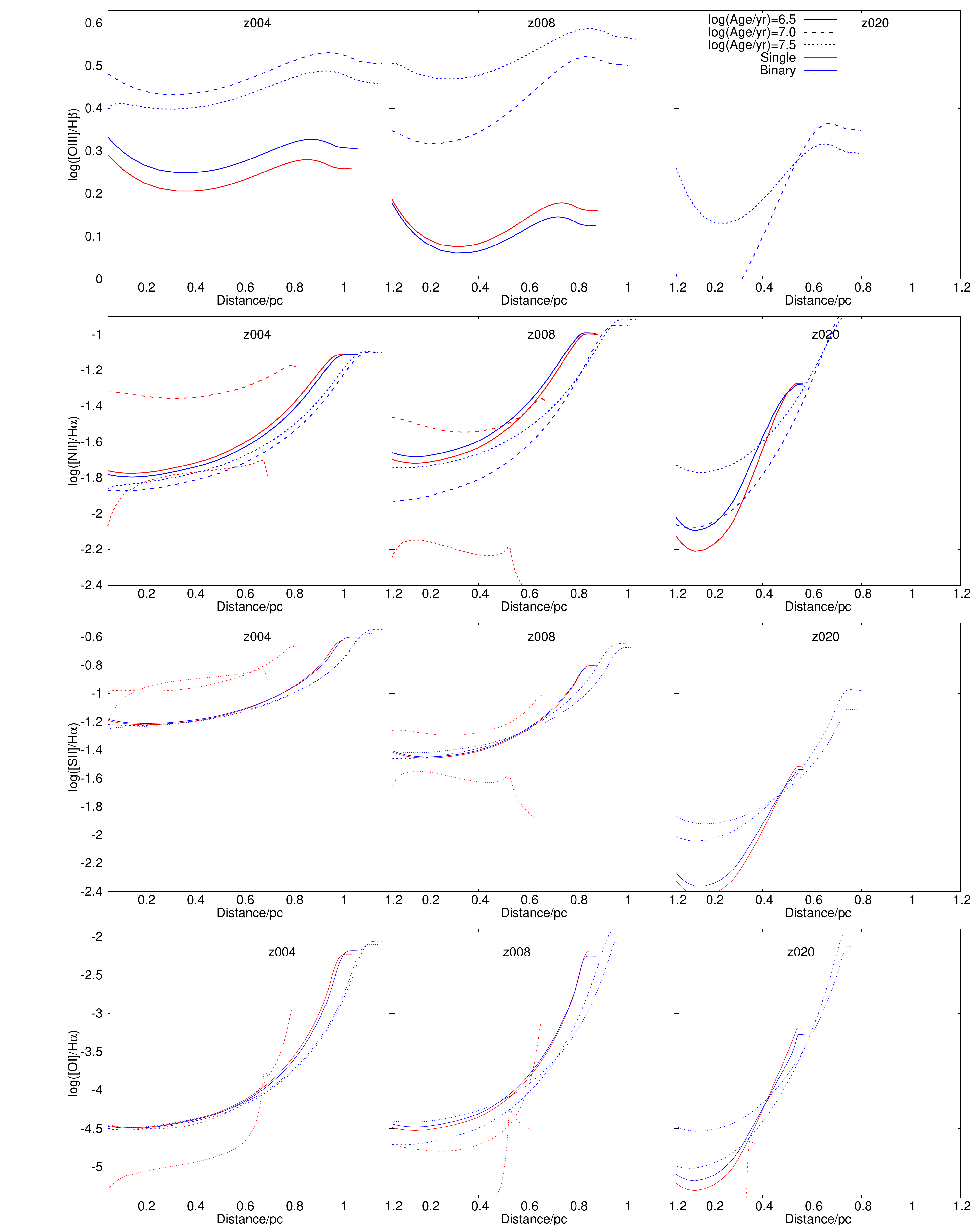}
\centering
\caption[\textsc{cloudy} model emission line ratios structure of population models of metallicity $ Z = $ 0.04, 0.08, 0.020]{\textsc{cloudy} model emission line ratios structure of population models of metallicity $ Z = $ 0.04, 0.08, 0.020, with $ {\rm \log(U) = -2.5 } $ at three different ages, 3\,Myr (solid lines), 10\,Myr (dashed lines) and 30\,Myr (doted lines). Red lines are from single-star models and blue lines are from binary-star models. } \label{fig:cloudy_lines}
\end{figure*}
\begin{figure*}
\centering
\hspace*{-1.0cm}\includegraphics[width=18cm]{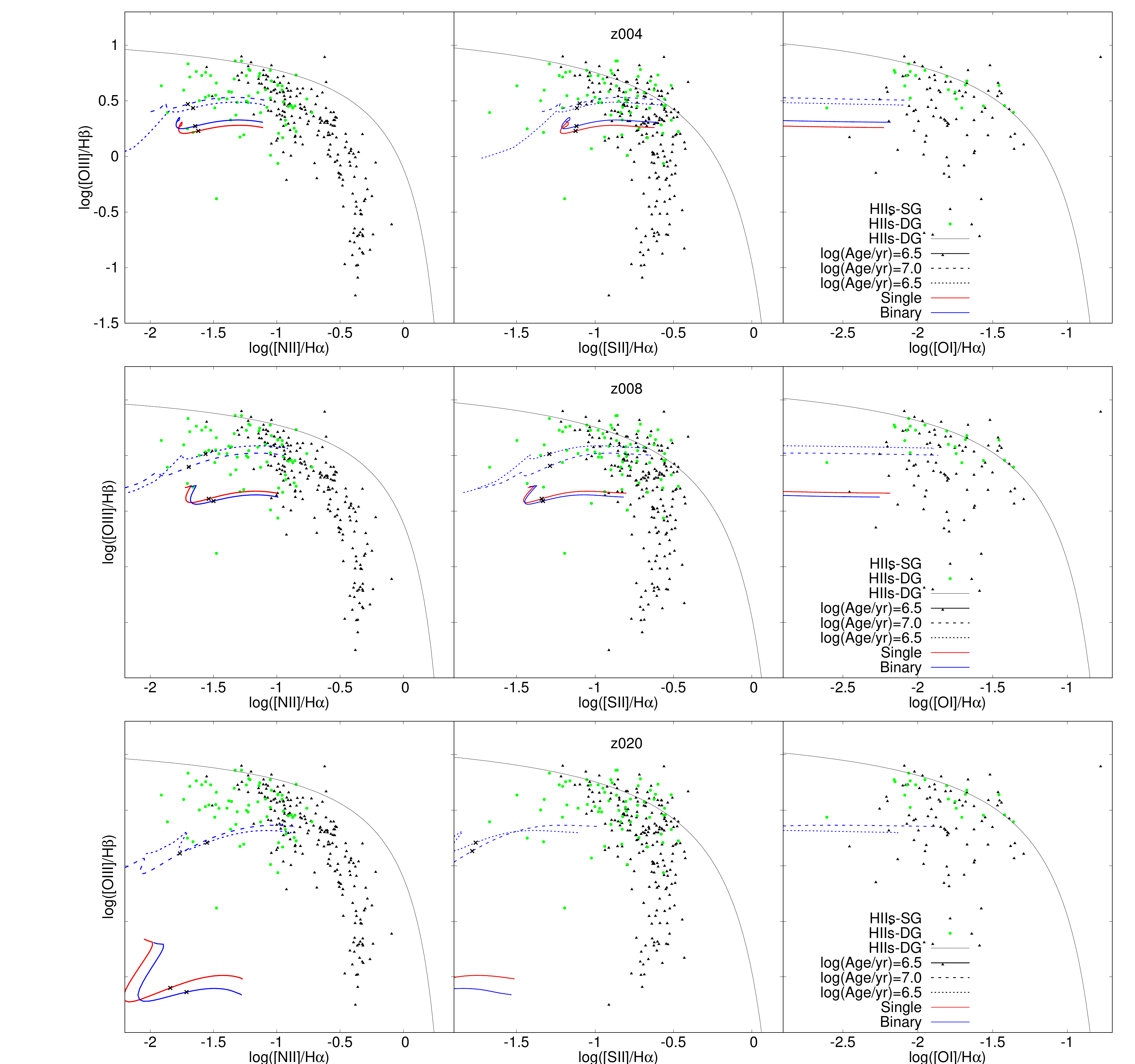}
\centering
\caption{BPT diagrams for model H\,II regions as a function of inner structure distance at metallicity $ Z = $ 0.04, 0.08, 0.020, with $ {\rm \log(U) = -2.5 } $ at three different ages, 3\,Myr (solid lines), 10\,Myr (dashed lines) and 30\,Myr (doted lines). Red lines are from single-star models and blue lines are from binary-star models. The H\,II regions from spiral galaxies are represented by black triangles and those from dwarf galaxies by green circles. The model tracks extend from left to right as the CLOUDY model computes from inner to outer layers. The black crosses on the model track mark the middle position in the cloud.} \label{fig:cloudy_BPTs_dis}
\end{figure*}

\subsubsection{General Trends in Emission Line Strength}
Figure \ref{fig:lines} shows the evolution of the line strength as a function of age for the strongest optical emission lines\footnote{Note that these can be compared directly to the similar figure 14 in \citet{2017ApJ...840...44B}, derived for models which incorporate stellar rotation but not binary interactions}: H$\beta \, \lambda$4861, [O\,{\sc III}]$ \, \lambda $5007, H$ \alpha \, \lambda$6563. [N\,{\sc II}]$ \, \lambda $6584, and [S\,{\sc II}]$ \, \lambda $6713 and [O\,{\sc I}]$ \, \lambda $6300. As we have already discussed the strength of these lines are set by the number of ionizing photons and the hardness of the ionizing spectrum which are determined in turn by the age, metallicity and binary interaction of the stellar populations.

The population age sets the overall intensity of the emission lines. Young populations are much brighter and produce many more ionizing photons than older populations and thus have stronger emission lines. Binary interactions produce more hot WR or helium stars at even later than 5\,Myr, so compared to single models these older binary models looks brighter and have stronger emission lines. We note the binary interactions have a stronger effect on the four metal lines $ - $ [N\,II], [S\,{\sc II}], [O\,{\sc III}] and [O\,{\sc I}] $ - $ at late times than the hydrogen recombination lines. 

When we vary the metallicity in our models this changes the gas phase abundances of the nebular cloud. These are reflected in changes in our model emission lines. Firstly, we see H$ \alpha $ and H$ \beta $ line strengths vary with metallicity. This is because low-metallicity models have more ionizing photons,  resulting in stronger hydrogen recombination lines. Secondly, the strength of the four forbidden metal lines $ - $ [N\,II], [S\,{\sc II}], [O\,{\sc III}] and [O\,{\sc I}], do not simply scale with metallicity but vary non-lineally. Their strength is dependent not only on the amount of each element but also on the temperature and density of the HII region. The lines initially increase with increasing metallicity and then after a maximum the emission line strength decreases with increasing metallicity. This is because at low metallicities the line is most sensitive to the abundance. Then at higher metallicities the radiation field cannot excite lines with high ionization potentials. 

Different emission lines are strongest at different metallicity depending on the changes in the ionization states. For [N\,{\sc II}], [S\,{\sc II}], the emitting atoms are singly ionized with no requirement for high temperatures. The lines are strongest around solar metallicity. But [O\,{\sc III}] emission can only be produced if oxygen is doubly ionized so higher temperatures are required implying lower metallicities. This causes their strongest emission to occur at lower metallicity around $ Z = 0.004 $. The weaker [O\,{\sc I}] emission line is strongest at LMC-like metallicity ($ Z = 0.008 $), but is is less sensitive to the metallicity as other factors contribute to this low ionization line \citep{2006ApJ...644L..29V}. 

Our main result here is that binary interactions increase the ionizing photon production and also enhance emission line strength especially at ages between 3 and 30\,Myr. It slows the decaying trends of H$ \alpha $ and H$ \beta $ as a population ages. Especially at the lower metallicities $ Z \leq 0.04 $, QHE significantly extends the main-sequence lifetime of massive stars and leads to the bump of the emission lines after 10\,Myr that separate these low metallicity models from higher metallicity ones. 

Note that the evolution in line strength is not smooth with age for either single or binary star populations. Single stars show a `bump' in the strength of certain lines at high metallicities at around 5\,Myr, due to the formation of hot, massive WR stars. This bump is weak and short-lived. Binary star models at high metallicities experience the same effect, but this is followed by a second, stronger, bump in line emission. This occurs after 10\,Myr and marks the production of more massive WR stars, followed by less massive helium stars, which extend the emission to ages of 30\,Myrs. 

\subsubsection{Ionization Structure of Nebula Model}
The emission properties of the model H\,II regions are also sensitive to their internal structure, with radial variations in ionization state and electron temperature. Fig. \ref{fig:cloudy_temp} depicts the variation of electron temperatures with radial distance from the radiation source at SMC ($ Z = 0.004 $), LMC ($  Z = 0.008 $) and Solar ($ Z = 0.020 $) metallicities with constant ionization parameter of $ \log(U) = -2.5 $. The internal temperature rises by 0.1 to 0.7 dex to crest near the reionization surface and then decreases dramatically after. At lower metallicity, stars with given age are hotter and a deeper ionized gas shell is formed by the stronger radiation. In the same metallicity environment, the internal temperature structure of single-star and binary star H\,II region models also evolve differently. The temperature of single-star models falls with increasing age so that weaker emission lines are produced at later age. However, binary-star models can push the temperature structure to a higher level at ages up to 30\,Myr due to hot stars being produced in binary interactions. 

To study this in more detail we use line ratios of a metal line to a hydrogen recombination line: $ {\rm \log([O\,III]/H\beta)} $ and ${\rm \log([N\,II]/H\alpha)} $. Fig. \ref{fig:cloudy_lines} depicts the line ratios as a function of radial distance. The ratio of [O\,{\sc III}] emission to H$ {\rm \beta} $ varies slowly with radial distance in binary-star models, rising to a peak just inside the ionization edge that reflects its strong temperature dependence. The same temperature dependence means that this line ratio is strongly metallicity and age dependent. In single star populations, it drops with increasing metallicity and is negligible at ages above 1\,Myr. In binary populations it strengthens slightly from Z=0.004 to Z=0.008, before dropping sharply at Solar metallicity, and can remain high for up to 30\,Myr at the lowest metallicities.

We see similar trends for the $ {\rm \log([N\,II]/H\alpha)} $ ratio structure in Figure \ref{fig:cloudy_lines}. In binary population it keeps approximately same strength at different ages while in single models it decreases in older populations. The ratio is also metallicity dependent such that in single-star models it reaches the highest value at LMC metallicity ($ Z = 0.008 $) and in binary models it shows a increasing trend at a higher metallicity to ($ Z = 0.020 $) as binary models are hotter at high metallicity than single models. Therefore the low-ionization emission line [N\,{\sc II}] production is more affected by chemical abundance than temperature. When it comes to $ {\rm \log([S\,II]/H\alpha)} $ ratio structure, this ratio is affected by the distance similarly and reaches the maximum near the outer surface. However, the $ {\rm \log([S\,II]/H\alpha)} $ ratio shows strong dependence on metallicity such that it reaches a lower maximum ratio more quickly at high metallicities. This is mainly due to the fact that increasing metals increase the surface opacity and make stars cooler. As for the ratio of $ {\rm \log([O\,I]/H\alpha)} $, it varies over 2 magnitude with distance from inner to outer surface of the cloud, but shows less variance with metallicity. 

In Figure \ref{fig:cloudy_BPTs_dis} we show the variation of the line ratios with cloud distance in the BPT diagrams. The model tracks extend from left to right as the {\sc CLOUDY} model computes to the outer layers where the calculation is density-bounded until the radiation-bounded final position on the rightmost position of the tracks. The line ratios from the middle of the cloud can reach the observed ratio of $ {\rm \log([O\,III]/H\beta)} $, $ {\rm \log([N\,II]/H\alpha)} $ and $ {\rm \log([S\,II]/H\alpha)} $, but fail in the case of $ {\rm \log([O\,I]/H\alpha)}$.

\subsection{Metallicity Calibrations}
\begin{table*}
\caption{The strong-line metallicity estimate calibrations, N2-vanZee \citep{1998AJ....116.2805V}, N2-Marino \citep{2013AA...559A.114M}, O3N2-Marino \citep{2013AA...559A.114M}, D16 \citep{2016ApSS.361...61D}, PP04-N2 \citep{2004MNRAS.348L..59P}, PP04-O3N2\citep{2004MNRAS.348L..59P}, Pilyugin-ONS \& ON \citep{2010ApJ...720.1738P}, and Pilyugin-R, S, R2D \& S2D \citep{2016MNRAS.457.3678P}, discussed in Section 3.5 and Figure \ref{fig:OH_models_Page}}
\begin{center} 
                      \begin{tabular}{l @{\hskip 0.15in} | @{\hskip 0.15in} c @{\hskip 0.15in} |@{\hskip 0.15in} c @{\hskip 0.15in} | @{\hskip 0.15in} c @{\hskip 0.15in} | @{\hskip 0.15in} c}
                        \hline
                        \hline
                        \rule{0pt}{3.ex}
                        Number  & ID & Emission Lines & [O/H] Validity range &  Match with BPASS\\
                        
                        \hline 
                         
                       1.......... & N2-vanZee & [N\,{\sc II}], H$ \alpha $ & $ < 8.2 ,\,\,\, 8.8-9.1 $
                       & overestimate at [O/H] $ \leq $ 8.8\rule{0pt}{3.ex}\\  
                       2.......... & N2-Marino & [N\,{\sc II}], H$ \alpha $ & $ 7.0 - 8.9 $  & poor \rule{0pt}{3.ex}\\ 
                       3.......... & O3N2-Marino & [O\,{\sc III}],H$ \beta $,[N\,{\sc II}], H$ \alpha $ & $ 7.0 - 8.9 $ &  poor \rule{0pt}{3.ex}\\  
                       4.......... & D16 & [N\,{\sc II}],[S\,{\sc II}],H$ \alpha $ & $ < 9.05 $ & underestimate at [O/H] $ \leq $ 8.8 \rule{0pt}{3.ex}\\
                       5.......... & PP04-O3N2 & [O\,{\sc III}],H$ \beta $,[N\,{\sc II}],H$ \alpha $ & $ 8.1 - 9.1 $ &  poor \rule{0pt}{3.ex}\\ 
                       6.......... & PP04-N2 & [N\,{\sc II}],H$ \alpha $ & $ 7.2 - 8.9 $ & poor \rule{0pt}{3.ex}\\ 
                       7.......... &  Pilyugin-ONS &  [N\,{\sc II}], [O\,{\sc II}], [S\,{\sc II}], [O\,{\sc III}],H$ \beta$  & $7.0-9.0$ & underestimate at [O/H] $ \leq $ 8.5 \\
                       8.......... &  Pilyugin-ON & [N\,{\sc II}], [O\,{\sc II}], [S\,{\sc II}], [O\,{\sc III}],H$ \beta $ & $7.0-9.0$ & poor\\
                       9.......... &  Pilyugin-R & [N\,{\sc II}], [O\,{\sc II}], [O\,{\sc III}],H$ \beta $ & $7.0-8.8$ & poor\\
                       10.......... &  Pilyugin-S& [S\,{\sc II}], [O\,{\sc II}], [O\,{\sc III}],H$ \beta $ & $7.0-8.8$ & poor\\
                       11.......... &  Pilyugin-R2D & [N\,{\sc II}], [O\,{\sc II}], [S\,{\sc II}], [O\,{\sc III}],H$ \beta $ & $7.0-8.8$ & underestimate at [O/H] $ \leq $ 8.5 \\
                       12.......... &  Pilyugin-S2D & [S\,{\sc II}], [O\,{\sc II}], [O\,{\sc III}],H$ \beta $ & $7.0-8.8$ & poor \\
                        \hline
 
\end{tabular}
\end{center}
\label{tab:6calibrations}
\end{table*}

\begin{figure*}
\centering
\subfloat[][]{ \includegraphics[width=18cm]{./figs/OH_models_ToT1_Page.pdf}}\\
\centering
\caption{Oxygen abundance [O/H] (given here as ${\rm = 12 + \log(O/H)}$) in different strong-line calibrations compared to the input value from BPASS models with weighting for cluster age included. The shaded areas show the maximum and minimum variation with the red for single star models and blue for binary-star models. Violet zones indicate overlap regions. \emph{(cont.)}} \label{fig:OH_models_Page}
\end{figure*}

\begin{figure*}
\centering
\ContinuedFloat
\subfloat[][]{ \includegraphics[width=18cm]{./figs/OH_models_ToT2_Page.pdf}}\\
\centering
\caption{Oxygen abundance [O/H] (given here as ${\rm = 12 + \log(O/H)}$) in different strong-line calibrations compared to the input value from BPASS models with weighting for cluster age included.  The shaded areas show the maximum and minimum variation with the red for single star models and blue for binary-star models.  Violet zones indicate overlap regions.} \label{fig:OH_models_Page}
\end{figure*}

Since oxygen is the most abundant metal in the gas phase and exhibits very strong nebular lines at optical wavelengths, it is often chosen as a metallicity indicator in ISM studies. As discussed above, this often requires use of a strong line ratio diagnostic calibrated through models or empirical observations .However, this method is strongly affected by the choice of which strong-line abundance calibrations are used as discussed in \cite{2008ApJ...681.1183K}. 

For the van Zee observational sample introduced in section \ref{sec:vanzee}, an estimate of the oxygen abundance was determined by the $ {\rm [N\,{\sc II}]/H\alpha} $ relation \citep{1998AJ....116.2805V}:
\begin{eqnarray}
   {\rm [O/H]} = 12 + \log({\rm O/H}) = 1.02\log({\rm [N\,{\sc II}]/H\alpha}) + 9.36
\end{eqnarray}
where the [N\,{\sc II}]/H$\alpha $ ratio increases with increasing oxygen abundance and is only valid as a metallicity  estimator for $ 12 + \log({\rm O/H}) < 9.1 $ and in the absence of shock excitation. The typical uncertainties are of the order of 0.2 dex or more; it is not an accurate abundance estimator. This dependence of relative nitrogen abundance on total metallicity, oxygen abundance or some other parameter is common to all the nitrogen-based calibrations (see Table  \ref{tab:6calibrations}) and results in an increasing discrepancy between our models (which used fixed [N/Fe] or [N/O] abundance ratios) and nitrogen-based calibrations, particularly at the high metallicity end.

To account for uncertainties in metallicity calibration, we compare the [O/H] abundances inferred from our model spectra using twelve different strong line indicators (listed in Table \ref{tab:6calibrations}). In each case, the input metallicity (in terms of both [Fe/H] and [O/H]) of the BPASS spectral model is known, and the inferred metallicity is determined for each age, electron density and ionization parameter in our nebular emission library. In figure \ref{fig:OH_models_Page} we compare the input metallicity to that inferred for populations of single stars and binaries. At each input metallicity we calculate an age-weighted mean and standard deviation of the model metallicities inferred (given different nebular conditions) by each method. The weighting for age reflects the weaker line emission expected in old stellar populations (see section 4). We also indicate the full range of abundances (from maximum to minimum) inferred at each input metallicity with a shaded region. It should be noted that each calibration in both figure \ref{fig:OH_models_Page} and Table \ref{tab:6calibrations} was originally calibrated over a specific metallicity range, as given in the table, and failure outside that range is to be expected.  Nonetheless, it is not uncommon for other authors to use a calibration outside its proposed range, and we thus consider the comparison for the full range of our model metallicities.

The strong line calibrations considered are typically trained on local galaxy survey data, and are not strongly constrained at the lowest metallicities in our model set. They show considerable variation in this regime, ranging from large underestimates (Pilyugin-ON) to large overestimates (PP04-03N2). Single-star models show more variance than binary-star models, and a smaller range between minimum and maximum metallicity estimates, although both sets show similar trends against inferred [O/H] in the six calibrations. 
The N2-vanZee, D16, Pilyugin-ONS and Pilyugin-R2D method show good correlations reproducing our model metallicity. The other methods are poor estimators for our model metallicity. The two O3N2 methods and the Pilyugin-ON Method perform most poorly, with the former two providing relatively constant estimations in different metallicity models and the later one providing a far more underestimation. The N2-Marino and PP04-N2 methods have a similar and gentle slope and fails to follow the trend of our model metallicities. All the six Pilyugin methods tend to underestimate the abundances of our higher metallicity models with [O/H] $>$ 8.6.

\section{Interpreting the Observed Nebular Emission}
\begin{figure*}
\centering
\hspace*{-1.5cm}\includegraphics[width=20cm]{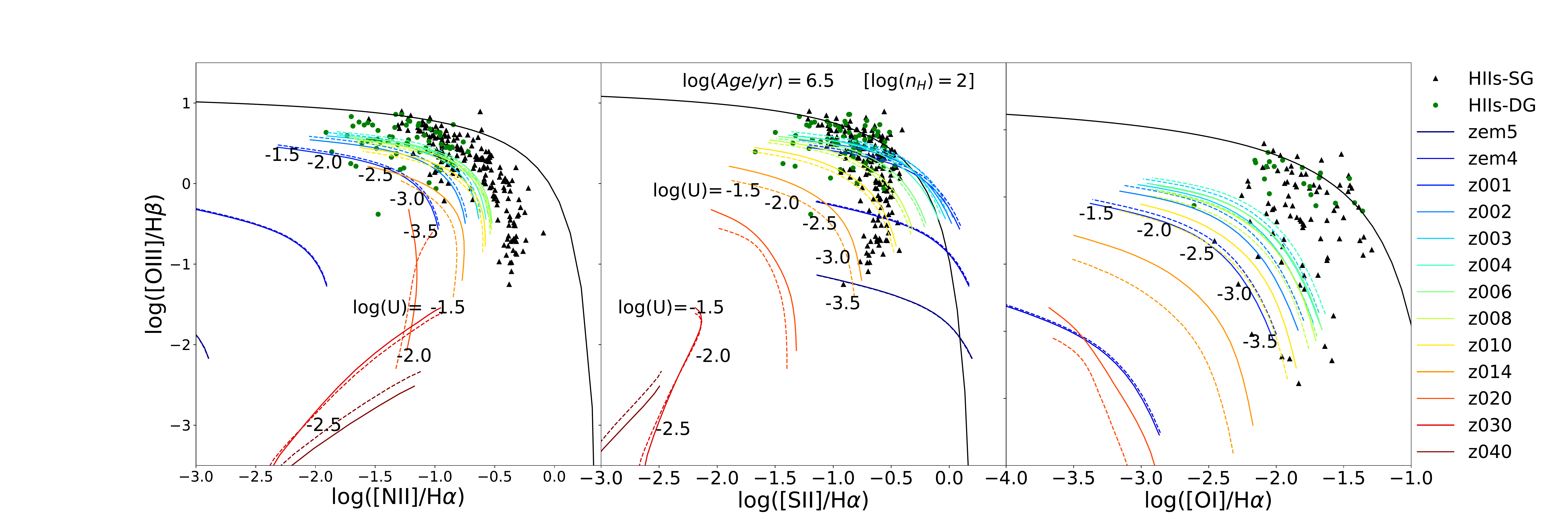}
\centering
\caption[The BPT diagrams of \textsc{bpass} models compared to the observed HIIs-SG and HIIs-DG from van Zee sample]{The BPT diagrams of \textsc{bpass} single-star models (solid lines) and binary-star models (dashed lines) in the 13 metallicities from Z = 0.00001 to 0.04, at $  \log(n_H)=2 $ and $ {\rm \log(Age/yr)=6.5} $, comparing with the observed HII-SG (black triangles) and HIIs-DG (green circles). The black curves show maximum starburst lines from \cite{2001ApJ...556..121K}. The colored tracks from models extend from upper left to lower right as ionization parameter decreases from $ \log(U) $ = -1.5 to -3.5.} \label{fig:BPT_Zout}
\end{figure*}


In this section we use BPT diagrams which map the ionizing state of a H\,II region and therefore allow us to test the agreement of theoretical emission lines with those from observations. We highlight the best-fitting models that match each individual H\,II regions and identify parameters which most influence the properties of the observed nebular emissions. 

\subsection{BPT Diagrams of model H\,II Regions}
\begin{figure*}
\centering
\hspace*{-0cm}\includegraphics[width=18cm]{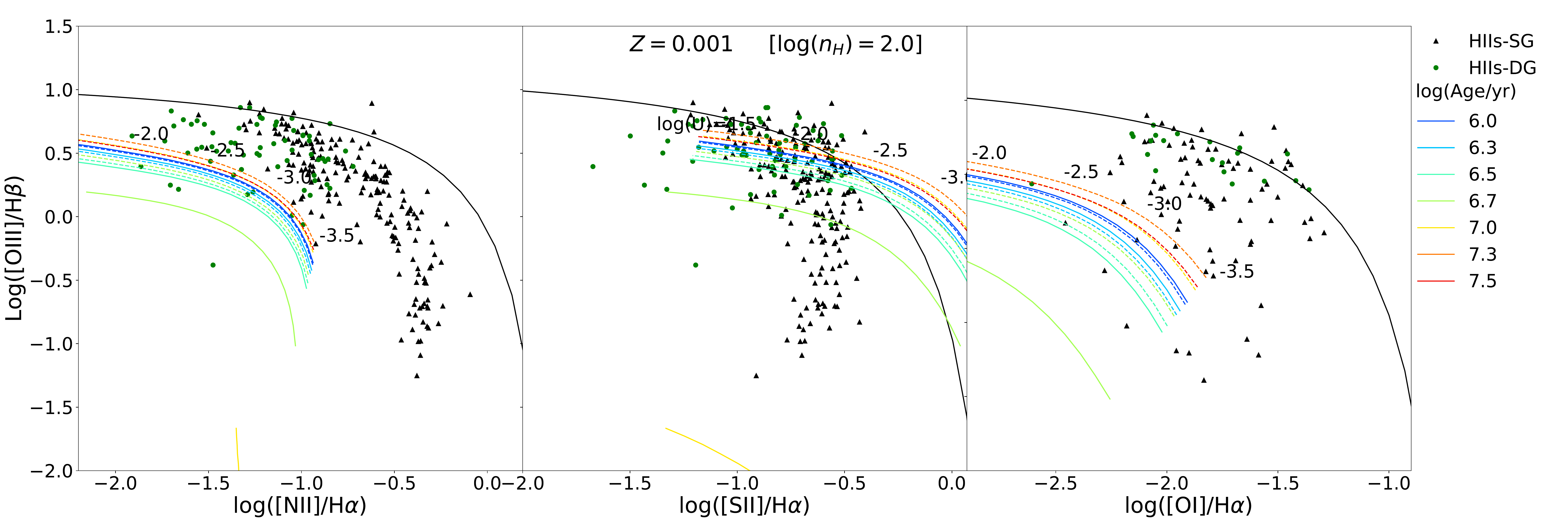}\vspace*{-0.39cm}
\hspace*{-0cm}\includegraphics[width=18cm]{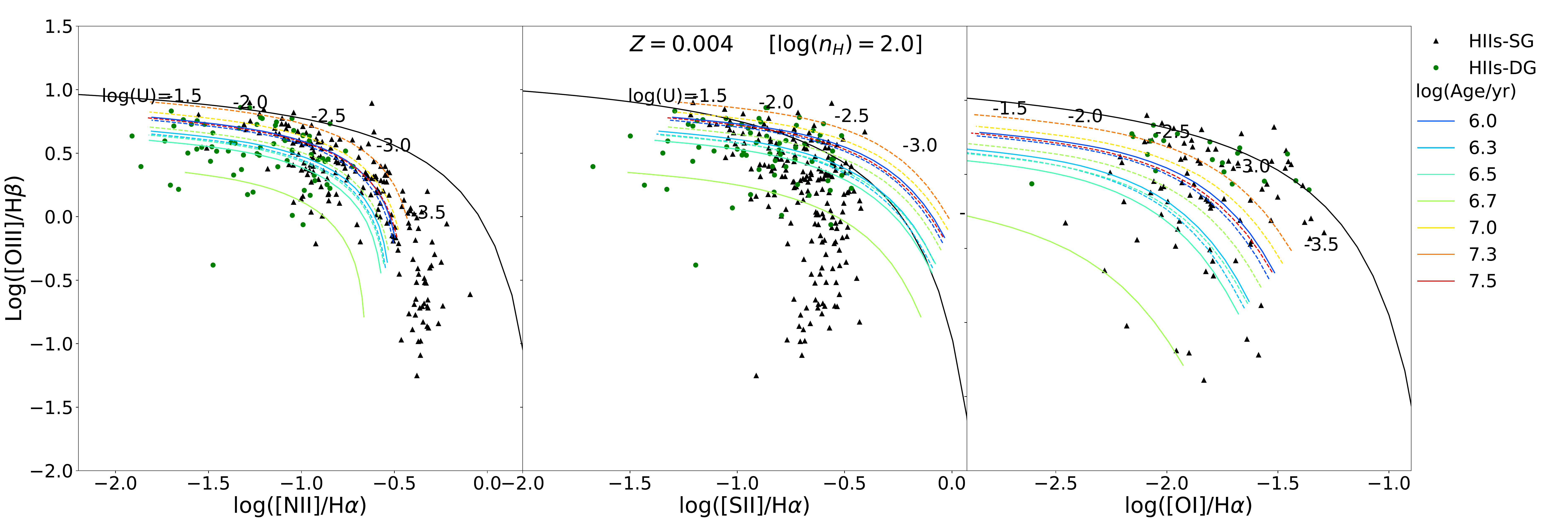}\vspace*{-0.39cm}
\hspace*{-0cm}\includegraphics[width=18cm]{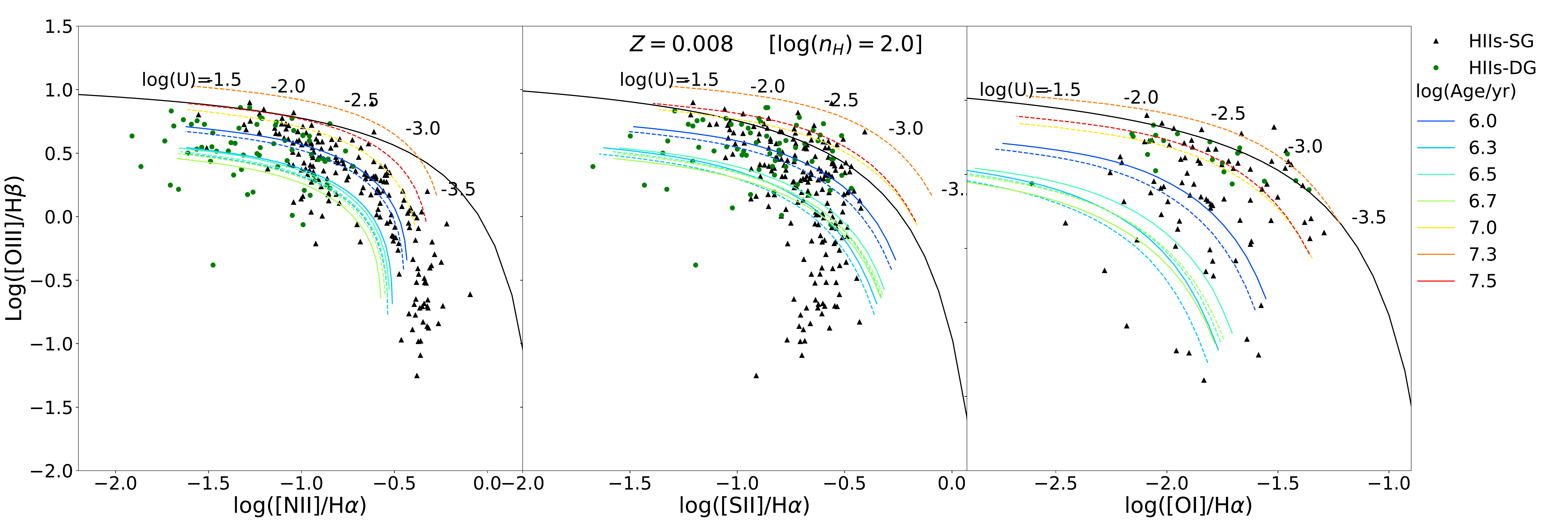}\vspace*{-0.39cm}
\hspace*{-0cm}\includegraphics[width=18cm]{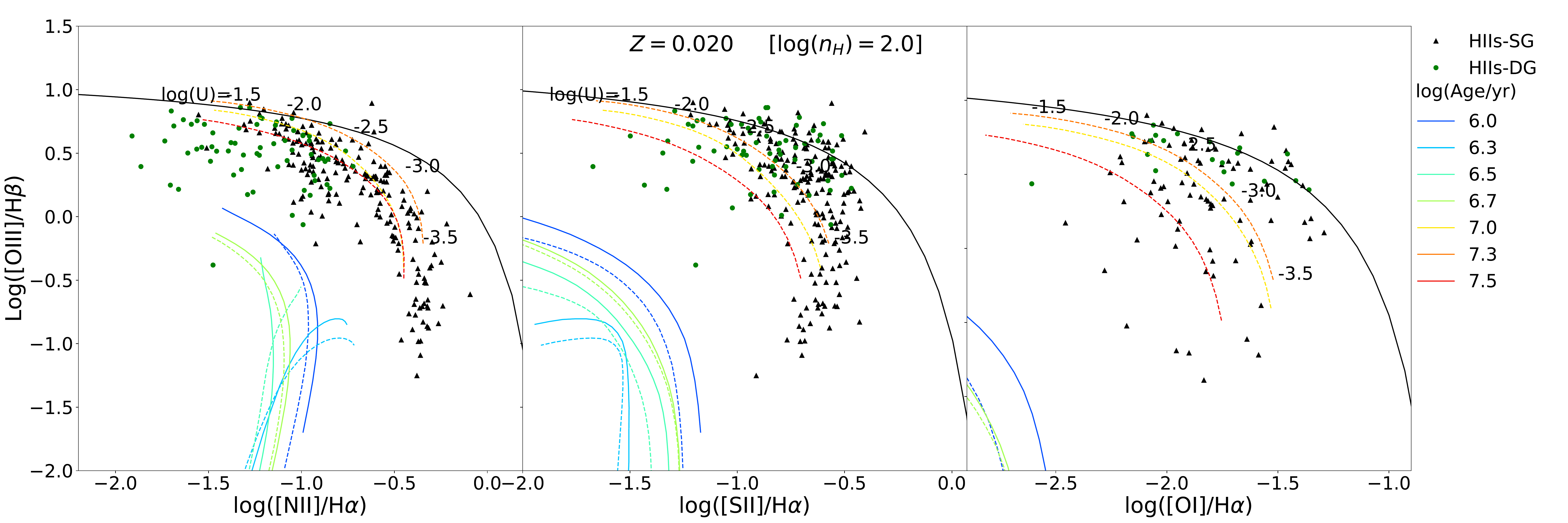}
\centering
\caption[The BPT diagrams of \textsc{bpass} models of $ {\rm \log(n_H)=2} $ varying with age from $ {\rm \log(Age/yr)} =$ 6.0 to 7.5]{The BPT diagrams of \textsc{bpass} models of $ {\rm \log(n_H)=2} $ varying with age from $ {\rm \log(Age/yr)} =$ 6.0 to 7.5: single-star models (solid lines) and binary-star models (dashed lines). The observed HIIs-SG are in the black triangles and HIIs-DG in the green circles. The black curves show the maximum starburst lines defined by \cite{2001ApJ...556..121K}.} \label{fig:BPT_Zin}
\end{figure*}
\begin{figure*}
\centering
\hspace*{-0cm}\includegraphics[width=18cm]{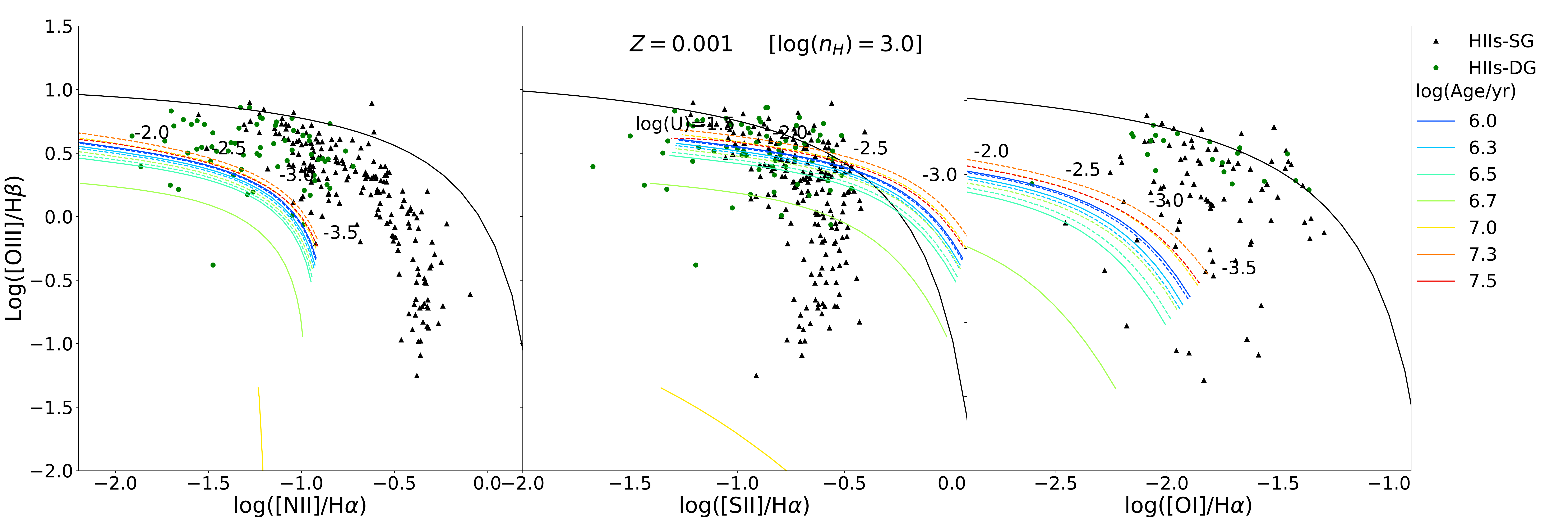}\vspace*{-0.39cm}
\hspace*{-0cm}\includegraphics[width=18cm]{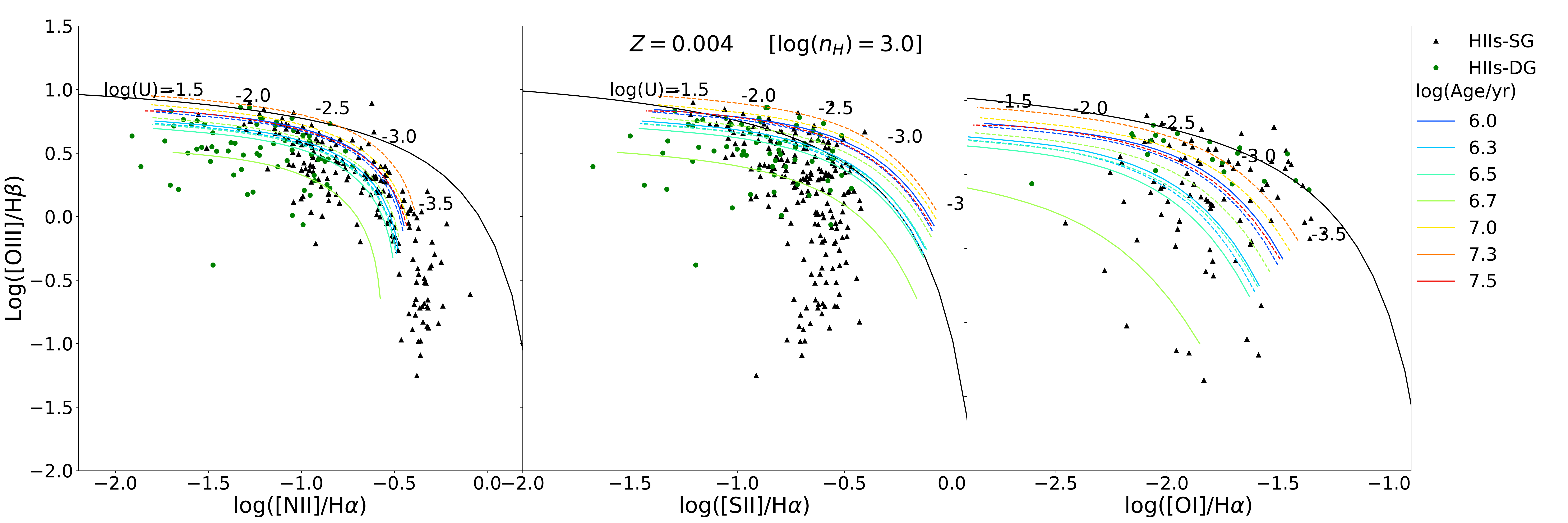}\vspace*{-0.39cm}
\hspace*{-0cm}\includegraphics[width=18cm]{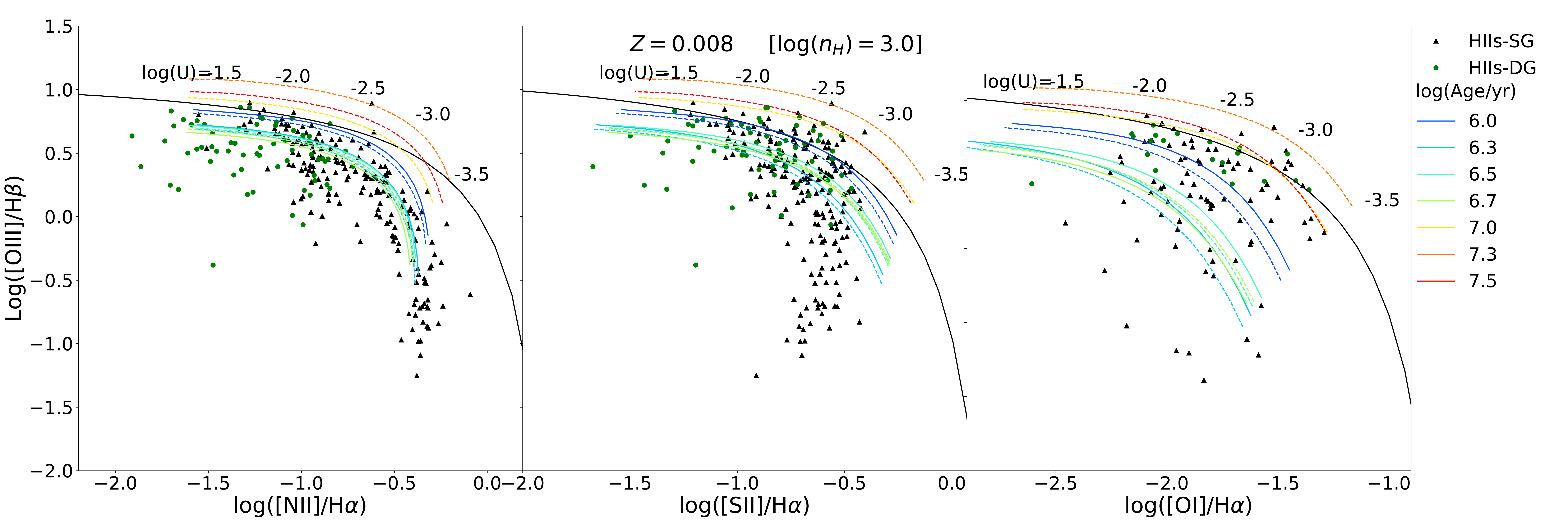}\vspace*{-0.39cm}
\hspace*{-0cm}\includegraphics[width=18cm]{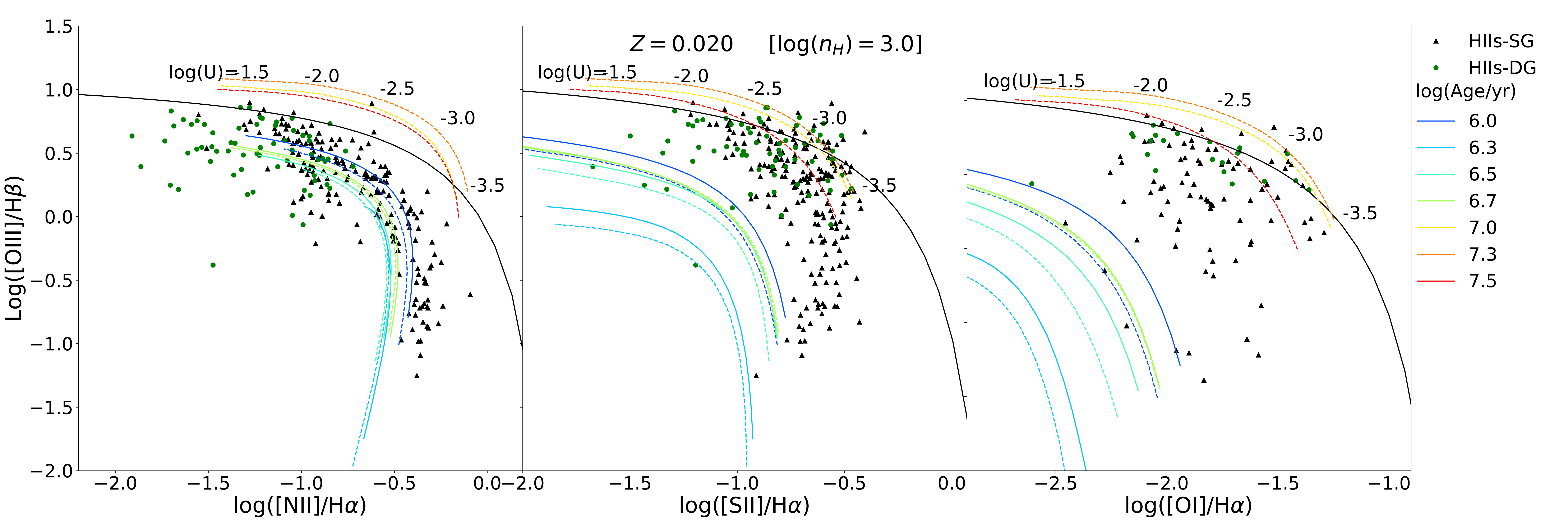}
\centering
\caption[The BPT diagrams of \textsc{bpass} models of $ {\rm \log(n_H)=3} $ varying with age from $ {\rm \log(Age/yr)} =$ 6.0 to 7.5]{As in figure \ref{fig:BPT_Zin} but with $ {\rm \log(n_H)=3}$.} \label{fig:BPT_Zin_Hd3}
\end{figure*}
An example of our model BPT diagrams are presented in Fig. \ref {fig:BPT_Zout}. For convenience, we name them by their characteristic x-axis line ratios, BPT-NII, BPT-SII and BPT-OI respectively for the BPT diagrams with x-axis, $ {\rm [N\,{\sc II}]\lambda 6584/H\alpha} $, $ {\rm [S\,{\sc II}]\lambda [6716+6731]/H\alpha} $ and $ {\rm [O\,{\sc I}]\lambda 6300/H\alpha} $. At constant hydrogen density $ {\rm \log(n_H)}=2 $ and constant age $ {\rm \log(Age/yr)}=6.5 $, our models gave arise to 13 separated tracks corresponding to 13 metallicities, extending from high log([O\,III]/H$ \beta $) to low log([N\,II]/H${\rm \alpha }$), log([S\,II]/H${\rm \alpha }$) and log(OI/H${\rm \alpha }$) as the ionization parameter is varied from $ {\rm \log(U) } = $ -1.5 to -3.5 along each track. In general, [N\,{\sc II}], [S\,{\sc II}] and [O\,{\sc I}] are strongest at low $ {\rm \log(U) } $, while [O\,{\sc III}] decreases with increasing $ {\rm \log(U) } $. Compared to observed H\,II regions our models form a sequence from upper left to lower right on the diagram as their $ {\rm \log(U) } $ decreases. 

We note at this young age (around 3\,Myr), single-star and binary-star models predict very similar trend which is consistent with both cases having a large number of OB stars and Wolf-Rayet stars. Binary interaction only make changes at older ages. The models show the expected variation with respect to metallicity. In these diagrams, the tracks approach the observed H\,II regions as metallicity increases up to about Z $ \sim $ 0.006. At higher metallicities the tracks fall away with the decreasing amount of ionizing radiation. The temperature-sensitive line [O\,{\sc III}] to H${\rm \beta }$ ratio falls off when model metallicity is greater than Z= 0.006 where the ratio is most strong. All of our model tracks at this age are below the maximal starburst line from \cite{2001ApJ...556..121K}, and both single and binary models only go though a small region of the observed regions in the BPT-NII and BPT-OI planes and with a larger overlap in BPT-SII plane. However, the models with extremely low metallicity, Z = 0.00001 \& 0.0001, cannot support a high [O\,{\sc III}]/H${\rm \beta }$, [N\,{\sc II}]/H${\rm \alpha }$ and [O\,{\sc I}]/H${\rm \alpha }$ ratios and therefore appear in the lower left-hand corner or lie outside the plotted range. We note that the low ionization lines, particularly the [O\,{\sc I}] 6300\AA\ feature are particularly susceptible to contamination from emission in the diffuse interstellar medium, boosting the line ratio relative to that expected from the H\,II region alone. However, as discussed in section \ref{sec:vanzee}, the 2 arcsec slit width used in the observations was well matched to the physical size of the H\,II regions in this sample. While in some cases, the observed spectra are derived from the H\,II region centre, and hence we expect underestimation of the low excitation lines (especially [S\,II] and [O\,I]), in others we might expect the [O\,{\sc I}] to be overestimated due to poor resolution. On average however, we expect this to be a minor effect, with the typical region lying close to the model locus. Without integral field unit spectroscopy, it is difficult to evaluate any offset in the low ionization lines further. However, we note that as no dust is included in our models, the depletion effect of dust on metal abundance is underestimated and strength of emission lines and corresponded gas density estimation may be discrepant as a result. Again we expect this to manifest most strongly in the low ionization species since they can be excited in low density regions where the effects of metal depletion are strongest. 
 
In Fig.\ref{fig:BPT_Zin} we investigate the effect of binary interaction, age and metallicity on the model ionization conditions across the BPT diagrams. For the first 3\,Myr the emission line ratios are reduced by  increasing age, with the line ratio tracks evolving away from the locus of observed H\,II regions, but from 5\,Myr the line ratio tracks of binary-star model stay close to the observed H\,II regions while single-star models still gradually decrease. This difference between single-star and binary-star models grows as the model ages and an abrupt difference occurs at ages beyond $ {\rm \log(Age/yr)}=7.0 $, where single-star models cannot produce the H\,II regions. Binary star models (with more hot post-main sequence stars) are even hotter than at early ages, emitting more strong lines. These can match most of the observed H\,II regions at same time in the three BPT diagrams and do a lot better in the BPT-OI diagram. 

At low metallicities (Z=0.002-0.008), BPASS binary tracks exceed the `maximal' line ratio line defined by \cite{2001ApJ...556..121K} based on their models. This is particularly true at ages $ \gtrsim 10$\,Myr and is most clearly shown in the BPT-SII diagrams in Figure \ref{fig:BPT_Zin}. This ability to exceed the \cite{2001ApJ...556..121K} lines at low metallicity is consistent with observations both of rare metal-poor H\,II regions in the local universe, and the metal-poor star forming galaxies observed at high redshift \citep[e.g. ][]{2016ApJ...826..159S}.
 
In denser H\,II regions with $ {\rm \log(n_H/cm^{-3}) = 3.0} $, as shown in Fig. \ref{fig:BPT_Zin_Hd3}, we see a similar result to that in Figure \ref{fig:BPT_Zin}. The key difference to low-density H\,II region models is that older binary models can rise up beyond the Kewley et al. limit not only at lower metallicities, but also at Solar metallicity. Therefore, this suggests that the observed galaxies above the Kewley line at high-redshift may be due to additional factors -- older populations beyond 10\,Myr and higher gas density $-$ not just metallicity. For future reference, we calculate new BPASS binary maximal starburst lines as a function of metallicity, and for the total model parameter distribution, and present these in the Appendix.

\subsection{Method to Find the Best-fitting Models}
We can now attempt to constrain the parameters of the observed H\,II region sample by calculating the likelihood that each model in our library accurately fits the observations. The quantities we consider are the 4 emission line ratios, $ {\rm [O\,{\sc III}]\lambda 5007/H\beta} $, $ {\rm [N\,{\sc II}]\lambda 6584/H\alpha} $, $ {\rm [S\,{\sc II}]\lambda 6724/H\alpha} $, $ {\rm [O\,{\sc I}]\lambda 6300/H\alpha} $, and one line strength $  {\rm EW[H\beta]} $. We assume the errors are Gaussian in the likelihood computation. The fitting probability for each model is given by, 
\begin{eqnarray}
   P_{\rm fit} =\prod_{i=1}^{5} \dfrac{1}{\sqrt{2\sigma_{i}^{2}\pi}}\exp{-\dfrac{(x_{i}-\mu_{i})^{2}}{2\sigma_{i}^{2}}}
\end{eqnarray}
where $ x_{i} $ is our model, $ \mu_{i} $ is observed value and $ \sigma_{i} $ is observation error. For each model defined by metallicity, stellar age, hydrogen density and ionization parameter, we assume {\it a priori} that these parameters are independent. 

We also consider that clusters do not all survive as they age. A combination of internal dynamical evolution, ejection of massive stars, relaxation and tidal effects means that fewer clusters survive to old ages than are formed \citep[e.g. ][]{2003MNRAS.340..197D}. The number of the clusters decreases by approximately an order of magnitude with every order of magnitude increase in age:
\begin{eqnarray}
   P_{\rm cluster} \propto 10^{(1-\log({\rm t/yr}))\times0.1}
\end{eqnarray} 
taking this into account, the final probability, $ P_{\rm total}= P_{fit} \times P_{cluster}$, and is then weighted toward younger clusters. We determine the best-fitting parameters, $ \overline{Y} $, for each H\,II region, as well as population averages, with $ \Phi$ = [O/H] , age($t$), $ \log(n_H) $ and $ \log(U) $:
\begin{eqnarray}
\overline{\Phi}=\dfrac{\sum_{i=1}^{n}P_{\rm total}^{i} \times \Phi}{\sum_{i=1}^{n}P_{\rm total}^{i}}
\end{eqnarray}

\begin{figure}
\centering
\includegraphics[width=8.3cm]{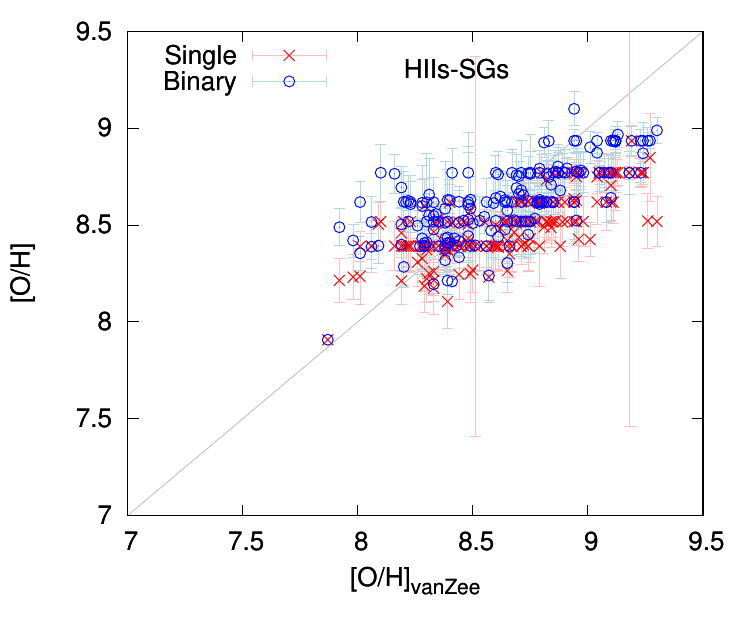}
\includegraphics[width=8.3cm]{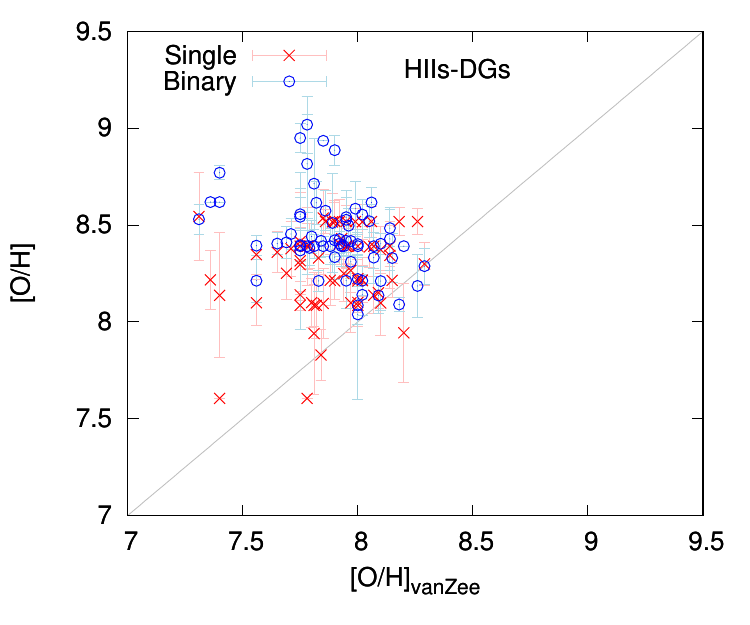}
\centering
\vspace*{-0.2cm}
\caption{Best-fitting Oxygen abundances [O/H] ($ = 12 + \log({\rm O/H} $ ) from our models, compared to those previously derived for the same sample. The red crosses with error bars are single-star models and those blue circles are binary-star models.}
\label{fig:MeanEW}
\end{figure}

\subsection{Oxygen Abundance Measurement}

\begin{figure}
\centering
\includegraphics[width=8.3cm]{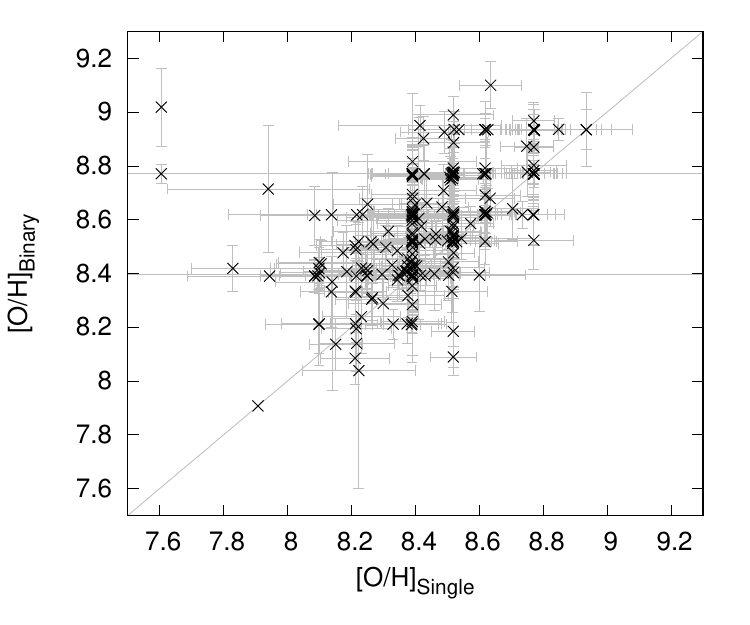}
\centering
\vspace*{-0.2cm}
\caption{[O/H] comparison between best-fitting models derived from single and binary populations.} \label{fig:[O/H]_sb}
\end{figure}

In Fig. \ref{fig:MeanEW} we compare our best-fitting models to the oxygen abundance [O/H] derived by \cite{1998AJ....116.2805V} and \cite{2006ApJ...636..214V} from observed emission lines. Our best-fitting models, when allowed to match over the full range of hydrogen density, provide a generally good match to the abundances derived by \cite{1998AJ....116.2805V} for HIIs-SG, all of which show 12+log(O/H)$\ge8.0$, deviating slightly at both the high and low metallicity ends of the distribution.  
This may reflect a variety of differences between the empirically calibrated nebular line metallicities and those in our stellar models. The model metallicity determines the metal mass fractions at a zero main sequence age. There is no guarantee that this will match the abundance in the observed stellar atmosphere at a later time, due to a combination of gravitational settling of heavy elements and transport of others to the surface through dredge-up and convection or rotational mixing, as well as through the effects of surface temperature and conditions on line strength, however since we are observing reprocessed nebular emission, this is unlikely to be the dominant factor.

A second possibility is that it is not necessary for the abundance pattern of the nebular gas in a HII region to perfectly match that of the irradiating stellar population (which is an assumption in our current model set). The very short lifetimes of the most massive stars means that a nebular gas cloud can be enriched by supernovae and ejected material while the population of O and B stars remains relatively robust, particularly if the main sequence lifetime of any significant fraction of the stars has been extended through binary interaction pathways.

Finally, and likely most significantly, it is worth noting that the comparison being made here is against a nitrogen-based abundance estimator.  This abundance estimator assumes N/O varies with O/H based on empirical relations, which themselves show substantial scatter. However, it is not clear whether the local relation between these terms is valid in all star formation environments, at high redshifts, or in young starbursts where supernova feedback produces more heavy element enrichment than stellar winds. Our models currently assume a constant N/O abundance ratio, independent of the overall metallicity mass fraction. The metallicity-dependent offset between our best fitting models and the previous estimate shown in Fig \ref{fig:MeanEW} may therefore be rooted more with the assumed constant N/O ratio in the calibration than the actual properties of the galaxies. This highlights the need to understand the underlying assumptions in the calibrations. Nonetheless, the differences between the single star populations and BPASS binary models remain significant. 

For the HIIs-DG, our models suggest significantly higher metallicities than derived by \cite{2006ApJ...636..214V}, particularly when binary models are included. This discrepancy can be reduced if we restrict our model library to high electron densities ($ {\rm \log(n_H)=3} $), and in this case we also derive lower metallicities for binary models (while also finding slightly higher [O/H] in HIIs-SG). However the higher hydrogen density has little effect on single star models, for which the best fits cluster tightly around [O/H]$\sim$8.5, independent of density. 

Fig. \ref{fig:[O/H]_sb} describes more clearly the [O/H] correlation between between single-star and binary-star models. We find the [O/H] predictions from binary-star models basically agree with those from single-star models, with scatter around the 1:1 line. In general the binary models suggest slightly higher metallicities than the single star models, since they are able to achieve a harder ionizing spectrum at a given metallicity.

\subsection{[O/H] variation $ U $, $ n_{H} $ and age}\label{sec:oh_variation}
\begin{figure}
\centering
\hspace*{-0cm}\includegraphics[width=8.3cm]{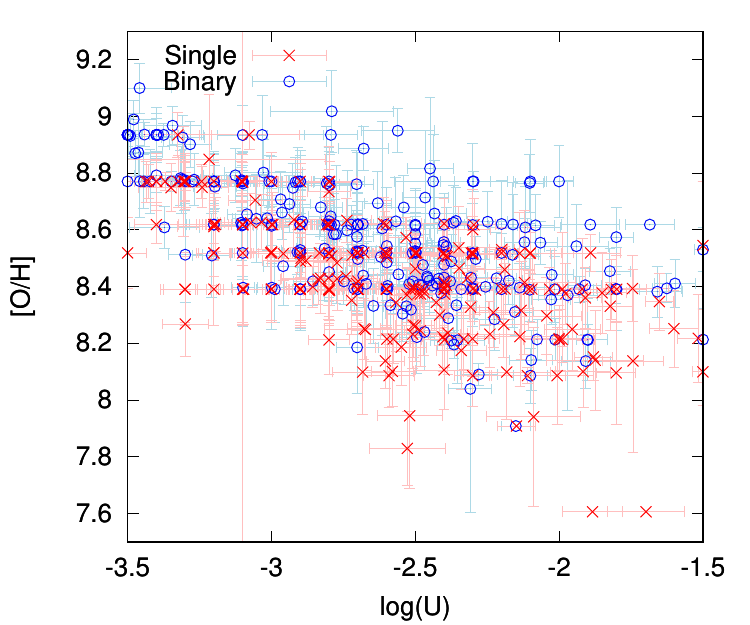}\\
\hspace*{-0cm}\includegraphics[width=8.3cm]{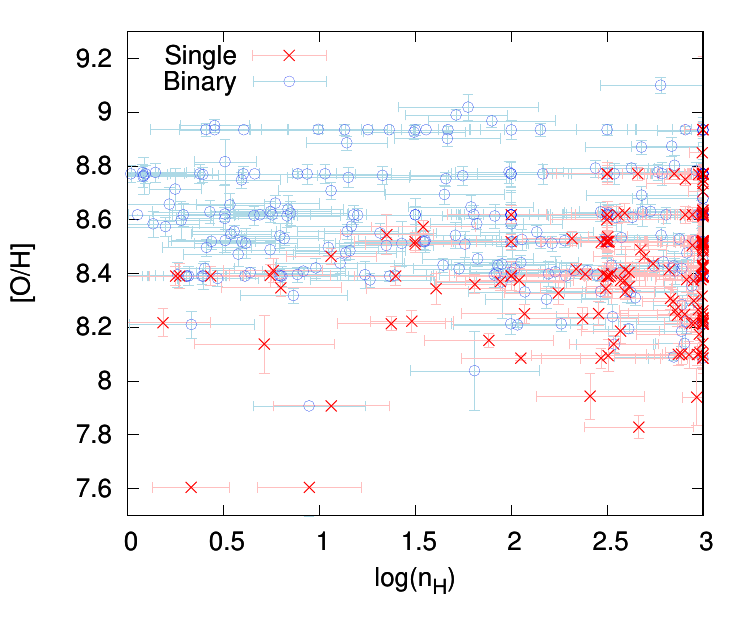}\\
\hspace*{-0cm}\includegraphics[width=8.3cm]{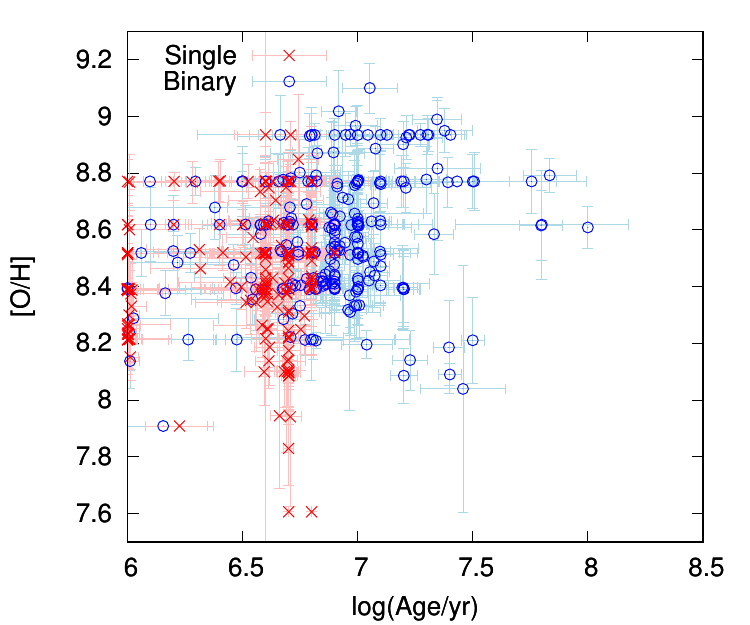}\\
\centering
\caption{The relationship of best-fitting [O/H] ($ = 12 + \log({\rm O/H} $ ) with derived ionization parameter $ \log(U) $ in the top panel, $ n_H $ in the middle and age in the bottom panel from \textsc{bpass} best-fitting models. Single models are in red crosses with error bars and binary models in blue circles with error bars.} \label{fig:MeanEW_Ps}
\end{figure}

In Figure \ref{fig:MeanEW_Ps}, we explore the properties of our best fit models, and
determine how the derived [O/H] correlates with other derived properties, namely the ionization parameter $ \log(U) $, $ {\rm n_H} $ and age, showing results for both single-star and binary-star model fits. These parameters reflect the properties of ionizing gas and irradiating stellar populations. We notice that the best-fitting models from both single- and binary-star populations show a metallicity, measured through 12+log(O/H), that increases as the value of the ionization parameter $ \log(U) $ decreases. This result is consistent with the fact that low-metallicity H\,II regions are hotter with a higher ionization state and as metallicity decreases the H\,II region cools down and lowers its ionization state. Indeed, the inferred ionization parameter from the strong line ratios is likely providing the strongest constraint on the metallicity. At a given value of $ \log(U) $, binary-star model fits have a higher [O/H] than single-star models, as expected given their harder radiation fields.

Single-star models favour the highest hydrogen densities in our model suite, $\log(n_H)=3$. Binary-star models vary over the full range from $\log(n_H)$ = 0.0 to 3.0, and cluster around 0.5. These differences between single-star and binary-star models are driven by the best-fit stellar ages. Best fitting single-star models have to be younger than 5\,Myr, while the age of binary-star models can
extend to 100\,Myr and are mostly around 10\,Myr. Therefore, binary interactions have a significant effect to extend the lifetime of a H\,II regions, and when measuring the age of the observed H\,II regions binary evolution need to be considered. 

We note that the H\,II regions in the van Zee sample show a relatively low electron density, $n_e \leq 2.0$, as measured by the observed [S\,II]$\lambda$6717/6731 ratio. We suggest that the discrepancy with our best-fitting models (which are typically higher in electron density) may be caused, at least in part, by the effects of depletion of metals onto interstellar dust grains, which can increases significantly with average gas density along the line of sight \citep[see][for a recent review]{2017arXiv171107434G}. However we note that without integral field spectroscopy, it is also possible that we are seeing aperture effects. As figure \ref{fig:cloudy_lines} demonstrated, low ionization lines such as [S\,II] and [O\,I] are dominated by emission in the fringes of a H\,II region, where the gas is likely to be more diffuse, and the ionization state is lower, than in the core. With a fixed aperture slit, the [S\,II] measurement may be affected by emission in the low ionization fringes of the H\,II regions and also by emission in diffuse interstellar gas lying outside the region itself, while our overall spectral fit is more sensitive to the highly irradiated core of each H\,II region. The two measurements are thus sensitive to different parts of the H\,II region ionization structure and potentially different electron number density regimes.

\subsection{Inner Radius and Mass of H\,II Regions}
\begin{figure}
\centering
\includegraphics[width=9cm]{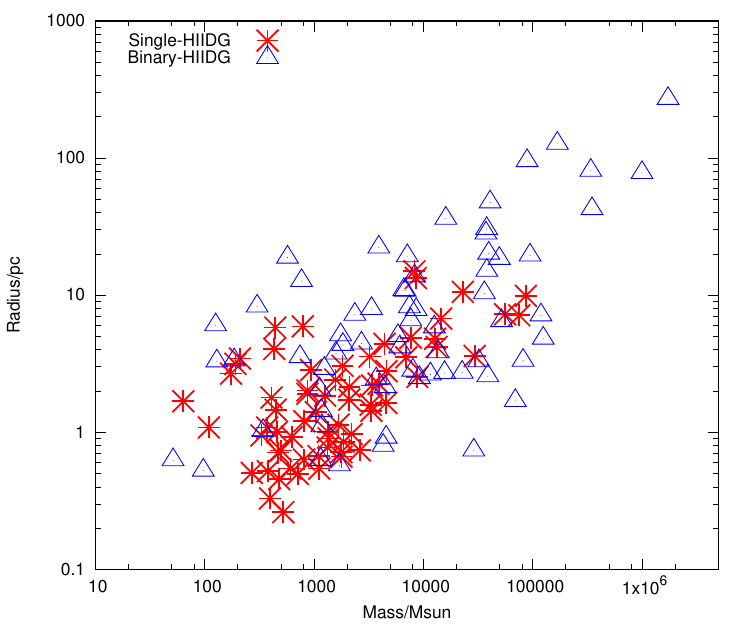}\\
\includegraphics[width=9cm]{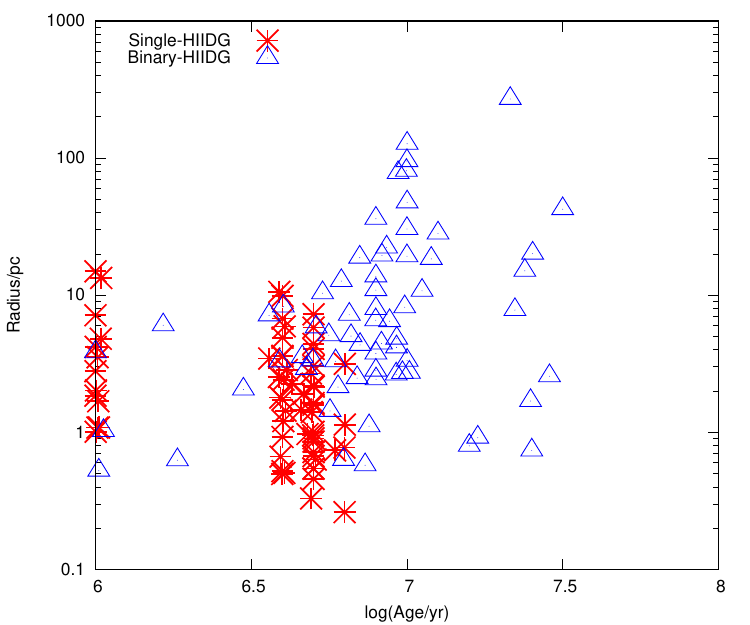}
\centering
\caption[The predicted radius and mass of van zee sample H\,II regions from \textsc{bpass} best-fitting models]{The predicted radius and mass of H\,II regions drawn from \cite{2006ApJ...636..214V} using \textsc{bpass} best-fitting models in the top panel and their corresponding ages in the bottom panel. The red stars are from single-star models and blue triangles are from binary-star models.} \label{fig:RadiMass}
\end{figure}
It is convenient to model the geometry and spectral property of a H\,II region using ionization parameter $ U $, but the importance of understanding the exact physical inner radius and mass of a H\,II regions drives us to the following test.

Based on the definition of ionization parameter $ U $ we can derive the inner radius $ R  = \sqrt{Q(H)/4\pi U n_{H} c} $  of a cloud. Here the ionizing photon rate $ Q(H) $, hydrogen density $ n_{H} $ and $ U $ depends on the metallicity and age of our best fitting models. Using this, we can estimate radius and mass of the HIIs-DG. For each observed H\,II regions, we have a measured $ {\rm H\beta} $ luminosity which was corrected for internal extinction using Balmer line decrememnt data, and which is in good agreement with the photometric and narrowband data for each source. This can be converted to an ionizing photon number using our best fitting models. We can then estimate the radius of H\,II regions based on the definition of $ U $ as above. We can also calculate the mass of the cluster from the H$ \beta $ flux.

In Fig. \ref{fig:RadiMass} we show the radius of these H\,II region as a function of their mass and how it differs when using single-star and binary-star models. In general, all the masses of HIIs-DG are less than $ 10^{6} {\rm M_{\odot}} $ and their radius grows linearly with increasing mass from around 0.5pc and 100 $ {\rm M_{\odot}} $ to 10pc and $ 10^{5} {\rm M_{\odot}} $ in single-star models, while in binary-star models the radius tends to increase toward 300pc as masses increase up to $ 10^{6} {\rm M_{\odot}} $. Combining their age information as shown in Fig. \ref{fig:RadiMass}, we suggest that binary-star models are able to produce more massive, larger and older H\,II regions than single-star models in the low metallicity environment of dwarf galaxies. The radius ranges produced from our models are consistent with the work of \cite{2014MNRAS.442.3711G} in which they measured the size of star formation regions to vary from 2 to 100 pc.  \\

\section{The Effect of Ionizing Photon Loss}
So far we have assumed that all the ionizing photons emitted from the inner irradiating source are used to heat and ionize the nebular gas. In general, this may be true for observation of a whole galaxy where even if some ionizing photons escape from an individual H\,II region they still contribute to the whole galaxy's diffuse nebula emission \citep{2009ApJ...692.1476C}. For individual H\,II regions, one has to consider if a fraction of ionizing photons have escaped from the circumstellar environment without interacting with the gas cloud. Observationally, this produces a decrease in inferred ionizing photon numbers and so weaker emission lines. 

Another factor that can reduce the total number of ionizing photon is dust reddening of the intrinsic ionizing radiation. This again leads to the lower strength in both nebular line emission and continuum emission. 
We consider both of these two effects in our best-fitting model selection and find that there is little impact on emission line ratios but a significant impact on the equivalent width of the H$ {\rm \beta} $ emission line. With more ionizing photon leakage the equivalent width of H$ {\rm \beta} $ decreases.

We note that, as discussed in section \ref{sec:vanzee}, the ${\rm {H\beta}}$ measurements used in this analysis are drawn from the same spectra as the line ratios used to determine the best fitting models. Therefore they encompass emission from an identical region of nebular gas to that which determines the line ratios. While the fixed slit width of the original observations means that the outskirts of a H\,II region may be underrepresented (i.e. omitted in the spectral direction, but not in the spatial direction along a slit), the ${\rm {H\beta}}$ emission line strength should represent the number of photons reprocessed in a region spatially coincident with the line emission and so any discrepancy in their properties will represent the same leakage as a spectrum precisely matched to the H\,II region size. In fact, there is some evidence from section \ref{sec:oh_variation} that the low ionization lines such as [S\,II] in a larger part of the van Zee sample may be contaminated by low density, diffuse interstellar gas emission. Thus any estimate of leakage drawn from these spectra will represent lower bounds on the leakage from the H\,II regions themselves.

We recalculate the likelihood of the emission line ratios as above but add the possibility that the observed $ EW_{\rm H\beta} $ can be less than predicted in our models. We use $ P({\rm EW_{H\beta}}) = 1  $ if model $ EW_{\rm H\beta} $ $ \geq $ observed $ EW_{\rm H\beta} $, otherwise $ P({\rm EW_{H\beta}}) $ is calculated as before. In this section, we highlight the variations in the inferred properties of the observed regions due to this new fitting prescription and the effect of ionizing photon leakage.

\subsection{[O/H] Measurement}
\begin{figure}
\centering
\includegraphics[width=8.5cm]{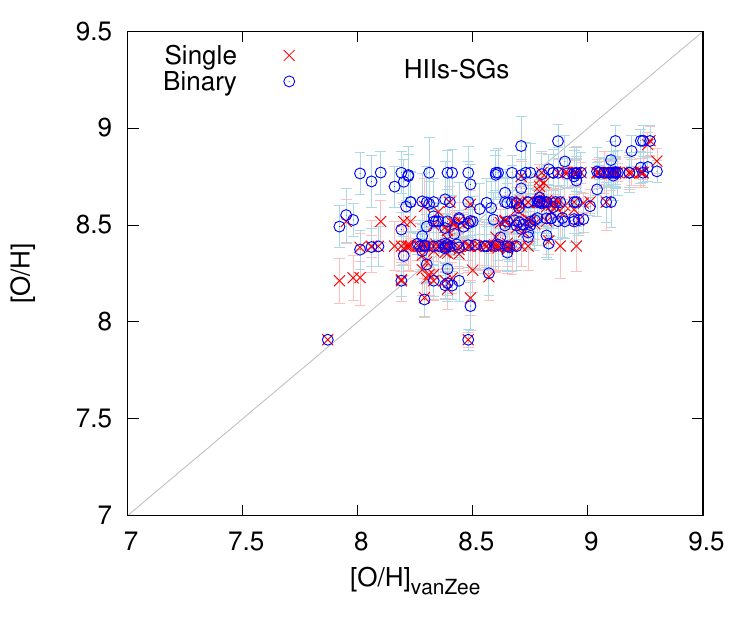}
\includegraphics[width=8.5cm]{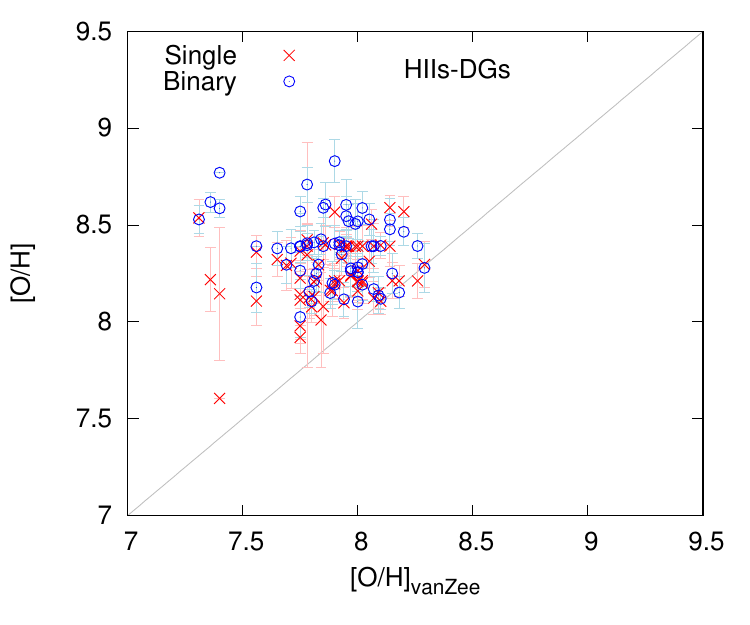}
\centering
\vspace*{-0.5cm}
\caption{Oxygen abundances [O/H] ($ = 12 + \log({\rm O/H} $ ) from our models when allowing for ionizing photon loss. The red crosses with error bars are single-star models and those blue ones are binary-star models.} \label{fig:[O/H]_leakage}
\end{figure}
\begin{figure}
\centering
\includegraphics[width=8.3cm]{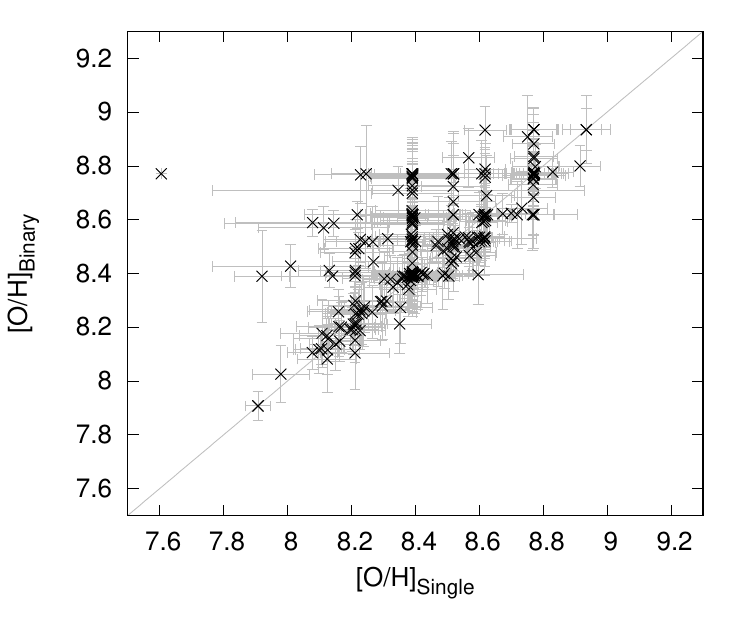}
\centering
\caption{Similar to Figure \ref{fig:[O/H]_sb} but including ionizing photon loss, the [O/H] ($ = 12 + \log({\rm O/H} $ ) correlation between single-star and binary-star models.} \label{fig:[O/H]_sb_leakage}
\end{figure}
In Fig. \ref{fig:[O/H]_leakage}, we present the oxygen abundance derived for each of the van Zee sample H\,II regions derived from our models when allowing for the possibility of ionizing photon escape from the region. The ionizing photon loss has only a slight effect on the [O/H] estimation and the results are similar to those described above. This is to be expected, since the metallicity is more strongly constrained by the ionizing radiation field, and hence line ratios, than the ongoing star formation rate, which determines the strength of the H$\beta$ line. Comparing the results from single-star and binary-star models, the single stars, unsurprisingly match the reports previously reported for the HIIs-SG, themselves based on single star calibrations, better. Both single star and binary models still estimate a higher [O/H] value for HIIs-DG than that derived by \cite{2006ApJ...636..214V}. Binary models find higher [O/H] values especially in the lower abundance cases.

We also present the comparison of [O/H] estimated from single-star and binary-star models in Fig. \ref{fig:[O/H]_sb_leakage}. The similar results show that allowing photon leakage reduces the differences between the model sets, but binary-star models are more likely to yield higher metallicity estimates and have a larger variance at low inferred metallicites.

\subsection{[O/H] variation with $ U $, $ n_{H} $ and age}
\begin{figure}
\centering
\hspace*{-0cm}\includegraphics[width=8.5cm]{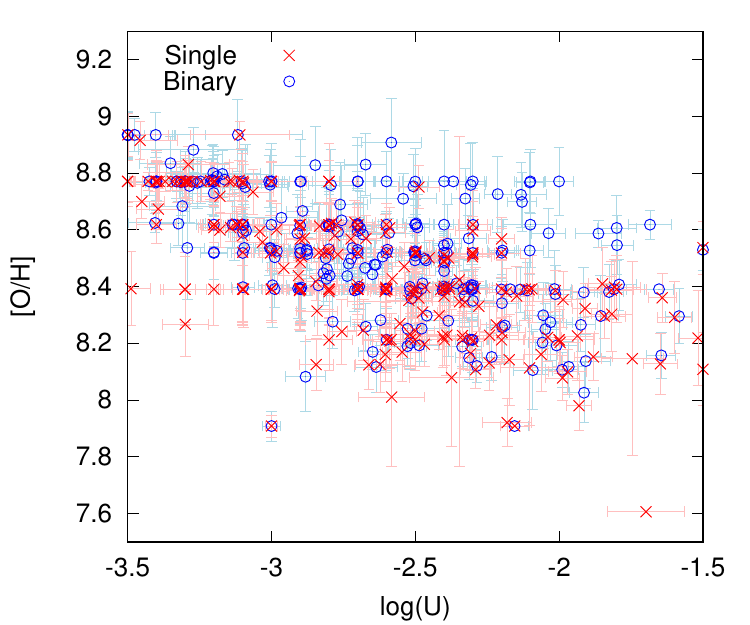}\\
\hspace*{-0cm}\includegraphics[width=8.5cm]{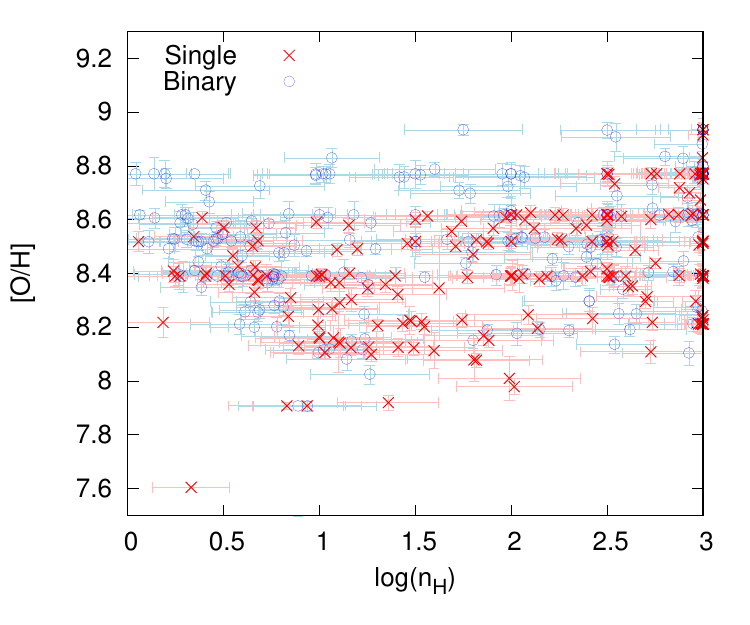}\\
\hspace*{-0cm}\includegraphics[width=8.5cm]{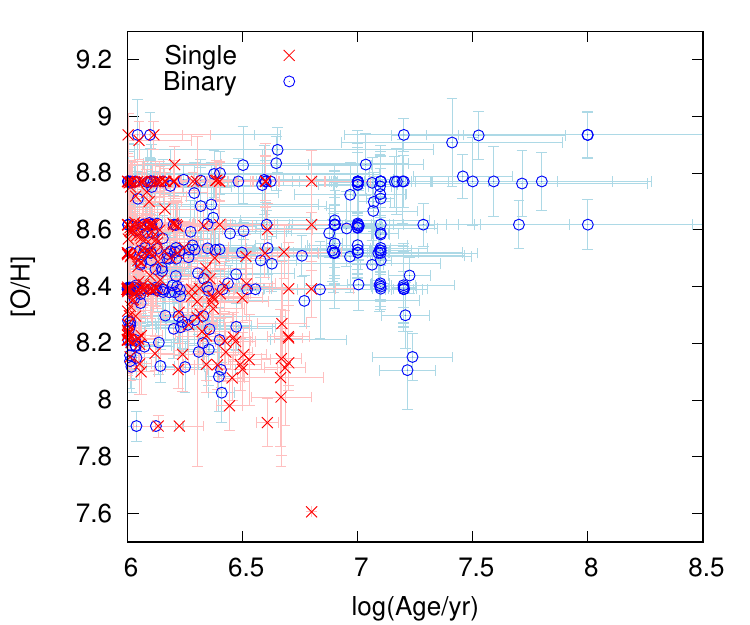}\\
\centering
\caption[The relationship of predicted oxygen abundance with ionization parameter $ \log(U) $, $ {\rm n_H} $ and age from \textsc{bpass} best-fitting models, allowing for the consideration of ionzing photon leakage]{Similar to Figure \ref{fig:MeanEW} but including ionizing photon loss, the relationship of [O/H] ($ = 12 + \log({\rm O/H} $ ) with ionization parameter $ \log(U) $ is shown in the top panel, with $ {\rm n_H} $ in the middle and with age in the bottom panel from \textsc{bpass} best-fitting models. Single models are in red crosses with error bars and binary models in blue circles with error bars.} \label{fig:MeanEW_leakage}
\end{figure}

The most important parameters that affect the ionization conditions of a model are its $ \log(U) $, $ {\rm n_H} $ and age as we discussed above. Figure \ref{fig:MeanEW_leakage} demonstrates the variations of these parameters from the new best-fitting models allowing for photon leakage, which can be compared to Figure \ref{fig:MeanEW_Ps}. The $ \log(U) $ distribution tends to shift slightly toward  lower values for both single-star and binary-star models. Since we use $ \log(U) $ to derive the inner radius for the H\,II regions, lower $ \log(U) $ means a smaller radius of the gas cloud which corresponded to a hotter inner region. In addition, we note that ionizing photon loss pushes both single and binary star models toward lower hydrogen density estimates. Significant numbers of best-fitting models have $ {\rm \log(n_H)} \leq 1.0 $, which is consistent with the fact that ionizing photons are more likely to escape without reprocessing by nebular gas in less dense clouds. This adjustment brings the best fitting model gas densities closer into alignment with those derived from the [S\,II] line ratios, which supports the idea that photon escape from this sample is relatively high.

In the bottom panel of Fig. \ref{fig:MeanEW_leakage}, we note that all our single-star models become very young with most having age $ \leq $ 3\,Myr and a fit is not possible for single-star models older than 5\,Myr. Binary-star models cover a wide range of age, but also shift towards younger populations. Most of them are found within 20\,Myr with a fraction of them even below 5\,Myr. The selection of younger models is made possible in this analysis because leaking H\,II regions can be young and still show low observed H$ {\rm \beta} $ equivalent widths. The preference for these models in both single- and binary-star populations suggests that the ionizing spectrum in these regions is typically comparatively hard, and also that a consistent solution matching the hardness of the spectrum and its Balmer line strength requires high loss fractions of ionizing photons.

\subsection{How much ionizing photon loss is required?}
\begin{figure}
\centering
\includegraphics[width=9cm]{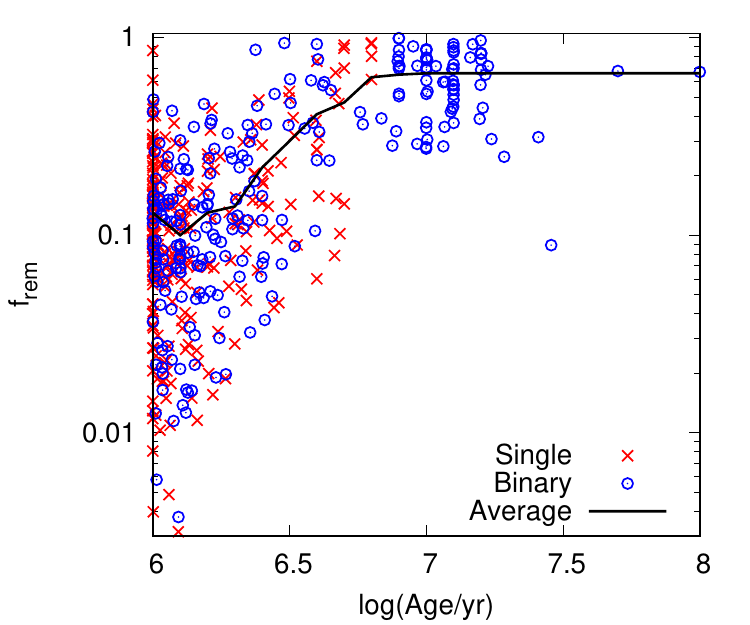}
\centering
\caption[The averaged value of $ f_{rem}$ as a function of age derived from best-fitting \textsc{bpass} models]{The derived value of $ f_{rem}$, the remaining ionizing photon fraction reprocessed by nebular emission after photon escape, from best-fitting \textsc{bpass} models. The red crosses are the distribution of best fitting values from single-star models and blue circles are from binary-star models. A running average of the derived values with stellar population age is given by the black line.} \label{fig:cover_factor}
\end{figure}

\noindent We now explore how the fraction of ionizing photons lost from the H\,II regions without exciting nebular emission varies with cluster age for our best-fitting models. To simplify the calculation we assume the strength of H$ {\rm \beta} $ emission line is decreased by the same fraction as H$ {\rm \beta} $ equivalent width. The fraction of remaining ionizing photons $ f_{\rm rem}$ (i.e. those that are not lost) is shown in Figure \ref{fig:cover_factor}. This remaining ionizing photon fraction is derived as follows:
\begin{equation}
EW_{obs} = \dfrac{F_{H\beta} \times f_{\rm rem}}{F_{*} + \bigtriangleup F \times f_{\rm rem} }
\end{equation}
where $ EW_{obs} $ is the observed H$ {\rm \beta} $ EW, $ F_{H\beta} $ is the H$ {\rm \beta} $ emission line flux, $ F_{*} $ is the stellar population continuum that is incident onto the nebulae, and $ \bigtriangleup F = F - F_{*}$ is the net increase of nebular continuum relative to the stellar input spectra. Therefore, we can derive the fraction of remained ionizing photons - $ f_{\rm rem}$ as a relation between observed H$ \beta $ EW and our model H$ \beta $ EW,
\begin{equation}
f_{\rm rem} = \dfrac{q \times EW_{obs}}{EW_{H\beta} - (1-q)\times EW_{obs}}
\end{equation}
where $ q = F_{*}/F$ is the ratio of input stellar continuum to the nebular continuum and $ EW_{H\beta} $ are model values for the H$ {\rm \beta} $ equivalent width. 

Fig. \ref{fig:cover_factor} illustrates the fraction of the remaining ionizing photons $ f_{\rm rem} $ as function of age. We also calculate the average fraction $ \bar{f}_{\rm rem} $ of the models as a function of age. The fraction gradually increases with increasing age from about 15 per cent up to about 70 per cent within the first 10\,Myr. Due to our selection method, there are more younger models with a large variations than the older ones. 
We present the exact value of the averaged remaining fraction in each model age bin in Table \ref{tab:leak_ave}, which quantifies the effect of ionizing photon leakage in theoretical models for a better understanding of observations. We note that we assumed that nebular continuum and emission lines are suppressed equally by photon loss. Since continuum emission can emerge from a larger region, with lower gas densities, than line emission, this is not necessarily true. If the continuum is less suppressed than the line emission, our fraction of photon loss may be slightly underestimated. 
\begin{table}
\caption{Average fraction of remaining ionizing photons $ \bar{f}_{\rm rem}  $ with 1-$ \sigma $ standard deviation, as a function of stellar population age.}

\begin{center}
                      \begin{tabular}{l @{\hskip 0.3in} | @{\hskip 0.3in} c}
                        \hline
                        \hline
                        \rule{0pt}{3.4ex}
                         log(Age/yr)    & $ \bar{f}_{\rm rem}  $  \\
                        
                        \hline 
                         
                       6.0 & 0.13 $ \pm $ 0.12 \rule{0pt}{2.4ex} \\ 
                       6.1 & 0.10 $ \pm $ 0.09  \rule{0pt}{2.4ex} \\ 
                       6.2 & 0.13 $ \pm $ 0.13  \rule{0pt}{2.4ex} \\ 
                       6.3 & 0.15 $ \pm $ 0.08 \rule{0pt}{2.4ex} \\ 
                       6.4 & 0.22 $ \pm $ 0.17 \rule{0pt}{2.4ex} \\ 
                       6.5 & 0.30 $ \pm $ 0.20 \rule{0pt}{2.4ex} \\ 
                       6.6 & 0.41 $ \pm $ 0.23 \rule{0pt}{2.4ex} \\ 
                       6.7 & 0.47 $ \pm $ 0.23 \rule{0pt}{2.4ex} \\ 
                       6.8 & 0.61 $ \pm $ 0.25  \rule{0pt}{2.4ex} \\     
                       6.9 & 0.64 $ \pm $ 0.26 \rule{0pt}{2.4ex} \\      
                       7.0 & 0.65 $ \pm $ 0.25  \rule{0pt}{2.4ex} \\ 
                       7.1 - 8.0 & 0.66 $ \pm $ 0.26 \\
                       
                        \hline
 
\end{tabular}
\end{center}
\label{tab:leak_ave} 
\end{table}

\section{Discussion \& Conclusions}
We have undertaken a detailed investigation into the physical conditions giving rise to diagnostic optical emission line ratios in the nearby H\,II region sample from \cite{1998AJ....116.2805V} and \cite{2006ApJ...636..214V}. We combine \textsc{bpass} population synthesis code with the photonization model \textsc{cloudy} to calculate the nebular component (line and continuum emission) of model H\,II regions and study the effect of interacting binary evolution pathways on observed properties of the H\,II regions. We use BPT diagrams to show the comparison between our models and observed H\,II regions in emission line ratios. Then we find the best-fitting models for each individual H\,II region after a selection based on the quantities of the four line ratios in the BPT diagrams and $ {\rm EW_{H{\rm \beta}}} $. These best-fitting models allow us to place constraints on the physical properties of the observed H\,II regions, such as oxygen abundance [O/H], age, $ U $ and hydrogen density. We summarize and discuss our findings as below:

\begin{enumerate}

\item Single-star and binary-star models occupy similar regions in the BPT diagrams at early ages of $ {\rm \log(Age/yr) \leq 6.5}$. At constant age, our models form separated tracks varying with metallicity. As metallicity increases toward Z=0.008 the model tracks overlap more with observed H\,II regions in the BPT diagrams. At young ages the model H\,II regions lay under the maximal starburst lines derived by \cite{2001ApJ...556..121K}, however certain combinations of age, metallicity and gas parameters push the models above these lines. 

At older ages of $ {\rm \log(Age/yr)} = 7.0$ and above, binary-star and single-star model behaviour diverges, with single-star models quickly dying away and binary-star models becoming hotter and moving up to provide a better match to most of the observed H\,II regions. At $ {\rm \log(n_H)} = 2 $, all the model tracks are still under the Kewley et al maximal curve, but when $ {\rm \log(n_H)}  $ increases to 3, the binary-star model tracks can over pass above this curve and forms an offset locus in BPT parameter space, as high-redshift star-forming galaxies also do. We define a new set of maximal starburst line relations and present these in Appendix \ref{sec:maximal}.

\item We investigate the effect of the choice of metallicity calibration on the measurement of oxygen abundance from strong line indicators. We find that two methods $ - $ N2-vanZee and D16 $ - $ show good agreement to the known oxygen abundance in our synthetic single and binary-star models, while another two methods $-$ Pilyugin-R2D and Pilyugin-ONS $ - $  show a good match when binary-star models are used. We also consider a direct comparison between the metallicities of our best fitting models and those previously inferred for this sample.

As a result of the good match between the N2-vanZee calibration and our model calibration, both our single and binary star models provide a fairly good estimation for the H\,II regions with medium [O/H], while failing to agree for those of [O/H] $ \geq $ 9.0 and $ \leq $ 8.2. In these significantly metal-poor or metal-rich H\,II regions, the metallicity calibrations for strong-line methods vary over a large range and cause difficulties in determining the accurate metal abundance for any given region. 

\item In addition, we have constrained the ionization parameter and age of the H\,II regions from the best-fitting models as these two parameters are very important for our understanding of the ionization state of the gas cloud and properties of the inner ionizing source.   Ionization parameters $ {\rm \log(U)} $ between -3 and -2 are most often selected by the best-fitting models and we find that $ {\rm \log(U)} $ is inversely proportional to [O/H]. This result is consistent with the fact that H\,II regions are hot at lower [O/H] and therefore the gas cloud is highly ionized with a larger value of $ {\rm \log(U)} $ and vice versa. At given $ {\rm \log(U)} $, binary-star models have a relatively higher [O/H] than those from single-star models which is caused by the variation of model age between these two populations. Single-star models are younger, with best fit values predominantly around 5\,Myr while binary-star models have older age from 5 to 10\,Myr and some binary-star models can give an age prediction extending to 50\,Myr. From $ {\rm \log(U)} $, we also derived the inner radius of the illuminating surface of the gas cloud with respect to the mass of enclosed stars in HIIs-DG, and find that the radius gradually increases with increasing mass. Compared to single-star models, binary-star models are more massive and hotter and push the illuminating surface further away to 300pc. Our estimated values are roughly consistent with size of star formation regions which scale from 2 to 100 pc \citep{2014MNRAS.442.3711G}. 

\item Finally we discussed the effect of ionizing photon leakage caused by photons escaping from their H\,II regions, with no interaction with gas or dust attenuation. This acts to reduce the line strength and equivalent width of a given region. It affects our selection for best-fitting models due to the variation in $ {\rm EW_{H\beta}} $, so that younger models with ionizing photon leakage can reach same level of $ {\rm EW_{H\beta}} $ as older H\,II regions in which no leakage is permitted. Our new best-fitting models are shifted toward younger ages, lower [O/H] and lower $ {\rm \log(U)} $ which overall suggests that the models are hotter, younger and push the gas cloud further away. Under the new selection method, we find single-star models predict values of [O/H] for HIIs-SG that are closer to those of the vanZee-N2 calibrations. However both single star and binary models still estimate a higher [O/H] value for HIIs-DG than that derived by \cite{2006ApJ...636..214V}. We also determine the fraction of remaining ionizing photons $ f_{\rm rem} $ as a function of age. The youngest models have more ionizing photon leakage with only about 15 per cent of photons processed by the nebular gas and this fraction gradually rises up to about 70 per cent as age increases beyond 10\,Myr. 

We note that the modelling of ionizing radiation fields is an active and evolving area. The work of \citet[2018 submitted]{2017ApJ...840...44B} has demonstrated that incorporation of stellar rotation has similar effects to the incorporation of binary evolution pathways and can produce similarly good fits to observational data. On the other hand, the effects of binary interaction are stronger at low metallicities and tend to prolong the lifetime of ionizing photon production further. Similarly the nebular emission models of \citep{2016MNRAS.462.1757G} have demonstrated the effectiveness of single star spectral synthesis incorporating some of the most recent stellar atmosphere models, and with evolutionary tracks extending to 350\,M$_\odot$ (above the upper mass limit of BPASS). Rotating star evolution is not considered in detail in BPASS, although some of its effects are simulated in our binary models. Given that stellar rotation is often a consequence of binary mass transfer, it is reassuring that these effects push models in the same direction. Nonetheless, given the known prevalence of multiple stellar systems, particularly amongst massive stars, it is clear that binary evolution pathways provide an additional modification of any output spectra, on top of that introduced by pure rotating models. As a result, any model which does not incorporate mass tranfer, merger and other binary evolution effects is necessarily missing important physical behaviour in a stellar population.
\end{enumerate}

\section*{ACKNOWLEDGEMENTS}
\addcontentsline{toc}{section}{Acknowledgement}
This work would not be possible without the NeSI high-performance computing facilities and the staff at the Centre for eResearch at the University of Auckland. New Zealand's national facilities are provided by the New Zealand eScience Infrastructure (NeSI) and funded jointly by NeSI's collaborator institutions and through the Ministry of Business, Innovation and Employment's Infrastructure programme. URL: \url{http://www.nesi.org.nz}. 

LX would like to thank the China Scholarship Council for its funding her PhD study at the University of Auckland and the travel funding and support from the University of Auckland. JJE acknowledges travel funding and support from the University of Auckland. ERS acknowledges travel funding and support from the University of Warwick. 

We thank the anonymous reviewer of this paper for valuable input, and we also thank Nell Byler and coauthors for contacting us with their comments, discussions and for making their 2018 paper draft available to the authors.

\appendix

\section{Defining the BPASS Maximal Starburst Relations}
\label{sec:maximal}

Comparisons have been made throughout this paper to the `maximal' starburst lines defined by Kewley et al (2001) based on the combination of {\sc PEGASE2} and {\sc STARBURST99} stellar population models with the {\sc MAPPINGS III} radiative transfer code. 

For future reference, we define a similar set of metallicity-dependent maximal starburst lines derived using {\sc BPASS} v2.1 and the CLOUDY radiative transfer models described in this paper.  These are a set of equations for an upper locus which encloses all possible star formation powered H\,II regions with $0<$log(n$_\mathrm{H}$)$<3$,  $-3.5<$log(U)$<-1.5$ and $6.0<$log(Age/yr)$<7.5$, assuming the inclusion of binary star populations.

  For each of the three BPT diagrams, and at each metallicity in our model grid, we define a line using the same functional form as Kewley et al (2001):
  \begin{equation}
    \hspace{0.5cm} \log(\mathrm{[O\,{III}]/H}\beta) = \frac{A_i}{\log(X_i)+B_i}+C_i,
  \label{eqn:maxlines}
  \end{equation}
  where $X_i$=[N\,{\sc II}]/H$\alpha$, [S\,{\sc II}]/H$\alpha$ or [O\,{\sc I}]/H$\alpha$, while $A$, $B$ and $C$ are constants.  We also derive a line encompassing all metallicities. Derived values for all constants are given in table \ref{tab:maxlines}
  
 \begin{figure}
\vspace*{-0.5cm}
  \includegraphics[width=7.5cm]{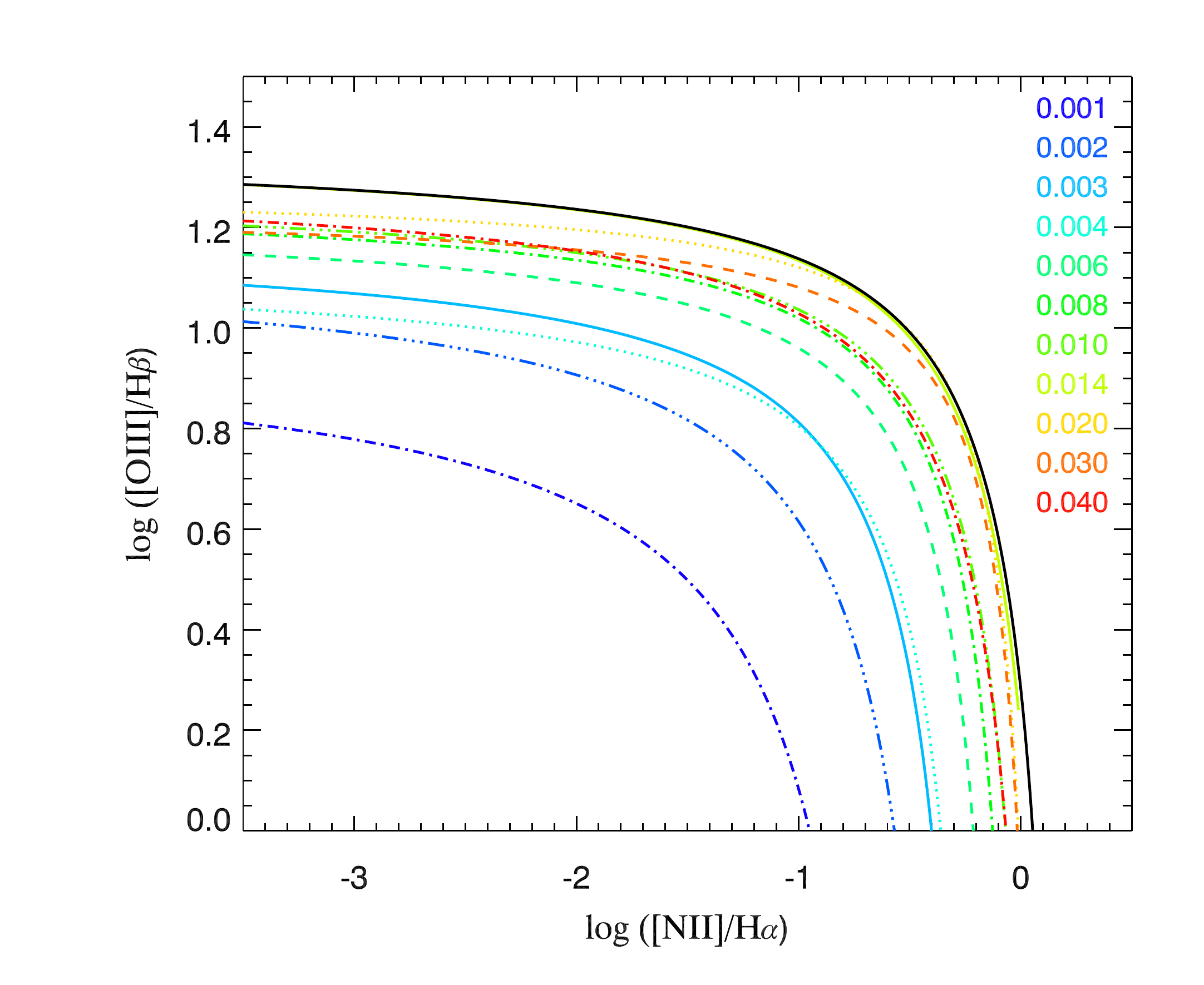}
\vspace*{-0.3cm}
\caption{The metallicity dependence of the BPASS binary maximal starburst in the [N\,{\sc II}]/H$\alpha$ vs  [O\,{\sc III}]/H$\beta$ BPT diagram. Metallicity dependent lines are shown in colour, while the solid black line gives the overall maximal starburst line for all BPASS binary models\label{fig:maxlines1}.}
\end{figure}

\begin{figure}
\vspace*{-0.7cm}
  \includegraphics[width=7.5cm]{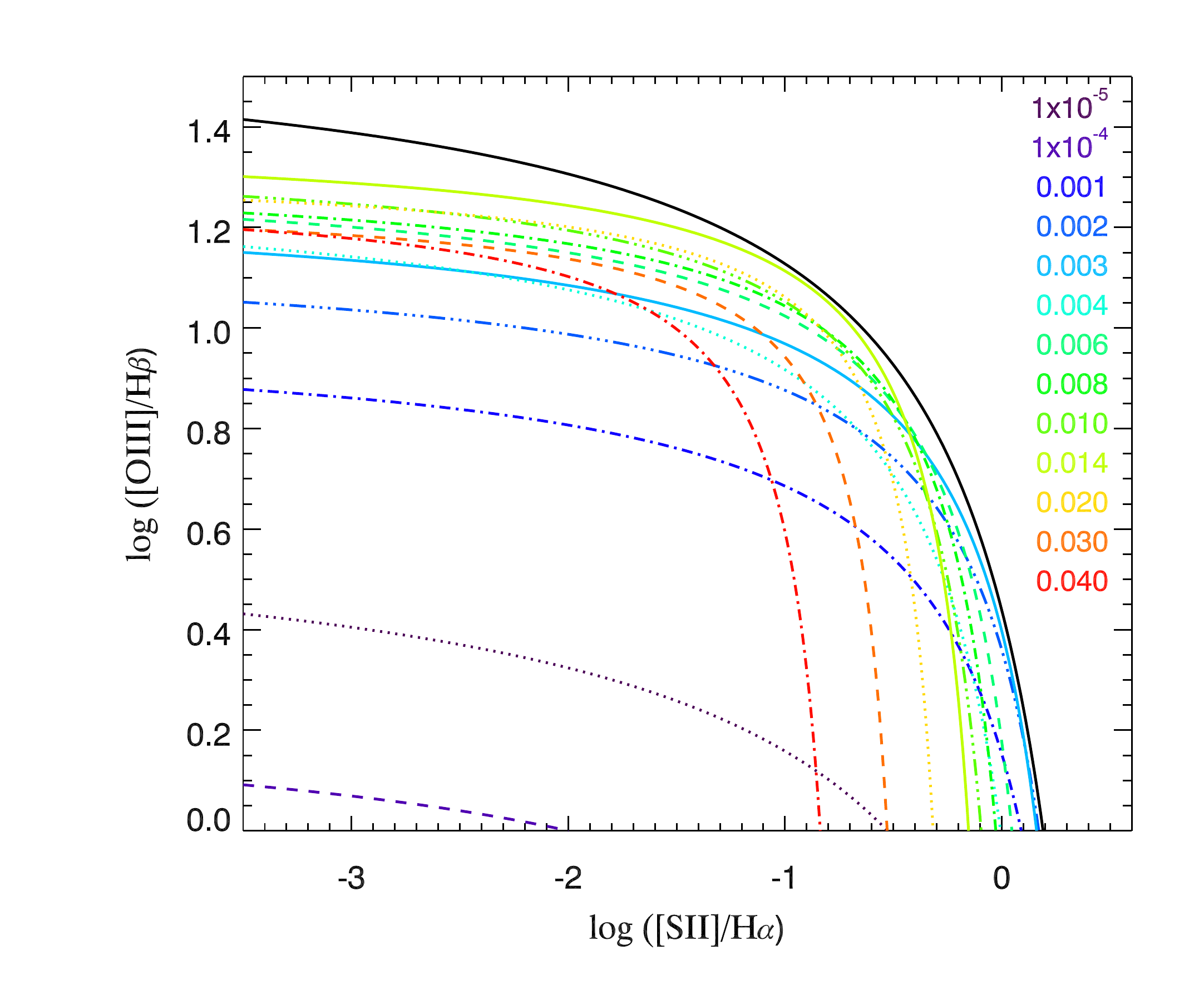}
  \vspace*{-0.3cm}
\caption{The metallicity dependence of the BPASS binary maximal starburst in the [S\,{\sc II}]/H$\alpha$ vs  [O\,{\sc III}]/H$\beta$ BPT diagram. Metallicity dependent lines are shown in colour, while the solid black line gives the overall maximal starburst line for all BPASS binary models\label{fig:maxlines2}.}
\end{figure}

\begin{figure}
\vspace*{-0.7cm}
  \includegraphics[width=7.5cm]{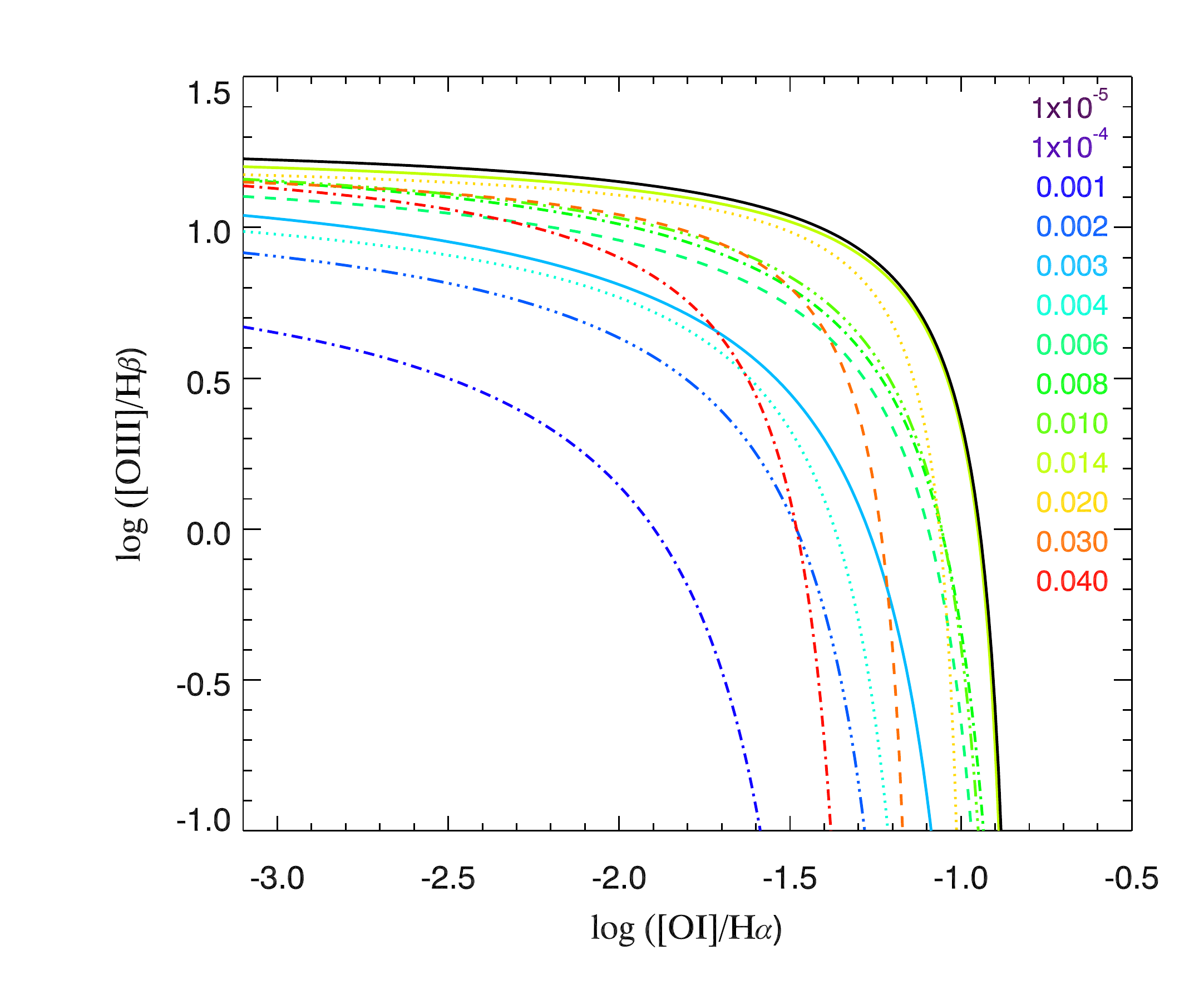}
  \vspace*{-0.3cm}
\caption{The metallicity dependence of the BPASS binary maximal starburst in the [O\,{\sc I}]/H$\alpha$ vs  [O\,{\sc III}]/H$\beta$ BPT diagram. Metallicity dependent lines are shown in colour, while the solid black line gives the overall maximal starburst line for all BPASS binary models\label{fig:maxlines3}.}
\end{figure} 

We also illustrate the dependence of these maximal lines on metallicity in Figures \ref{fig:maxlines1},  \ref{fig:maxlines2} and  \ref{fig:maxlines3} respectively. \\

\begin{table*}
  \begin{tabular}{cccc|ccc|cccc}
    Z  & \multicolumn{3}{c}{$X_1$=[N\,{\sc II}]/H$\alpha$} & \multicolumn{3}{c}{$X_2$=[S\,{\sc II}]/H$\alpha$} & \multicolumn{3}{c}{$X_3$=[O\,{\sc I}]/H$\alpha$}\\
       &\hspace{0.5cm} $A_1$ & $B_1$ & $C_1$ &\hspace{0.5cm} $A_2$ & $B_2$ & $C_2$ & \hspace{0.5cm} $A_3$ & $B_3$ & $C_3$\\
    \hline \hline
  $1\times10^{-5}$ & \hspace{0.5cm}   0.74 &   2.31 &  -0.72  &\hspace{0.5cm}  0.92   &  -0.91  & -0.64 &\hspace{0.5cm}   1.04&  3.01  &  -0.62\\
  $1\times10^{-4}$ & \hspace{0.5cm}   0.78 &   1.27 &   0.26  &\hspace{0.5cm}  0.72   &  -0.77  &  0.26 &\hspace{0.5cm}   1.00&  2.03  &   0.32\\
            0.001 &  \hspace{0.5cm}  0.52 &   0.42 &   0.98  &\hspace{0.5cm}   0.50  &  -0.59 &   1.00 & \hspace{0.5cm}  0.64&   1.27 &   1.02\\
            0.002 &  \hspace{0.5cm}  0.42 &   0.20 &   1.14  &\hspace{0.5cm}   0.44  &  -0.55 &   1.16 &\hspace{0.5cm}   0.50&   1.05 &   1.16\\
            0.003 &  \hspace{0.5cm}  0.32 &   0.13 &   1.18  &\hspace{0.5cm}   0.44  &  -0.51 &   1.26 &\hspace{0.5cm}   0.54&   0.85 &   1.28\\
            0.004 &  \hspace{0.5cm}  0.28 &   0.11 &   1.12  &\hspace{0.5cm}   0.54  &  -0.41 &   1.30 &\hspace{0.5cm}   0.40&   1.03 &   1.18\\
            0.006 &  \hspace{0.5cm}  0.26 &   0.00 &   1.22  &\hspace{0.5cm}   0.40  &  -0.35 &   1.32 &\hspace{0.5cm}   0.36&   0.81 &   1.26\\
            0.008 &  \hspace{0.5cm}  0.26 &  -0.08 &   1.26  &\hspace{0.5cm}   0.34  &  -0.23 &   1.32 & \hspace{0.5cm}  0.38&   0.77 &   1.32\\
            0.010 &  \hspace{0.5cm}  0.28 &  -0.15 &   1.28  &\hspace{0.5cm}   0.36  &  -0.17 &   1.36 &\hspace{0.5cm}   0.32&   0.81 &   1.30\\
            0.014 &  \hspace{0.5cm}  0.28 &  -0.24 &   1.36  &\hspace{0.5cm}   0.28  &  -0.05 &   1.38 & \hspace{0.5cm}  0.18&   0.81 &   1.28\\
            0.020 &  \hspace{0.5cm}  0.18 &  -0.13 &   1.28  &\hspace{0.5cm}   0.22  &   0.15 &   1.32 & \hspace{0.5cm}  0.14&   0.95 &   1.24\\
            0.030 &  \hspace{0.5cm}  0.18 &  -0.13 &   1.24  &\hspace{0.5cm}   0.20  &   0.37 &   1.26 &\hspace{0.5cm}   0.18&   1.09 &   1.24\\
            0.040 &  \hspace{0.5cm}  0.32 &  -0.18 &   1.30  &\hspace{0.5cm}   0.24  &   0.65 &   1.28 &\hspace{0.5cm}   0.30&   1.25 &   1.30\\
  \hline
  All &   \hspace{0.5cm}   0.28  &  -0.26 &    1.36 &  \hspace{0.5cm} 0.82 & 0.70 & 1.61 &   \hspace{0.5cm} 0.19 & 0.80 & 1.31\\
  \end{tabular}
  \caption{Maximal starburst relation parameters as defined in equation \ref{eqn:maxlines} for $X_1$=[N\,{\sc II}]/H$\alpha$,  $X_2$=[S\,{\sc II}]/H$\alpha$  and $X_3$=[O\,{\sc I}]/H$\alpha$, vs [O\,{\sc III}]/H$\beta$.\label{tab:maxlines}}
  \end{table*}


\section{Photoionization Diagnostics in the Ultraviolet}\label{sec:uvphoto}

The focus of this paper has been on the optical BPT diagrasm as  photoionization diagnostics, both due to their widespread use as a discriminator between AGN and starburst galaxies, and due to the extensive extant data on local star forming regions which form the core of this analysis.

As near-infrared spectroscopy has become more sensitive and efficient, and particularly with the imminent advent of space-based spectroscopy using the {\it James Webb Space Telescope}, these optical strong line diagnostics are being used across an increasingly large redshift range, including results based on BPASS (e.g. Steidel et al 2014, Strom et al 2017).

However at the highest redshifts ($z>4$), the rest-frame optical is shifted to wavelengths currently inaccessible to spectroscopy, and which will be challenging to exploit even with {\it JWST}. For these distant objects, and for the large samples at intermediate redshifts ($z\sim2-4$) where no infrared spectroscopy is currently available, the rest-frame ultraviolet is likely to remain the spectral regime most often used for photoionization studies. Ultraviolet spectroscopy and emission lines also provide a more direct indication of the spectral shape close to the 912\,\AA\ Lyman limit and so can yield valuable information in extreme photoionization regions nearby, as long as ultraviolet spectroscopy (again requiring space-based observations) is available.

Driven by high redshift galaxy studies, particular interest has been paid recently to the use of rest-ultraviolet lines as AGN diagnostics, in a parallel manner to the use of the BPT diagrams in the optical. An AGN versus stellar photoionization diagnostic involving the C\,III] 1907 and C\,III] 1910\AA\ emission lines, the OIII] 1661+1666\AA\ line doublet and the He\,II 1640\AA\ feature has been proposed by Feltre et al (2016), and recently applied by Mainali et al (2017). Here we consider the location of our photoionization grid (as described in the main text) on this diagnostic diagram.

\begin{figure}
  \includegraphics[width=\columnwidth]{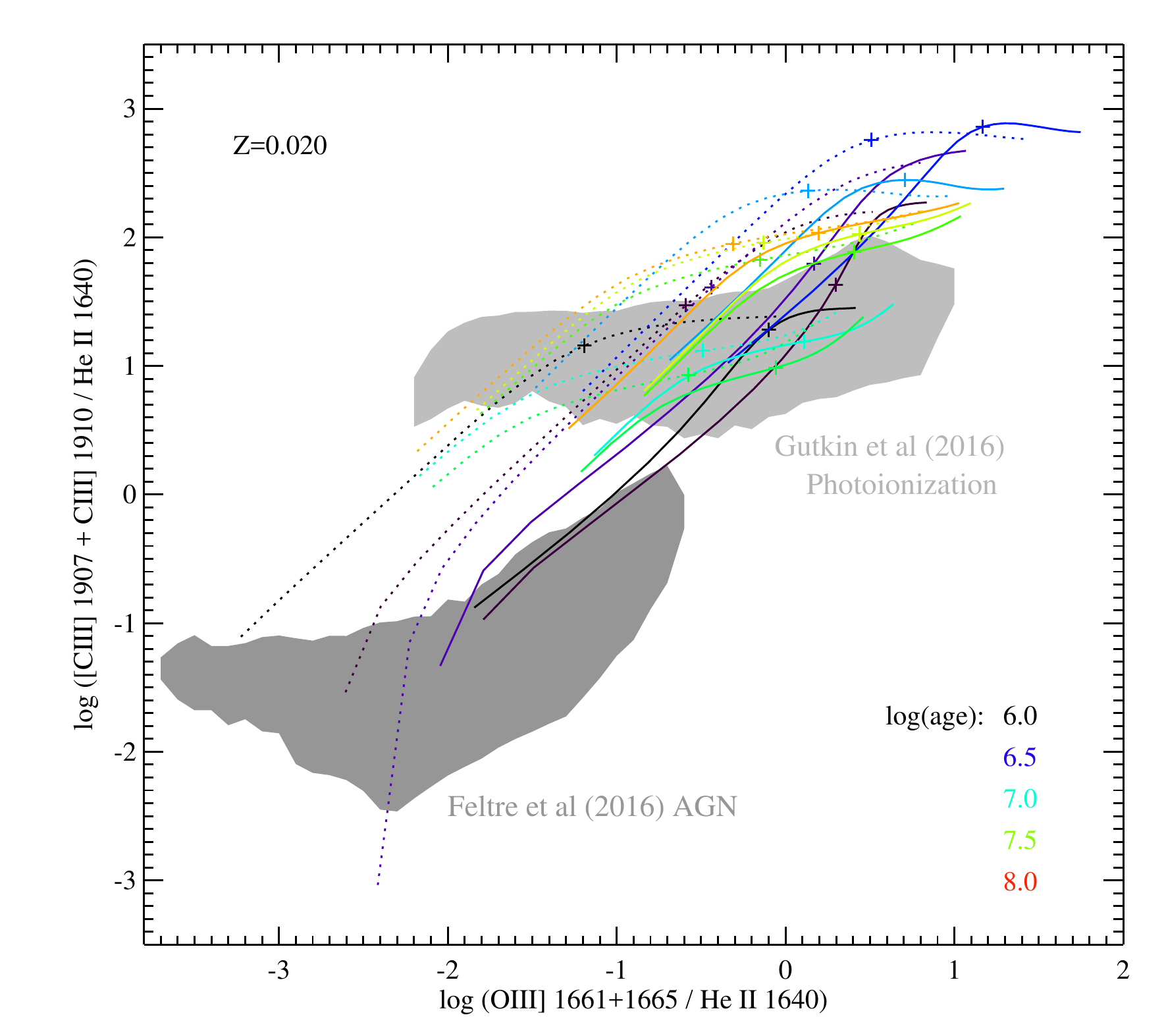}
  \includegraphics[width=\columnwidth]{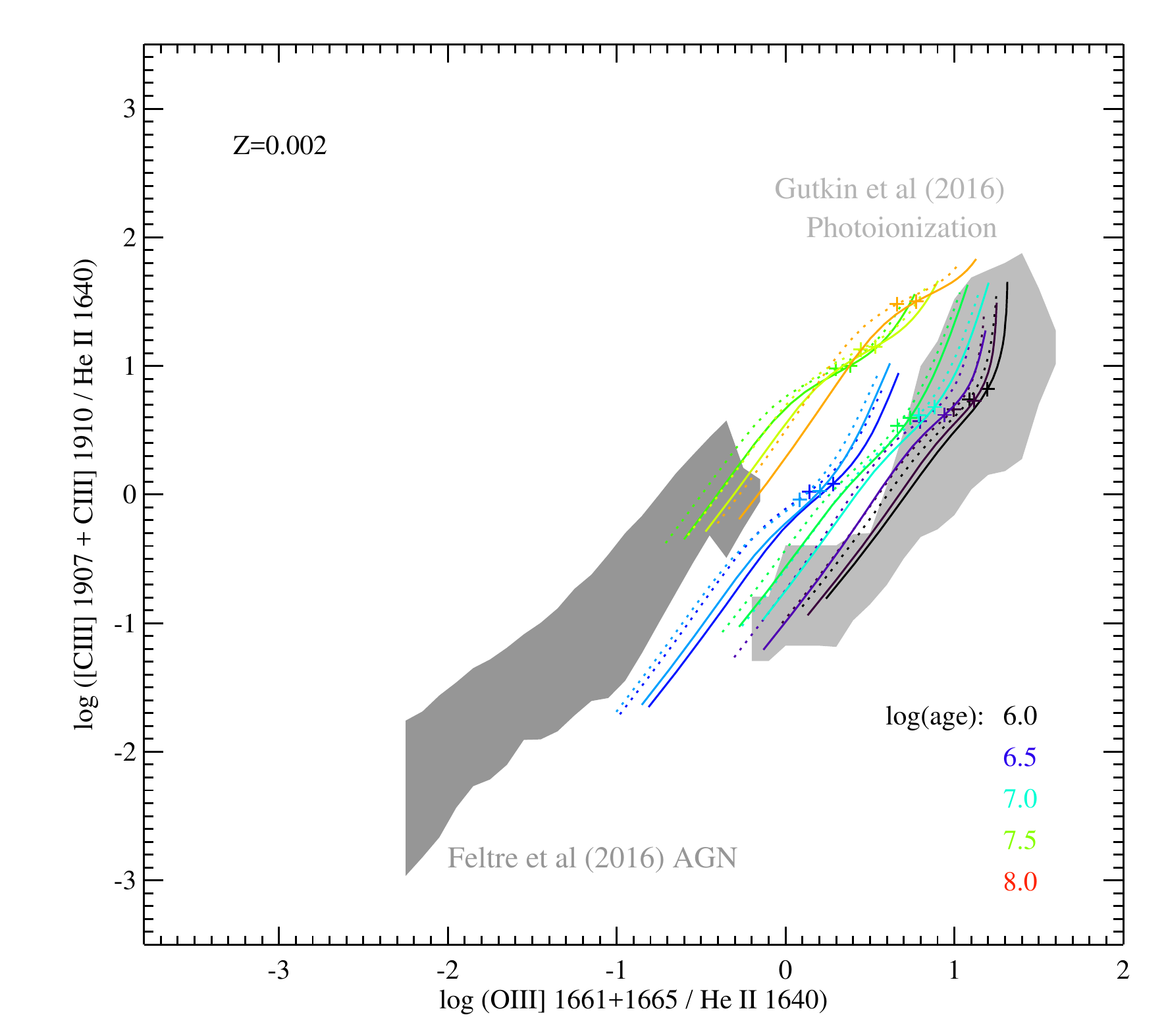}
\caption{The rest-frame ultraviolet line ratio photoionization diagnostic parameter space proposed by Feltre et al (2016), as described in the text. Coloured lines indicate the values measured for BPASS binary models with log($n_H$)=3.0 (solid lines) and log($n_H$)=2.0 (dotted lines). Ionization parameter varies along lines, with log($U$)=-2.0 indicated by a cross. Each line is plotted at a different stellar population age, ranging from log(age/yrs)=6.0 (black) to log(age/yrs)=8.0 (red). We show the distribution at Solar metallicity (top) and a tenth Solar (bottom) \label{fig:uv_feltre1}.}
\end{figure}

In figure \ref{fig:uv_feltre1} we illustrate how varying stellar age, gas phase electron density and ionization parameter affect location in the C\,III] vs O\,III] rest-ultraviolet parameter space, for BPASS binary models at two representative metallicities. Ionization parameter varies from upper right to lower left along tracks, each of which is at a different stellar population age. Models with log($n_H$)=3.0 are shown as solid lines, and those with log($n_H$)=2.0 as dotted lines. For comparison, we also show the regions occupied by the constant star-formation rate photoionization models reported by Gutkin et al (2016, $-4<$log($U$)$<0$)), and by the AGN photoionization models of Feltre et al (2016,$-5<$log($U$)$<-1$) at the same metallicity mass fraction.

Despite the narrower range of ionization parameters explored in our analysis ($-3.5<$log($U$)$<-1.5$), allowing for a larger range of stellar population ages, rather than requiring constant star formation, leads to the BPASS models presented in this paper spanning a wider range of line ratios than those of the Gutkin et al (2016) dataset. Importantly, there is no clear distinction in this parameter space between our models and the AGN photoionization models of Feltre et al (2016) at any given metallicity.

\begin{figure*}
  \includegraphics[width=1.9\columnwidth]{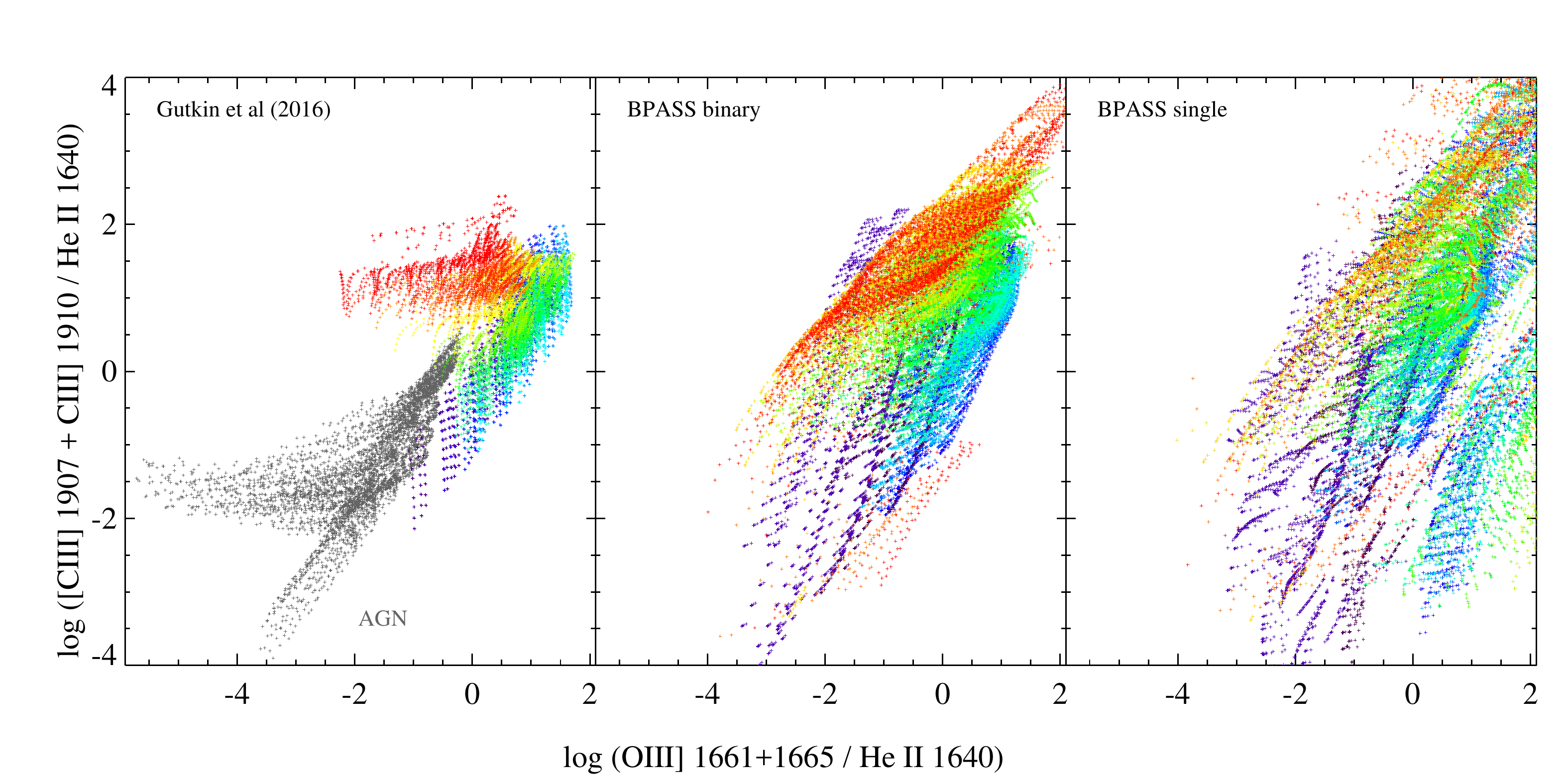}
\caption{As in figure \ref{fig:uv_feltre1}, but now showing the full distribution of the models presented in this paper in this parameter space. The central panel shows values for binary models. The right hand models give the same, but this time for single star models. The left hand panel shows the stellar photoionization models of Gutkin et al. (2016) and AGN models of Feltre et al (2016) for comparison. Stellar photoionization models are coloured from low metallicity (blue) to high metallicity (red). \label{fig:uv_feltre2}}
\end{figure*}

In figure \ref{fig:uv_feltre2}, we show the full range of BPASS models for both binary star populations (centre panel) and single stars (right hand panel). As comparison with the Gutkin et al (2016) and Feltre et al (2016) models (left hand panel) demonstrates, our analysis suggests that this parameter space may not be so clean an AGN versus stellar photoionization diagnostic as previously suggested.

\begin{figure*}
  \includegraphics[width=\columnwidth]{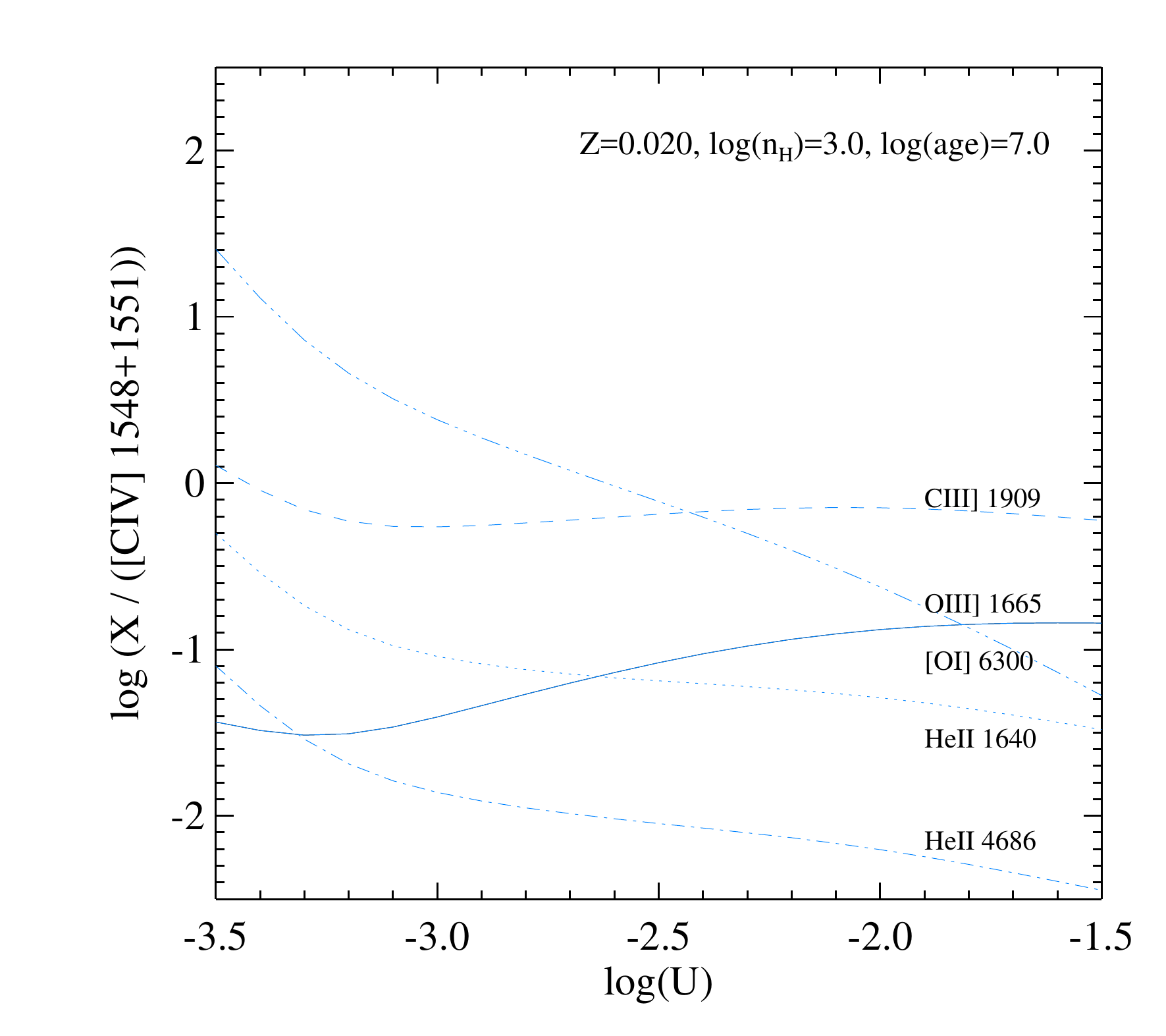}
  \includegraphics[width=\columnwidth]{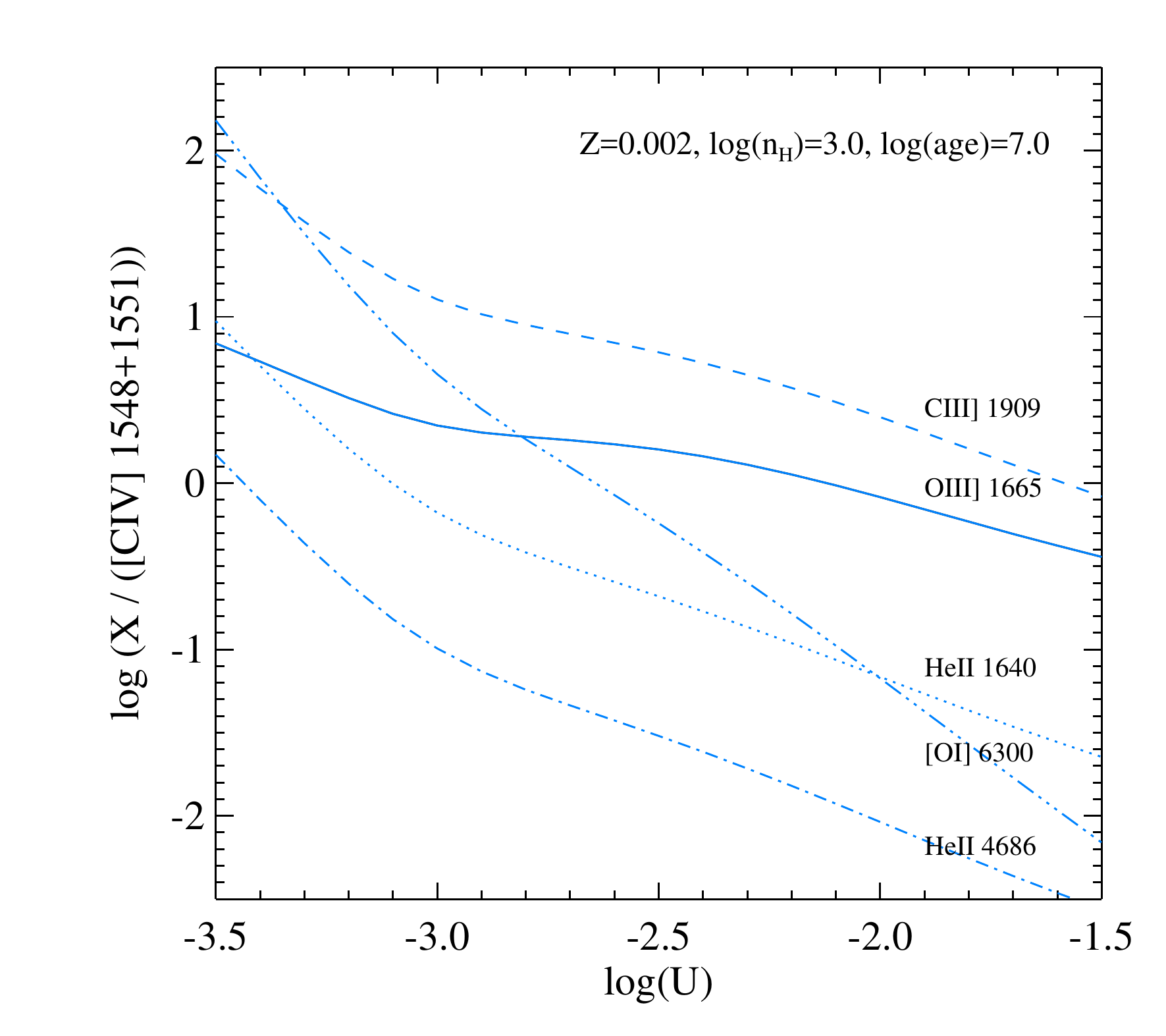}
  \includegraphics[width=\columnwidth]{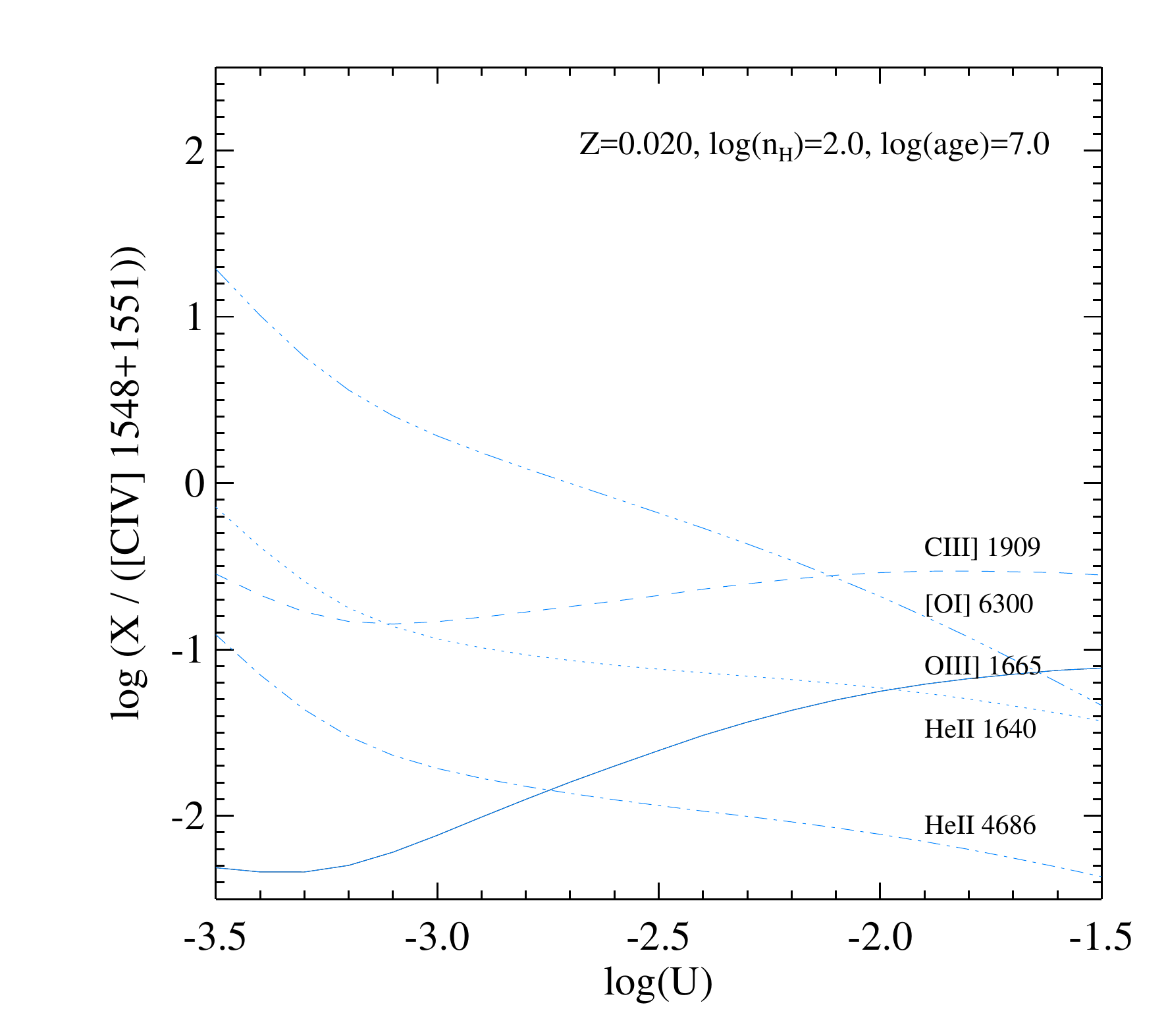}
  \includegraphics[width=\columnwidth]{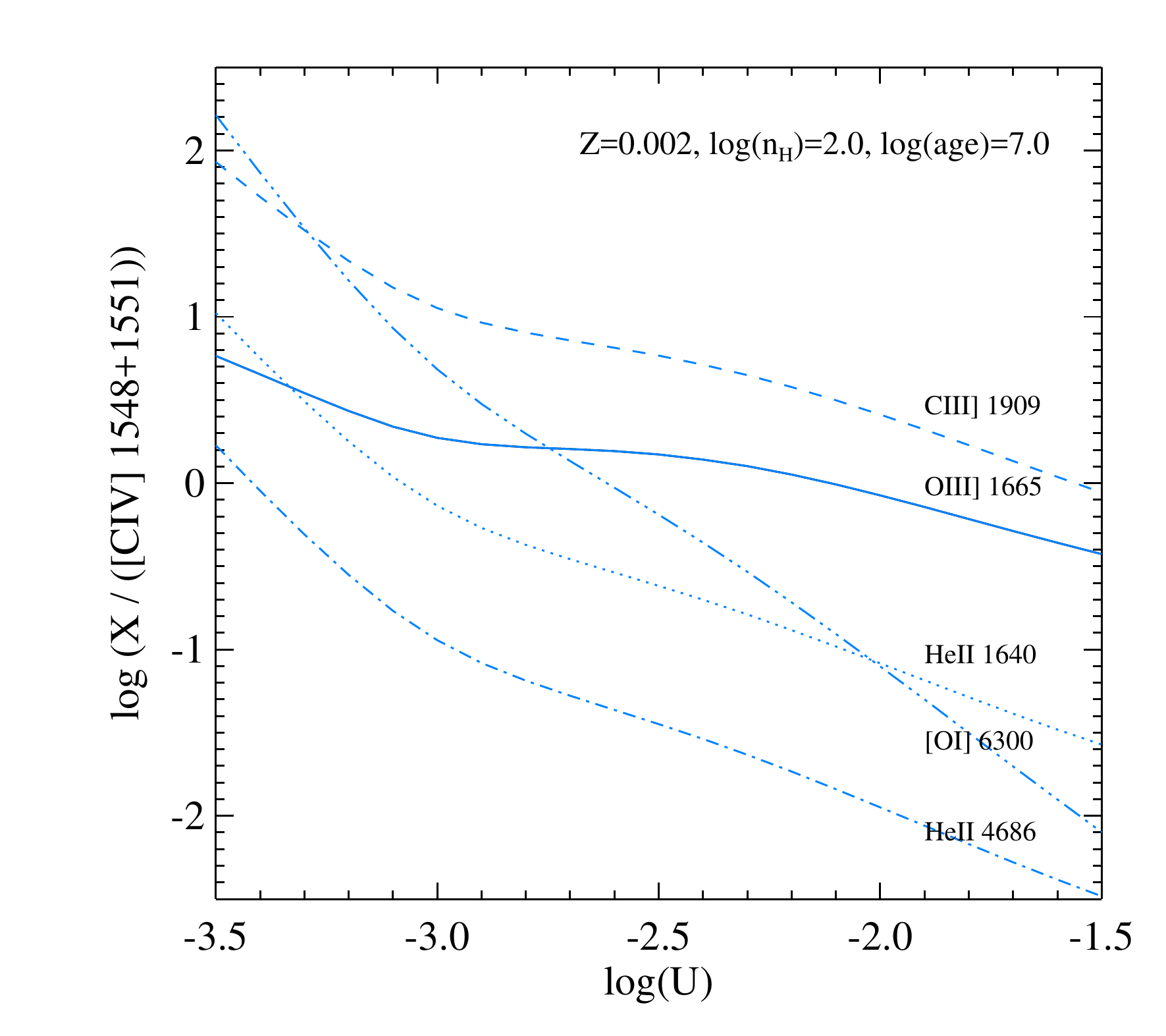}
  \caption{The strengths of a range of rest-frame ultraviolet and optical emission lines, as a function of ionization parameters. Line strengths are scaled to the [C\,IV] 1548+1551\AA\ doublet, and are shown for BPASS binary models at a stellar population age of 10\,Myr and electron densities  log($n_H$)=2.0 (bottom panels) and 3.0 (top). Where the lines plotted are doublets, the emission strength of the two elements are summed before the ratio to [C\,IV] is taken.
 We show the line strengths at Solar metallicity (left) and a tenth Solar (right) \label{fig:uv_others}.}
\end{figure*}

Clearly, this analysis can be extended to the full range of emission lines, and in figure \ref{fig:uv_others}, we illustrate the variation in a range of line ratios, relative to the strength of the rest-frame ultraviolet [C\,IV]\,1548+1551\AA\ doublet, against photoionization parameter \cite[see e.g.][for comparison with AGN broad line regions]{1982ApJ...262..564F}. As before we show results for two representative metallicities, and at log($n_H$)=2.0 and 3.0. Given any single diagnostic line or line ratio, the interpretation (in terms of ionization parameter, metallicity, electron density and, indeed, age) is far from unique. However, given a suitable combination of lines, or better still a full spectral fit, these degeneracies can be broken, giving a reasonable estimate of the gas and stellar parameters.

A full exploration of every possible line ratio combination is beyond the scope of this work (see e.g. Senchyna et al 2017 for other diagnostics) To facilitate further work by others in this area, we make the full set of line emission tables discussed in this paper, in both the optical and ultraviolet and using BPASS v2.1 binary and single star models, available from the BPASS website\footnote{see bpass.auckland.ac.nz/v2.1/nebular}.

\bsp

\label{lastpage}


\begin{thebibliography}{99} 


\bibitem[Allen et al.(2008)]{2008ApJS..178...20A} Allen, M.~G., Groves, B.~A., Dopita, M.~A., Sutherland, R.~S., \& Kewley, L.~J.\ 2008, {\rm ApJS}, 178, 20-55 

5\bibitem[Ando et al.(2005)]{2005PhRvL..95q1101A} Ando, S., Beacom, J.~F., {\ Y\& uuml}ksel, H.\ 2005, Physical Review Letters, 95, 171101 

\bibitem[Baldwin et al.(1981)]{1981PASP...93..817B} Baldwin, A., Phillips, M.~M., \& Terlevich, R.\ 1981, {\rm PASP}, 93, 817 

\bibitem[Binette et al.(1985)]{1985ApJ...297..476B} Binette, L., Dopita, M.~A., \& Tuohy, I.~R.\ 1985, {\rm ApJ}, 297, 476 

\bibitem[Boissier \& Prantzos(2009)]{2009A&A...503..137B} Boissier, S., \& Prantzos, N.\ 2009, {\rm A\&A}, 503, 137 

\bibitem[Bresolin et al.(2005)]{2005A&A...441..981B} Bresolin, F., Schaerer, D., Gonz{\'a}lez Delgado, R.~M., \& Stasi{\'n}ska, G.\ 2005, {\rm A\&A}, 441, 981 

\bibitem[Bresolin(2007)]{2007ApJ...656..186B} Bresolin, F.\ 2007, {\rm ApJ}, 656, 186

\bibitem[Byler et al.(2017)]{2017ApJ...840...44B} Byler, N., Dalcanton, J.~J., Conroy, C., \& Johnson, B.~D.\ 2017, {\rm ApJ}, 840, 44

\bibitem[Cantiello et al.(2007)]{2007A&A...465L..29C} Cantiello, M., Yoon, S.-C., Langer, N., \& Livio, M.\ 2007, {\rm A\&A}, 465, L29   

\bibitem[Charlot \& Fall(2000)]{2000ApJ...539..718C} Charlot, S., \& Fall, S.~M.\ 2000, {\rm ApJ}, 539, 718 

\bibitem[Conroy et al.(2008)]{2008ApJ...679.1192C} Conroy, C., Shapley, A.~E., Tinker, J.~L., Santos, M.~R., \& Lemson, G.\ 2008, {\rm ApJ}, 679, 1192-1203

\bibitem[Cowie et al.(2009)]{2009ApJ...692.1476C} Cowie, L.~L., Barger, A.~J., \& Trouille, L.\ 2009, {\rm ApJ}, 692, 1476 

\bibitem[de Grijs et al.(2003)]{2003MNRAS.340..197D} de Grijs, R., Bastian, N., \& Lamers, H.~J.~G.~L.~M.\ 2003, {\rm MNRAS}, 340, 197

\bibitem[de Mink et al.(2013)]{2013ApJ...764..166D} de Mink, S.~E., Langer, N., Izzard, R.~G., Sana, H., \& de Koter, A.\ 2013, {\rm ApJ}, 764, 166

\bibitem[Dopita et al.(2016)]{2016ApSS.361...61D} Dopita, M.~A., Kewley, L.~J., Sutherland, R.~S., \& Nicholls, D.~C.\ 2016, {\rm Ap\&SS}, 361, 61

\bibitem[\protect\citeauthoryear{Duarte Puertas et al.}{2017}]{2017A&A...599A..71D} Duarte Puertas S., Vilchez J.~M., Iglesias-P{\'a}ramo J., Kehrig C., P{\'e}rez-Montero E., Rosales-Ortega F.~F., 2017, A\&A, 599, A71 

\bibitem[Duch{\^e}ne \& Kraus(2013)]{2013ARA&A..51..269D} Duch{\^e}ne, G., \& Kraus, A.\ 2013, {\rm ARA \& A}, 51, 269 

\bibitem[Eldridge \& Tout(2004)]{2004MNRAS.348..201E} Eldridge, J.~J., \& Tout, C.~A.\ 2004, {\rm MNRAS}, 348, 201 

\bibitem[Eldridge { \& } Stanway(2009)]{2009MNRAS.400.1019E} Eldridge, J.~J., {\&} Stanway, E.~R.\ 2009, {\rm MNRAS}, 400, 1019
 
\bibitem[Eldridge et al.(2011)]{2011MNRAS.414.3501E} Eldridge, J.~J., Langer, N., \& Tout, C.~A.\ 2011, {\rm MNRAS}, 414, 3501

\bibitem[Eldridge \& Stanway(2012)]{2012MNRAS.419..479E} Eldridge, J.~J., \& Stanway, E.~R.\ 2012, {\rm MNRAS}, 419, 479

\bibitem[Eldridge, Stanway et al.(2017)]{arXiv:1710.02154} Eldridge, J.~J., Stanway, E., Xiao, L., L. A. S. McClelland, G. Taylor, M. Ng, S. M. L. Greis, J. C. Bray \ 2017, PASA, 34, e058 

\bibitem[\protect\citeauthoryear{Feltre, Charlot, \& Gutkin}{2016}]{2016MNRAS.456.3354F} Feltre A., Charlot S., Gutkin J., 2016, MNRAS, 456, 3354 

\bibitem[\protect\citeauthoryear{Feltre, Charlot, \& Gutkin}{2016}]{2016MNRAS.456.3354F} Feltre A., Charlot S., Gutkin J., 2016, MNRAS, 456, 3354 

\bibitem[\protect\citeauthoryear{Ferland \& Mushotzky}{1982}]{1982ApJ...262..564F} Ferland G.~J., Mushotzky R.~F., 1982, ApJ, 262, 564

\bibitem[Ferland et al.(1998)]{1998PASP..110..761F} Ferland, G.~J., Korista, K.~T., Verner, D.~A., et al.\ 1998, {\rm PASP}, 110, 761

\bibitem[Ferland et al.(2013)]{2013RMxAA..49..137F} Ferland, G.~J., Porter, R.~L., van Hoof, P.~A.~M., et al.\ 2013, {\rm RMxAA}, 49, 137 

\bibitem[Fernandes et al.(2003)]{2003MNRAS.340...29C} Fernandes, R.~C., Le{\~a}o, J.~R.~S., \& Lacerda, R.~R.\ 2003, {\rm MNRAS}, 340, 29 

\bibitem[Galliano et al.(2017)]{2017arXiv171107434G} Galliano, F., Galametz, M., \& Jones, A.~P.\ 2017, arXiv:1711.07434 

\bibitem[\protect\citeauthoryear{Gomes et al.}{2016}]{2016A&A...586A..22G} Gomes J.~M., et al., 2016, A\&A, 586, A22 

\bibitem[\protect\citeauthoryear{Grevesse \& Noels}{1993}]{1993oee..conf...15G} Grevesse N., Noels A., 1993, oee..conf, 15 

\bibitem[Grossi et al.(2010)]{2010A&A...521A..41G} Grossi, M., Corbelli, E., Giovanardi, C., \& Magrini, L.\ 2010, {\rm A\&A}, 521, A41

\bibitem[Gusev(2014)]{2014MNRAS.442.3711G} Gusev, A.~S.\ 2014, {\rm MNRAS}, 442, 3711 

\bibitem[\protect\citeauthoryear{Gutkin, Charlot, \& Bruzual}{2016}]{2016MNRAS.462.1757G} Gutkin J., Charlot S., Bruzual G., 2016, MNRAS, 462, 1757 

\bibitem[\protect\citeauthoryear{Gutkin, Charlot, \& Bruzual}{2016}]{2016MNRAS.462.1757G} Gutkin J., Charlot S., Bruzual G., 2016, MNRAS, 462, 1757 

\bibitem[Hunt \& Hirashita(2009)]{2009A&A...507.1327H} Hunt, L.~K., \& Hirashita, H.\ 2009, {\rm A\& A}, 507, 1327 

\bibitem[\protect\citeauthoryear{Iglesias-P{\'a}ramo et al.}{2013}]{2013A&A...553L...7I} Iglesias-P{\'a}ramo J., et al., 2013, A\&A, 553, L7 

\bibitem[\protect\citeauthoryear{Iglesias-P{\'a}ramo et al.}{2016}]{2016ApJ...826...71I} Iglesias-P{\'a}ramo J., et al., 2016, ApJ, 826, 71 

\bibitem[Izotov et al.(2006)]{2006A&A...448..955I} Izotov, Y.~I., Stasi{\'n}ska, G., Meynet, G., Guseva, N.~G., \& Thuan, T.~X.\ 2006, {\rm A\&A}, 448, 955

\bibitem[Kennicutt(1984)]{1984ApJ...287..116K} Kennicutt, R.~C., Jr.\ 1984, {\rm ApJ}, 287, 116

\bibitem[Kennicutt et al.(2003)]{2003ApJ...591..801K} Kennicutt, R.~C., Jr., Bresolin, F., \& Garnett, D.~R.\ 2003, {\rm ApJ}, 591, 801 

\bibitem[Kewley et al.(2001)]{2001ApJ...556..121K} Kewley, L.~J., Dopita, M.~A., Sutherland, R.~S., Heisler, C.~A., \& Trevena, J.\ 2001, {\rm ApJ}, 556, 121 

\bibitem[Kewley \& Dopita(2002)]{2002ApJS..142...35K} Kewley, L.~J., \& Dopita, M.~A.\ 2002, {\rm ApJS}, 142, 35 

\bibitem[Kewley \& Ellison(2008)]{2008ApJ...681.1183K} Kewley, L.~J., \& Ellison, S.~L.\ 2008, {\rm ApJ}, 681, 1183-1204

\bibitem[Kuncarayakti et al.(2013)]{2013AJ....146...30K} Kuncarayakti, H., Doi, M., Aldering, G., et al.\ 2013, {\rm aj}, 146, 30
 
\bibitem[Leitherer et al.(1999)]{1999ApJS..123....3L} Leitherer, C., Schaerer, D., Goldader, J.~D., et al.\ 1999, {\rm ApJS}, 123, 3 

\bibitem[Levesque et al.(2010)]{2010AJ....139..712L} Levesque, E.~M., Kewley, L.~J., \& Larson, K.~L.\ 2010, {\rm AJ}, 139, 712 

\bibitem[\protect\citeauthoryear{Mainali et al.}{2017}]{2017ApJ...836L..14M} Mainali R., Kollmeier J.~A., Stark D.~P., Simcoe R.~A., Walth G., Newman A.~B., Miller D.~R., 2017, ApJ, 836, L14 

\bibitem[\protect\citeauthoryear{Mainali et al.}{2017}]{2017ApJ...836L..14M} Mainali R., Kollmeier J.~A., Stark D.~P., Simcoe R.~A., Walth G., Newman A.~B., Miller D.~R., 2017, ApJ, 836, L14 

\bibitem[Marino et al.(2013)]{2013AA...559A.114M} Marino, R.~A., Rosales-Ortega, F.~F., S{\'a}nchez, S.~F., et al.\ 2013, {\rm A\&A}, 559, A114 

\bibitem[Mathis(1990)]{1990ARA&A..28...37M} Mathis, J.~S.\ 1990, {\rm ARA\&A}, 28, 37 

\bibitem[McCall et al.(1985)]{1985ApJS...57....1M} McCall, M.~L., Rybski, P.~M., \& Shields, G.~A.\ 1985, {\rm ApJS}, 57, 1 

\bibitem[Oey \& Shields(2000)]{2000ApJ...539..687O} Oey, M.~S., \& Shields, J.~C.\ 2000, {\rm ApJ}, 539, 687

\bibitem[Oey et al.(2000)]{2000ApJS..128..511O} Oey, M.~S., Dopita, M.~A., Shields, J.~C., \& Smith, R.~C.\ 2000, {\rm ApJS}, 128, 511 

\bibitem[\protect\citeauthoryear{Oey et al.}{2003}]{2003AJ....126.2317O} Oey M.~S., Parker J.~S., Mikles V.~J., Zhang X., 2003, AJ, 126, 2317 


\bibitem[\protect\citeauthoryear{Peimbert, Peimbert, \& Delgado-Inglada}{2017}]{2017PASP..129h2001P} Peimbert M., Peimbert A., Delgado-Inglada G., 2017, PASP, 129, 082001 

\bibitem[Pettini \& Pagel(2004)]{2004MNRAS.348L..59P} Pettini, M., \& Pagel, B.~E.~J.\ 2004, {\rm MNRAS}, 348, L59 

\bibitem[Pilyugin et al.(2010)]{2010ApJ...720.1738P} Pilyugin, L.~S., V{\'{\i}}lchez, J.~M., \& Thuan, T.~X.\ 2010, {\rm ApJ}, 720, 1738 

\bibitem[Pilyugin \& Grebel(2016)]{2016MNRAS.457.3678P} Pilyugin, L.~S., \& Grebel, E.~K.\ 2016, {\rm MNRAS}, 457, 3678

\bibitem[Proxauf et al.(2014)]{2014A&A...561A..10P} Proxauf, B., {\"O}ttl, S., \& Kimeswenger, S.\ 2014, {\rm A\&A}, 561, A10

\bibitem[Rigby \& Rieke(2004)]{2004ApJ...606..237R} Rigby, J.~R., \& Rieke, G.~H.\ 2004, {\rm ApJ}, 606, 237  

\bibitem[S{\'a}nchez et al.(2015)]{2015A&A...574A..47S} S{\'a}nchez, S.~F., P{\'e}rez, E., Rosales-Ortega, F.~F., et al.\ 2015, {\rm A\&A}, 574, A47 

\bibitem[S{\'a}nchez et al.(2012)]{2012A&A...546A...2S} S{\'a}nchez, S.~F., Rosales-Ortega, F.~F., Marino, R.~A., et al.\ 2012, {\rm A\&A}, 546, A2 

\bibitem[\protect\citeauthoryear{Senchyna et al.}{2017}]{2017MNRAS.472.2608S} Senchyna P., et al., 2017, MNRAS, 472, 2608 

\bibitem[Shields(1990)]{1990ARA&A..28..525S} Shields, G.~A.\ 1990, {\rm ARA\&A}, 28, 525

\bibitem[\protect\citeauthoryear{Spitzer}{1948}]{1948ApJ...107....6S} Spitzer L., Jr., 1948, ApJ, 107, 6 

\bibitem[Stanway et al.(2014)]{2014MNRAS.444.3466S} Stanway, E.~R., Eldridge, J.~J., Greis, S.~M.~L., et al.\ 2014, {\rm MNRAS}, 444, 3466

\bibitem[Stanway et al.(2016)]{2016MNRAS.456..485S} Stanway, E.~R., Eldridge, J.~J., \& Becker, G.~D.\ 2016, {\rm MNRAS}, 456, 485 

\bibitem[Stasi{\'n}ska(2006)]{2006A&A...454L.127S} Stasi{\'n}ska, G.\ 2006, {\rm A\&A}, 454, L127

\bibitem[Steidel et al.(2014)]{2014ApJ...795..165S} Steidel, C.~C., Rudie, G.~C., Strom, A.~L., et al.\ 2014, {\rm ApJ}, 795, 165 

\bibitem[\protect\citeauthoryear{Steidel et al.}{2014}]{2014ApJ...795..165S} Steidel C.~C., et al., 2014, ApJ, 795, 165 

\bibitem[Steidel et al.(2016)]{2016ApJ...826..159S} Steidel, C.~C., Strom, A.~L., Pettini, M., et al.\ 2016, {\rm ApJ}, 826, 159 

\bibitem[\protect\citeauthoryear{Strom et al.}{2017}]{2017ApJ...836..164S} Strom A.~L., Steidel C.~C., Rudie G.~C., Trainor R.~F., Pettini M., Reddy N.~A., 2017, ApJ, 836, 164 

\bibitem[\protect\citeauthoryear{Strom et al.}{2017}]{2017arXiv171108820S} Strom A.~L., Steidel C.~C., Rudie G.~C., Trainor R.~F., Pettini M., 2017, arXiv, arXiv:1711.08820 

\bibitem[Sutherland \& Dopita(1993)]{1993ApJS...88..253S} Sutherland, R.~S., \& Dopita, M.~A.\ 1993, {\rm ApJS}, 88, 253 

\bibitem[V{\'a}zquez \& Leitherer(2005)]{2005ApJ...621..695V} V{\'a}zquez, G.~A., \& Leitherer, C.\ 2005, {\rm ApJ}, 621, 695
 
\bibitem[van Zee et al.(1995)]{1995AJ....109..990V} van Zee, L., Haynes, M.~P., \& Giovanelli, R.\ 1995, {\rm AJ}, 109, 990

\bibitem[van Zee et al.(1996)]{1996ibms.conf..563V} van Zee, L., Haynes, M.~P., Salzer, J.~J., \& Broeils, A.~H.\ 1996, The Interplay Between Massive Star Formation, the ISM and Galaxy Evolution, 563
 
\bibitem[van Zee et al.(1998)]{1998AJ....116.2805V} van Zee, L., Salzer, J.~J., Haynes, M.~P., O'Donoghue, A.~A., \& Balonek, T.~J.\ 1998, {\rm AJ}, 116, 2805 

\bibitem[van Zee \& Haynes(2006)]{2006ApJ...636..214V} van Zee, L., \& Haynes, M.~P.\ 2006, {\rm ApJ}, 636, 214  

\bibitem[Voges \& Walterbos(2006)]{2006ApJ...644L..29V} Voges, E.~S., \& Walterbos, R.~A.~M.\ 2006, {\rm ApJ}, 644, L29 

\bibitem[Zhang et al.(2015)]{2015MNRAS.447L..21Z} Zhang, F., Li, L., Cheng, L., et al.\ 2015, {\rm MNRAS}, 447, L21 





\end{thebibliography}
\end{document}